\documentclass[aps,prl,twocolumn,showpacs,superscriptaddress,groupedaddress,acknowledgment]{revtex4-1}  
\usepackage{dcolumn}   
\usepackage{bm}        

\usepackage{hyperref}
\hypersetup{
}

\usepackage{amsmath}
\usepackage{braket}
\usepackage{amsfonts}
\usepackage{graphicx}
\usepackage{amsthm}
\usepackage{color}
\usepackage{setspace}
\usepackage{amssymb}
\usepackage{latexsym}
\usepackage{here,comment}

\theoremstyle{definition}

\newtheorem*{theorem*}{Theorem}

\newtheorem*{definition*}{Definition}

\newcommand{\av}[1]{\overline{#1}}
\newcommand{\mc}[1]{\mathcal{#1}}
\newcommand{\mr}[1]{\mathrm{#1}}

\newcommand{\lrl}[1]{\left[ #1 \right]}

\newcommand{\fracpd}[2]{\frac{\partial #1 }{\partial #2 }}

\newcommand{\aln}[1]{
\begin{align}
#1
\end{align}
}

\newcommand{\ra}{\rightarrow}

\newcommand{\Tr}{\mr{Tr}}

\newcommand{\hb}{\hat{b}}

\begin{document}
\title{
Non-Hermitian Many-Body Localization
}
\date{\today}
\author{Ryusuke Hamazaki$^1$, Kohei Kawabata$^1$, and Masahito Ueda$^{1,2}$}
\affiliation{
$^1$Department of Physics, University of Tokyo, 7-3-1 Hongo, Bunkyo-ku, Tokyo 113-0033, Japan\\
$^2$RIKEN Center for Emergent Matter Science (CEMS), Wako 351-0198, Japan
}

\begin{abstract}
Many-body localization is shown to suppress imaginary parts of complex eigenenergies for general non-Hermitian Hamiltonians having time-reversal symmetry. 
We demonstrate that a real-complex transition, which we conjecture occurs upon many-body localization, profoundly affects the dynamical stability of non-Hermitian interacting systems with asymmetric hopping that respect time-reversal symmetry.
Moreover, the real-complex transition is shown to be absent in non-Hermitian many-body systems with gain and/or loss that breaks time-reversal symmetry, even though the many-body localization transition still persists.
\end{abstract}

\maketitle

\paragraph{Introduction.}
The reality of eigenenergies of a Hamiltonian is closely related to the dynamical stability.
Even without Hermiticity, certain classes of non-Hermitian Hamiltonians are known to have real eigenenergies.
For example, a real-complex transition of eigenenergies of non-Hermitian systems featuring parity-time (PT) symmetry~\cite{Bender98,Bender02,Makris08,Graefe08,Musslimani08,Longhi09,Klaiman08,Lin11,Malzard15,Kawabata18P,Jin13,Tripathi16,Ashida17,Lourenco18,Nakagawa18} has attracted growing interest motivated by their experimental realization~\cite{Guo09,Ruter10,Regensburger12,Peng14,Feng14,Hodaei14,Zhen15,Zeuner15,Poli15,Li19,Doppler16,Weimann17,Hodaei17,Xiao17,Chen17,Zhou18,Bandres18,Konotop16,ElGanainy18}.
Another important class exhibiting a real-complex transition is non-Hermitian systems with disorder and time-reversal symmetry (TRS).
Hatano and Nelson~\cite{Hatano96,Hatano97,Hatano98,Nelson98} investigated a single-particle disordered model with asymmetric hopping, and found that real eigenenergies become complex when the Anderson localization is destroyed by strong non-Hermiticity.
However, it is highly nontrivial whether a real-complex transition due to localization and TRS found in noninteracting models~\cite{Efetov97,Feinberg97,Goldsheid98,Markum99,Feinberg99,Longhi15,Longhi15R} persists in non-Hermitian many-body systems.
Previous results~\cite{Hatano97,Kim01} are not conclusive because the many-body localization (MBL)~\cite{Basko06,Zunidaric08,Pal10,Gogolin11,Bardarson12,Iyer13,Huse13,Serbyn13,Huse14,Nandkishore14,Bera15,Ponte15,Schreiber15,Luitz15,Smith16,Choi16,Imbrie16D,Imbrie16O} was not well established at that time.
This problem is also relevant to the depinning transition  in type-II superconductors~\cite{Hatano96,Hatano97,Hatano98}.

Such non-Hermitian setups are relevant for  continuously measured quantum many-body systems.
Indeed, non-Hermitian dynamics is justified for individual quantum trajectories (i.e., pure states where measurement outcomes are postselected) with no quantum jumps~\cite{Daley14}.
It is nontrivial whether disorder, which prevents the system from thermalizing in Hermitian systems, affects the dynamics of such open systems.
The non-Hermitian treatment of disordered open systems describes physics different from the master-equation approach~\cite{Fischer16,Levy16,Medvedyeva16I,Luschen16,Nieuwenburg17},  where dissipative outcomes are averaged out.

\begin{figure}
\begin{center}
\includegraphics[width=\linewidth]{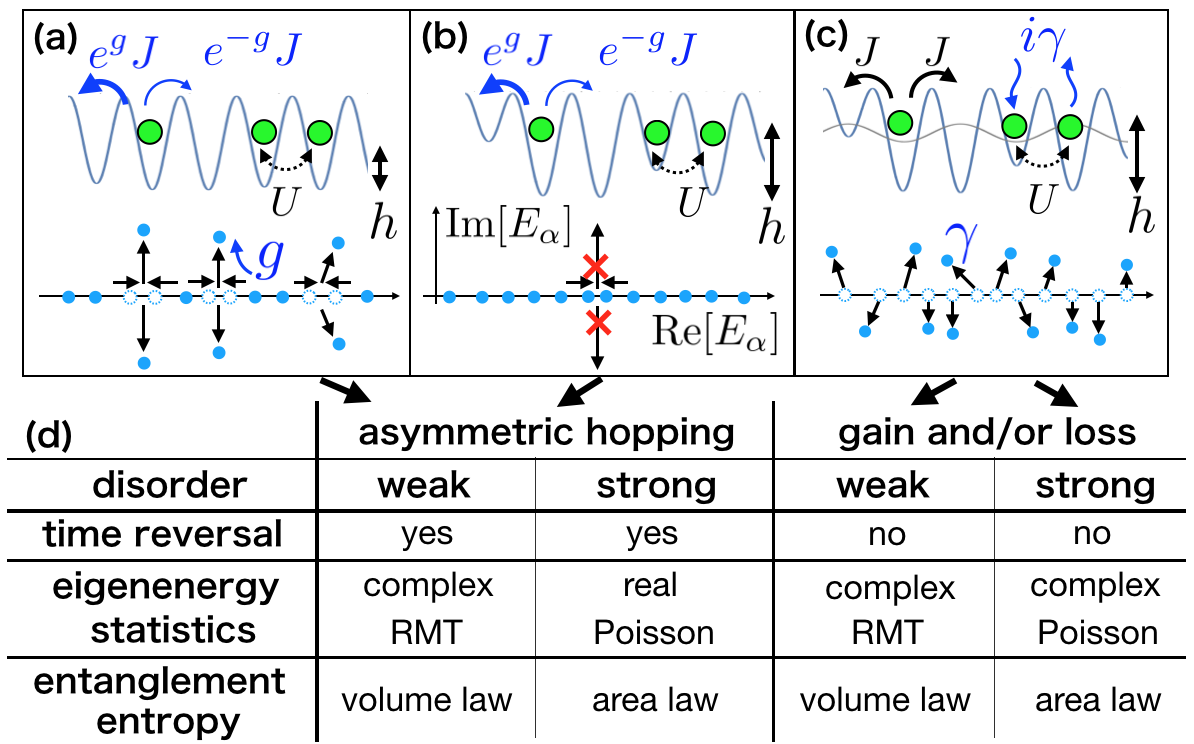}
\caption{
(a) Weakly disordered models with asymmetric hopping.
Two close eigenenergies in the real axis coalesce by a non-Hermitian perturbation and become complex-conjugate pairs
 due to  delocalization of eigenstates and TRS.
(b) Strongly disordered models with asymmetric hopping.
Coalescence of eigenenergies due to perturbation is prohibited by  localization.
(c) Disordered models with gain and/or loss. Eigenenergies acquire nonzero imaginary parts without coalescence due to the absence of TRS, irrespective of the presence of localization.
(d) Eigenenergy statistics (reality and level-spacing statistics) and entanglement entropy of eigenstates for an asymmetric-hopping model (Eq.~\eqref{model}) and a gain-loss model (Eq.~\eqref{modelwosym}) with varying disorder strength.
}
\label{fig1}
\end{center}
\end{figure}

In this Letter, we show that localization suppresses imaginary parts of many-body eigenenergies for non-Hermitian interacting Hamiltonians having TRS and can induce a real-complex transition.
Investigating disordered interacting particles with asymmetric hopping, we find a  real-complex transition at which almost all eigenenergies become real as disorder is increased (Fig.~\ref{fig1}(a) and (b)).
In the real-eigenenergy phase, energy absorption/emission disappears despite  non-Hermiticity, demonstrating that quantum states become dynamically stable.
We show that a non-Hermitian MBL occurs close to the real-complex phase transition point.
We conjecture that the two transitions coincide in the thermodynamic limit on the basis of an analytical discussion about the stability of eigenstates.
We also demonstrate that real-complex transitions are absent in non-Hermitian  systems with gain and/or loss that break TRS, though non-Hermitian MBL still survives (see Fig.~\ref{fig1}(c)).
The results are summarized in Fig.~\ref{fig1}(d).

\paragraph{Localization suppresses complex eigenenergies.}
We first show that localization suppresses imaginary parts of many-body eigenenergies for generic non-Hermitian Hamiltonians having TRS.
Let us decompose the local Hamiltonian into the unperturbed part and the non-Hermitian perturbation as $\hat{H}=\hat{H}_0+\hat{V}_\mr{NH}$ ($\hat{H}_0$ can be  non-Hermitian).
We consider a set of real eigenenergies $\{\mc{E}_a\}$ of $\hat{H}_\mr{0}$ and  the corresponding right (left) eigenstates $\ket{\mc{E}_a^R}\:(\ket{\mc{E}_a^L})$, which satisfy $\braket{\mc{E}_a^L|\mc{E}_b^R}=\delta_{ab}$~\cite{Brody13}.

To see why TRS is crucial for the reality of eigenenergies, we consider the first-order energy shift 
${\braket{\mc{E}_a^L|\hat{V}_\mr{NH}|\mc{E}_a^R}}$, which is in general complex (Fig.~\ref{fig1}(c)) but becomes real when TRS is imposed~\cite{TRS-NHMBL}.
On the other hand, even with TRS,  eigenenergies can coalesce and acquire imaginary parts~\cite{Bender98} for sufficiently large non-Hermiticity $\hat{V}_{\rm NH}$.
The coalescence occurs due to mixing of eigenstates, resulting from higher-order perturbations.
As detailed later, while the mixing of two adjacent eigenstates occurs for delocalized eigenstates, it does not for MBL eigenstates~\cite{Serbyn15}.
In fact, for delocalized phases, many complex eigenenergies appear via  coalescence of excited eigenstates~\cite{exception-NHMBL} (Fig.~\ref{fig1}(a)), whereas such coalescence is suppressed for MBL phases (Fig.~\ref{fig1}(b)).

\paragraph{Model.}
We consider hard-core bosons with  asymmetric hoppings on a one-dimensional lattice subject to the periodic boundary condition:
\aln{\label{model}
\hat{H}=\sum_{i=1}^L\lrl{-J(e^{-g}\hb_{i+1}^\dag\hb_i+e^{g}\hb_{i}^\dag\hb_{i+1})+U\hat{n}_i\hat{n}_{i+1}+h_i\hat{n}_i}.
}
Here, $\hat{n}_i=\hat{b}_i^\dag\hat{b}_i$ is the particle-number operator at site $i$ with the annihilation operator $\hat{b}_i$ of a hard-core boson, $g$ controls non-Hermiticity, and $h_i$ is randomly chosen from $[-h,h]$.
This model has TRS (complex conjugation) and can be realized experimentally in continuously measured ultracold atomic systems, where strong disorder has been realized~\cite{Choi16} and asymmetric hopping can also be implemented~\cite{Gong18T}.
In the following, we assume $J=1$, $U=2$, $g=0.1$, and a fixed-number subspace with $M=L/2$ (half filling).
See Supplemental Material~\cite{footNHMBL1} for other parameters and a Bose-Hubbard model with asymmetric hopping~\cite{Kim01}.

\begin{figure}
\begin{center}
\includegraphics[width=\linewidth]{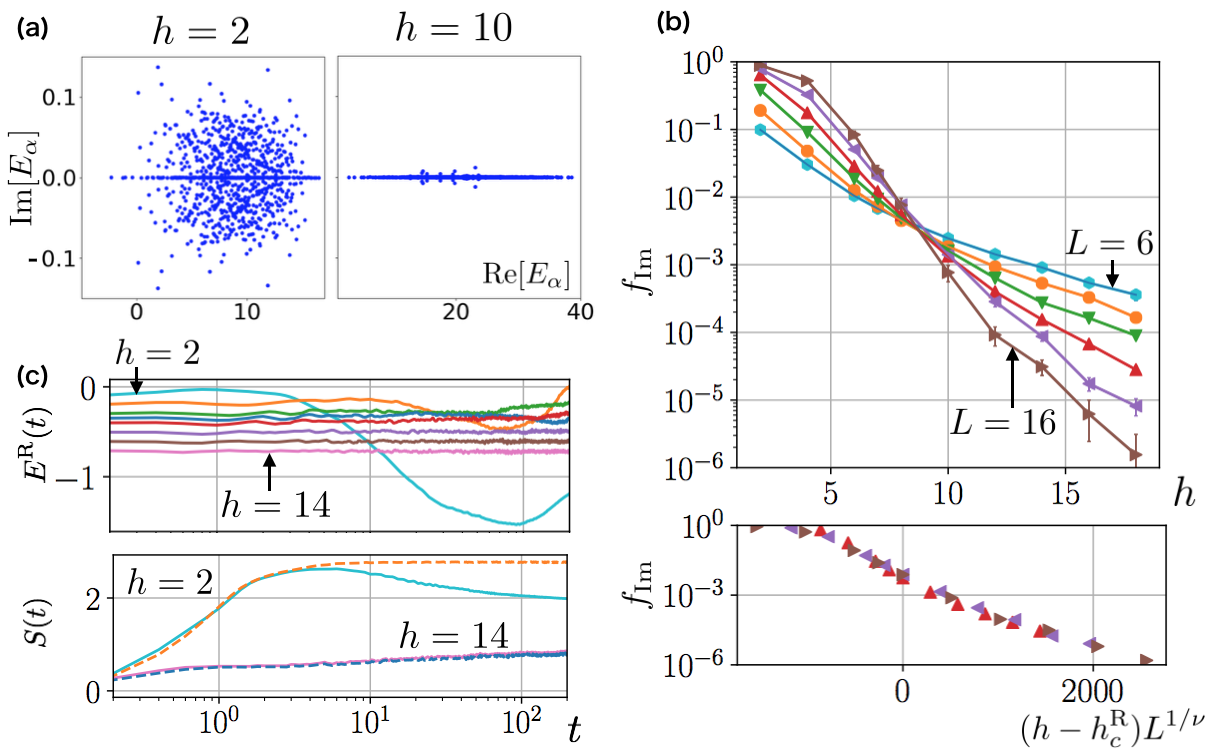}
\caption{
(a) Eigenenergies of the non-Hermitian Hamiltonian~\eqref{model} with $L=12$ for different values of $h$.
As $h$ increases, the fraction of complex eigenenergies  decreases.
(b) (top) Dependence of $f_\mr{Im}$ on $h$ for $L=6, 8, 10, 12, 14$, and $16$~\cite{average-NHMBL}.
As $L$ increases, $f_\mr{Im}$ increases for $h\lesssim h_c^\mr{R} (\simeq 8)$
and decreases for $h\gtrsim h_c^\mr{R}$.
(bottom) The critical scaling collapse of $f_\mr{Im}$ is found as a function of $(h-h_c^\mr{R})L^{1/\nu}$, where we use $h_c^\mr{R}=8.0$ and $\nu=0.5$.
The value $f_\mr{Im}$ is determined by setting a cutoff of the imaginary part $C=10^{-13}$, where pure complex eigenvalues  ($|\mr{Im}E_\alpha|\gg C$) and machine errors ($|\mr{Im}E_\alpha|\ll C$) are clearly separated.
(c) (top) Time evolution of the real part of the energy of the system for $h=2, 4, 6, 7, 8,10,12$, and $14$, where the
initial energy decreases with increasing $h$.
The energy changes in time for $h\lesssim h_c^\mr{R}$ but stays almost constant for $h\gtrsim h_c^\mr{R}$.
(bottom) The dynamics of the half-chain entanglement entropy for $g=0.1$ (solid) and $g=0$ (dotted) with different $h$.
For $h=2$, $S(t)$ first exhibits a linear growth for both $g$ but decreases in the long time only for $g=0.1$.
For $h=14$, $S(t)$ exhibits a logarithmic growth for both $g=0$ and 0.1 even in the long time.
We take $\ket{\psi_0}=\ket{1010\cdots}$ as an initial state.
}
\label{fig2}
\end{center}
\end{figure}

\paragraph{Real-complex transition.}
Fig.~\ref{fig2}(a) shows the eigenenergies of the Hamiltonian in Eq. \eqref{model}.
The spectrum is symmetric with respect to the real axis due to TRS.
The fraction of eigenenergies with nonzero imaginary parts decreases for large $h$.

 As a quantitative indicator of the reality of eigenenergies, we define
$
f_\mr{Im}=\av{\frac{D_\mr{Im}}{D}},
$
where $D_\mr{Im}$ is the number of eigenenergies with nonzero imaginary parts, $D$ is the total number of eigenenergies, and the overline denotes the disorder average~\cite{density-NHMBL}.
This quantity measures the fraction of eigenenergies with nonzero imaginary parts.

Figure~\ref{fig2}(b) shows the $h$-dependence of $f_\mr{Im}$ for different values of  $L$.
As the system size increases, $f_\mr{Im}$ increases for $h\lesssim h_c^\mr{R}\simeq 8$ and decreases for $h\gtrsim h_c^\mr{R}$.
We also confirm the critical scaling collapse of $f_\mr{Im}$  as a function of $(h-h_c^\mr{R})L^{1/\nu}$, indicating a real-complex phase transition of many-body eigenenergies at $h=h_c^\mr{R}$ in the thermodynamic limit ($L\ra \infty$):
almost all eigenenergies are complex for $h<h_c^\mr{R}$ and real for $h>h_c^\mr{R}$.
This real-complex transition is identified by the statistics of the spectrum (i.e., the average over disorder and eigenstates) in the thermodynamic limit (the maximum imaginary part among all eigenenergies also exhibit the same transition~\cite{footNHMBL1}).
This is in contrast to the conventional PT transition, which  is identified to be the point at which eigenstates coalesce  without an average over many eigenstates.

Our results show that highly-excited eigenenergies suddenly become almost real with increasing disorder.
This change significantly affects the dynamical stability of the system.
Figure~\ref{fig2}(c) shows the evolution of the real part of energy $E^\mr{R}(t)=\mr{Re}[\av{\braket{\psi(t)|\hat{H}|\psi(t)}}]$.
Here $\ket{\psi(t)}=\frac{e^{-i\hat{H}t}\ket{\psi_0}}{||e^{-i\hat{H}t}\ket{\psi_0}||}$ corresponds to the state at time $t$ for a quantum trajectory with no quantum jumps, which is microscopically justified for continuously measured quantum systems~\cite{Daley14}.
While the energy should be conserved  for any $h$ in the Hermitian Hamiltonian, it is sensitive to small non-Hermiticity for $h \lesssim h_{c}^\mr{R}$.
This sensitivity is due to the delocalized eigenstates with nonzero imaginary parts and signals the dynamical instability.
On the other hand, the system is stable for $h \gtrsim h_{c}^{\rm R}$, where the energy is conserved except for negligibly small oscillations  since almost all eigenenergies are real.
Moreover, the real-complex transition is relevant for the dynamics of other quantities~\cite{footNHMBL1}, e.g., the half-chain entanglement entropy $S(t)$ as shown in Fig.~\ref{fig2}(c).

While the non-Hermitian dynamics relevant for continuously measured systems is  different from isolated quantum systems, our results indicate that dynamics and stationary states are distinct in the two phases.
This provides a first step toward understanding statistical mechanics of such open quantum systems.
For example, recurrence does not occur  in the complex-eigenenergy phase but it does in the real-eigenenergy phase~\cite{therm-NHMBL}.

\paragraph{Non-Hermitian many-body localization.}

\begin{figure}
\begin{center}
\includegraphics[width=\linewidth]{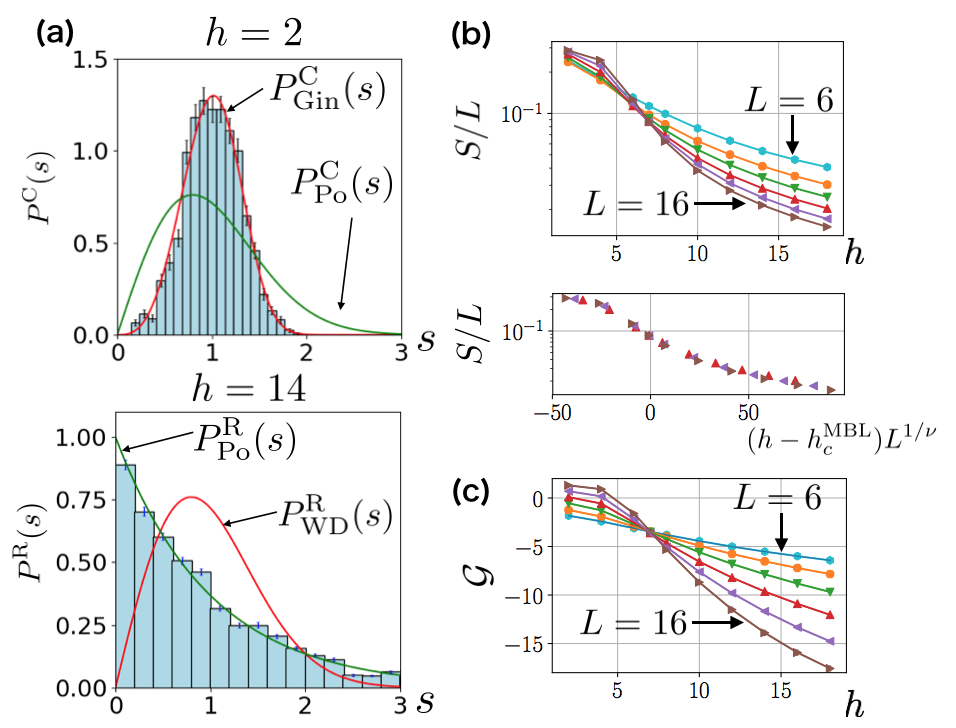}
\caption{
(a) Nearest-level-spacing distribution of (unfolded) eigenenergies on the complex plane for $h=2$ and that of eigenenergies on the real axis for $h=14$ (a single disorder realization with $L=16$).
For $h=2$, the distribution is a Ginibre distribution $P_\mr{Gin}^\mr{C}(s)$ rather than a Poisson distribution $P_\mr{Po}^\mr{C}(s)=\frac{\pi s}{2}e^{-\frac{\pi}{4}s^2}$ on the complex plane~\cite{Haake}.
For $h=14$, the distribution is a Poisson distribution on the real axis $P_\mr{Po}^\mr{R}(s)=e^{-s}$ rather than the Wigner-Dyson distribution $P^\mr{R}_\mr{WD}(s)=\frac{\pi s}{2}e^{-\frac{\pi}{4}s^2}$.
Statistics are taken from eigenenergies lying within $\pm 10\% $ of the real and imaginary parts from the middle of the spectrum.
(b) (top) Half-chain entanglement entropy $S/L$ obtained by averaging $S_{\alpha}/L$ over disorder~\cite{average-NHMBL} and eigenstates whose eigenenergies lie within $\pm 2\% $ from the middle of the real part of the spectrum.
With increasing $h$, $S/L$ shows a crossover from the volume to area law.
(bottom) The critical scaling collapse is confirmed as a function of $(h-h_c^\mr{MBL})L^{1/\nu}$, where we use $h_c^\mr{MBL}=7.1$ and $\nu=1.3$.
(c) Stability of eigenstates $\mc{G}$ for different values of $L$ and with $\hat{V}_\mr{NH}=\hat{b}_i^\dag\hat{b}_{i+1}$~\cite{average-NHMBL}.
With increasing $h$, $\mc{G}$ changes from $\sim \alpha L$ to $\sim -\beta L$ ($\alpha,\beta>0$) at $h_c^\mr{MBL}\simeq 7\pm 1$.
}
\label{fig3}
\end{center}
\end{figure}

We next discuss a delocalization-localization transition of the system.
While it is nontrivial how to characterize MBL in non-Hermitian systems, we show that some of the known machinery to characterize Hermitian MBL can be generalized to the non-Hermitian regime.
We first consider the nearest-level-spacing distribution of (unfolded) eigenenergies~\cite{Haake} on the complex plane for small $h$ and  on the real axis for large $h$ in Fig.~\ref{fig3}(a)~\cite{rare-NHMBL}.
Here, nearest-level spacings (before unfolding) for an eigenenergy $E_\alpha$ on the complex plane is given by the minimum distance $\min_\beta|E_\alpha-E_\beta|$.
For weak disorder, we numerically find that the distribution is a Ginibre distribution $P^\mr{C}_\mr{Gin}(s)=cp(cs)$, which describes an ensemble for non-Hermitian Gaussian random matrices~\cite{Haake,Schomerus17}.
Here $p(s)=\lim_{N\ra\infty}\lrl{\prod_{n=1}^{N-1}e_n(s^2)e^{-s^2}}  \sum_{n=1}^{N-1}\frac{2s^{2n+1}}{n!e_n(s^2)}
$ with $e_n(x)=\sum_{m=0}^n\frac{x^m}{m!}$ and $c=\int_0^\infty dssp(s)=1.1429\cdots$~\cite{Mehta,Haake,Markum99}.
Our result demonstrates a non-Hermitian generalization of the conjecture~\cite{Haake,Bohigas84,Pal10} that level-spacing statistics of delocalized Hermitian systems obey the Wigner-Dyson distribution.
This Ginibre-type phase is a novel delocalized phase that we conjecture emerges for generic non-Hermitian interacting systems.
For strong $h$, the level-spacing distribution becomes Poissonian on the real axis $P_\mr{Po}^\mr{R}(s)=e^{-s}$, indicating that neighboring eigenstates become uncorrelated.

We also consider the half-chain entanglement entropy for the right eigenstates $\ket{E_\alpha^R}$~\cite{left-NHMBL}:
$
S_{\alpha}=\Tr_{L/2}[\ket{E_\alpha^R}\bra{E_\alpha^R}],
$
where $\ket{E_{\alpha}^{\rm R}}$ is normalized to unity: $\braket{E_\alpha^R|E_\alpha^R}=1$~\cite{entanglement-NHMBL}.
In Fig.~\ref{fig3}(b), we show the $L$-dependence of $S_{\alpha}/L$ averaged over the eigenstates around the middle of the spectrum~\cite{middle-NHMBL}.
This figure shows that the entanglement entropy exhibits a crossover from the volume to area law as we enter the MBL phase around $h\simeq h_c^\mr{MBL}$.
We also confirm the critical scaling collapse as a function of $(h-h_c^\mr{MBL})L^{1/\nu}$.
These results show that the delocalized and MBL phases can be distinguished by the entanglement entropy even in non-Hermitian systems, in a manner similar to the  Hermitian case~\cite{Bauer13,Luitz15}.
For the delocalized phase, eigenstates of Hermitian/non-Hermitian systems are well described by those of Hermitian/non-Hermitian random matrices, which are regarded as random eigenvectors satisfying the volume law of entanglement.
For the localized phase,
eigenstates are characterized by quasilocal conserved quantities for both Hermitian and non-Hermitian cases (because non-Hermiticity does not significantly affect eigenstates for the localized phase as discussed below) and the area law of entanglement holds.

As discussed above, the stability of eigenstates of $\hat{H}_0$ under local perturbations $\hat{V}_\mr{NH}$ is important for the suppression of complex eigenenergies.
We consider
$
\mc{G}=\av{\ln\frac{|\braket{\mc{E}^L_{a+1}|\hat{V}_\mr{NH}|\mc{E}^R_{a}}|}{|\mc{E}'_{a+1}-\mc{E}'_a|}}
$
as a measure of the stability of the eigenstates.
The factor in the logarithm is nothing but the convergence factor for the standard perturbation theory of quantum mechanics~\cite{footNHMBL1}.
Here we only consider $\mc{E}'_a=\mc{E}_a+{\braket{\mc{E}_a^L|\hat{V}_\mr{NH}|\mc{E}_a^R}}$ that stays real and the labels of eigenstates $a$'s are taken such that $\mc{E}'_1\leq\mc{E}'_2\leq\cdots$.
This is a non-Hermitian generalization of the Hermitian counterpart introduced in Ref.~\cite{Serbyn15}, where $\fracpd{\mc{G}}{L}>0$ for the delocalized phase because level spacings (denominator) become much smaller than typical off-diagonal matrix elements (numerator)~\cite{Srednicki99,Khatami13,Beugeling15,Mondaini17,Hamazaki18A}, and $\fracpd{\mc{G}}{L}<0$ for the localized phase because   quasilocal conserved quantities drastically suppress the off-diagonal matrix elements~\cite{Serbyn13,Huse14,Imbrie16D,Imbrie16O}.

In Fig.~\ref{fig3}(c), we show the $h$-dependence of $\mc{G}$ for the non-Hermitian setting.
We find $\mc{G}\sim \alpha L\:(\alpha>0)$ for the delocalized phase and $\mc{G}\sim -\beta  L\:(\beta>0)$ for the localized phase, which is similar to the Hermitian case because of the same reason as the entanglement entropy.
From the point at which $\mc{G}$ is independent of $L$, we determine  the non-Hermitian MBL transition point to be $h_c^\mr{MBL}\simeq 7\pm 1$, which is close to $h_c^\mr{R}$.

We conjecture that the real-complex transition point $h_c^\mr{R}$ and the MBL transition point $h_c^\mr{MBL}$ coincide in the thermodynamic limit.
Indeed, from the above discussions of stability $\mc{G}$, the coalescence of adjacent eigenstates is suppressed due to the non-Hermitian MBL.
Furthermore, for our model in Eq.~\eqref{model}, we can show that the coalescence process  is suppressed even for non-adjacent eigenstates and hence the entirely real spectra are realized, as detailed in Supplemental Material~\cite{footNHMBL1}.
Note that in our numerics up to $L=16$, the two transition points are close ($h_c^\mr{R}\simeq 8\pm 1$ and $h_c^\mr{MBL}\simeq 7\pm 1$ for $g=0.1$) but slightly different.
On the basis of the analytical discussions of the stability, we conjecture that the small deviation is attributed to the finite-size effects.

\begin{figure}
\begin{center}
\includegraphics[width=\linewidth]{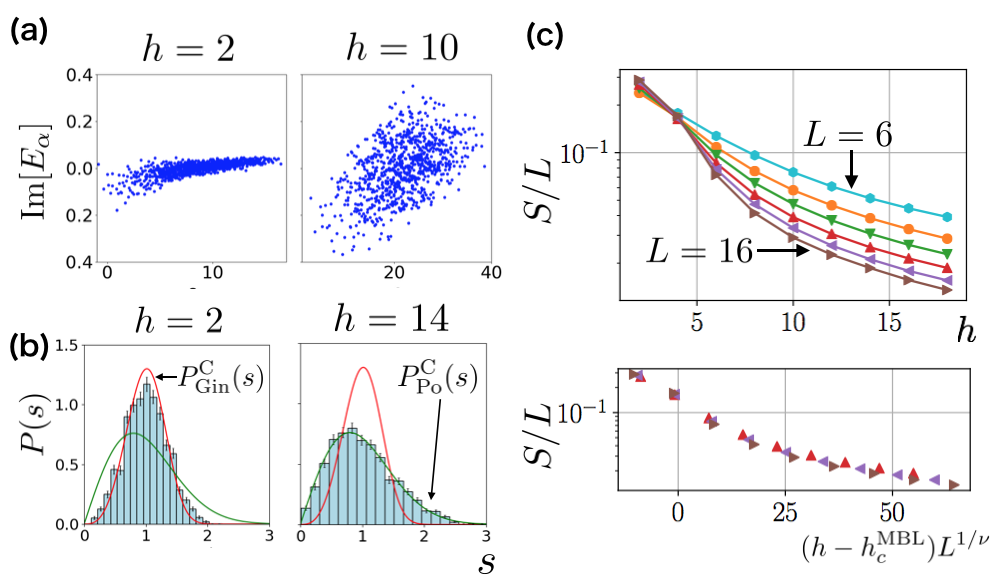}
\caption{
(a) Complex spectrum of the non-Hermitian Hamiltonian in Eq.~\eqref{modelwosym} with $\gamma=0.1$ for a single disorder realization.
The eigenenergies have nonzero imaginary parts irrespective of the disorder strength $h$.
(b) Nearest-level-spacing distributions on the complex plane for  eigenenergies having nonzero imaginary parts.
The distribution obeys random-matrix universality slightly different from the Ginibre distribution $P_\mr{Gin}^\mr{C}(s)$~\cite{hamazaki19T} for $h=2$ and
  a Poisson distribution $P_\mr{Po}^\mr{C}(s)$ for $h=14$.
The statistics are taken from the eigenstates of single disorder realization ($L=16$) whose eigenenergies lie within $\pm 10\% $ of real and imaginary parts from the
middle of the spectrum.
(c) (top) System-size dependence  of $S/L$ averaged over the eigenstates from the middle ($\pm 2\% $) of the real part of the spectrum for different values of $L$ with $\gamma=0.1$~\cite{average-NHMBL}.
The entanglement entropy decreases with increasing $L$ when the many-body localization sets in.
(bottom) The critical scaling collapse of $S/L$ is found as a function of $(h-h_c^\mr{MBL})L^{1/\nu}$, where we use $h_c^\mr{MBL}=4.2$ and $\nu=1.8$.
}
\label{fig4}
\end{center}
\end{figure}

\paragraph{Disordered model with gain and loss.}
Finally, we show that  without TRS, the real-complex transition does not occur upon the non-Hermitian MBL transition.
We focus on the model with gain and loss, which is experimentally feasible~\cite{Lee14,Lapp19}:
\aln{\label{modelwosym}
\hat{H}=\sum_{i=1}^L-J(\hb_{i+1}^\dag\hb_i+\mr{H.c.})&+U\hat{n}_i\hat{n}_{i+1}\nonumber\\
&+({h_i-i{\gamma}(-1)^i)}\hat{n}_i.
}
In this model with broken TRS, the particle gain/loss at odd/even sites ($+i\gamma\hat{n}_i/-i\gamma\hat{n}_i$) tends to decrease/increase the number of particles at the corresponding site.
Figure ~\ref{fig4}(a) shows that the eigenenergies have nonzero imaginary parts irrespective of $h$.
We confirm $f_\mr{Im}=1$ for any $h$ and $L$ (data not shown).

While the real-complex transition is absent, the delocalization-localization transition occurs in this model.
Figure~\ref{fig4}(b) shows nearest-level-spacing distributions on the complex plane for different values of $h$.
For $h \lesssim h_c^\mr{MBL}\: (h_c^\mr{MBL}\gtrsim 4)$, the distribution is a non-Hermitian random-matrix distribution with transposition symmetry~\cite{hamazaki19T}, which is slightly different from  $P_\mr{Gin}^\mr{C}(s)$;
for $h \gtrsim h_c^\mr{MBL}$, it is a Poisson distribution $P_\mr{Po}^\mr{C}(s)$.
This change in the level statistics indicates the non-Hermitian MBL transition.
While we find that the transition is a crossover for $L=16$ (data not shown), it is a future challenge to investigate whether it becomes sharp at the MBL transition in the thermodynamic limit.
The entanglement entropy of the eigenstates shows the volume law for $h\lesssim h_c^\mr{MBL}$ and the area law for $h\gtrsim h_c^\mr{MBL}$ (see Fig.~\ref{fig4}(c)).

\paragraph{Conclusion and Outlook.}
We have shown that non-Hermitian MBL suppresses complex eigenenergies for generic non-Hermitian interacting Hamiltonians having TRS.
We have demonstrated that a real-complex transition, which occurs upon MBL, profoundly affects the dynamical stability of interacting systems with asymmetric hopping.
We have also demonstrated that real-complex transitions are absent in systems with gain and/or loss that break TRS, though the non-Hermitian MBL persists.

The real-complex transition found in our work is conceptually new in that it  never occurs in isolated, few-body, or clean systems.
It is of interest to investigate other properties of the non-Hermitian MBL such as critical phenomena (which are implied by our critical scaling collapse), where all of the non-Hermiticity, disorder, and interaction come into play.
Our work is also relevant to quantum chaos~\cite{Bohigas84,Grobe88,DAlessio16} in open interacting systems described by non-Hermitian Hamiltonians, as indicated by its random-matrix level statistics.
Furthermore, the properties of non-Hermitian MBL such as the entanglement entropy
may offer a new approach to understanding the interaction effect in the depinning transition of type-II superconductors~\cite{Hatano96}.
A detailed investigation of these problems merits further study.

\begin{acknowledgments}
We are grateful to Zongping Gong, Yuto Ashida, Masaya Nakagawa, and Naomichi Hatano for fruitful discussions.
This work was supported by
KAKENHI Grant No. JP18H01145 and
a Grant-in-Aid for Scientific Research on Innovative Areas ``Topological Materials Science" (KAKENHI Grant No. JP15H05855)
from the Japan Society for the Promotion of Science.
R. H. was supported by the Japan Society for the Promotion of Science through Program for Leading Graduate Schools (ALPS) and JSPS fellowship (JSPS KAKENHI Grant No. JP17J03189).
K. K. was supported by the JSPS through Program for Leading Graduate Schools (ALPS).
\end{acknowledgments}

\bibliographystyle{apsrev4-1}
\bibliography{../../../../refer_them}

\begin{thebibliography}{102}%
\makeatletter
\providecommand \@ifxundefined [1]{%
 \@ifx{#1\undefined}
}%
\providecommand \@ifnum [1]{%
 \ifnum #1\expandafter \@firstoftwo
 \else \expandafter \@secondoftwo
 \fi
}%
\providecommand \@ifx [1]{%
 \ifx #1\expandafter \@firstoftwo
 \else \expandafter \@secondoftwo
 \fi
}%
\providecommand \natexlab [1]{#1}%
\providecommand \enquote  [1]{``#1''}%
\providecommand \bibnamefont  [1]{#1}%
\providecommand \bibfnamefont [1]{#1}%
\providecommand \citenamefont [1]{#1}%
\providecommand \href@noop [0]{\@secondoftwo}%
\providecommand \href [0]{\begingroup \@sanitize@url \@href}%
\providecommand \@href[1]{\@@startlink{#1}\@@href}%
\providecommand \@@href[1]{\endgroup#1\@@endlink}%
\providecommand \@sanitize@url [0]{\catcode `\\12\catcode `\$12\catcode
  `\&12\catcode `\#12\catcode `\^12\catcode `\_12\catcode `\%12\relax}%
\providecommand \@@startlink[1]{}%
\providecommand \@@endlink[0]{}%
\providecommand \url  [0]{\begingroup\@sanitize@url \@url }%
\providecommand \@url [1]{\endgroup\@href {#1}{\urlprefix }}%
\providecommand \urlprefix  [0]{URL }%
\providecommand \Eprint [0]{\href }%
\providecommand \doibase [0]{http://dx.doi.org/}%
\providecommand \selectlanguage [0]{\@gobble}%
\providecommand \bibinfo  [0]{\@secondoftwo}%
\providecommand \bibfield  [0]{\@secondoftwo}%
\providecommand \translation [1]{[#1]}%
\providecommand \BibitemOpen [0]{}%
\providecommand \bibitemStop [0]{}%
\providecommand \bibitemNoStop [0]{.\EOS\space}%
\providecommand \EOS [0]{\spacefactor3000\relax}%
\providecommand \BibitemShut  [1]{\csname bibitem#1\endcsname}%
\let\auto@bib@innerbib\@empty
\bibitem [{\citenamefont {Bender}\ and\ \citenamefont
  {Boettcher}(1998)}]{Bender98}%
  \BibitemOpen
  \bibfield  {author} {\bibinfo {author} {\bibfnamefont {C.~M.}\ \bibnamefont
  {Bender}}\ and\ \bibinfo {author} {\bibfnamefont {S.}~\bibnamefont
  {Boettcher}},\ }\href {\doibase 10.1103/PhysRevLett.80.5243} {\bibfield
  {journal} {\bibinfo  {journal} {Phys. Rev. Lett.}\ }\textbf {\bibinfo
  {volume} {80}},\ \bibinfo {pages} {5243} (\bibinfo {year}
  {1998})}\BibitemShut {NoStop}%
\bibitem [{\citenamefont {Bender}\ \emph {et~al.}(2002)\citenamefont {Bender},
  \citenamefont {Brody},\ and\ \citenamefont {Jones}}]{Bender02}%
  \BibitemOpen
  \bibfield  {author} {\bibinfo {author} {\bibfnamefont {C.~M.}\ \bibnamefont
  {Bender}}, \bibinfo {author} {\bibfnamefont {D.~C.}\ \bibnamefont {Brody}}, \
  and\ \bibinfo {author} {\bibfnamefont {H.~F.}\ \bibnamefont {Jones}},\ }\href
  {\doibase 10.1103/PhysRevLett.89.270401} {\bibfield  {journal} {\bibinfo
  {journal} {Phys. Rev. Lett.}\ }\textbf {\bibinfo {volume} {89}},\ \bibinfo
  {pages} {270401} (\bibinfo {year} {2002})}\BibitemShut {NoStop}%
\bibitem [{\citenamefont {Makris}\ \emph {et~al.}(2008)\citenamefont {Makris},
  \citenamefont {El-Ganainy}, \citenamefont {Christodoulides},\ and\
  \citenamefont {Musslimani}}]{Makris08}%
  \BibitemOpen
  \bibfield  {author} {\bibinfo {author} {\bibfnamefont {K.~G.}\ \bibnamefont
  {Makris}}, \bibinfo {author} {\bibfnamefont {R.}~\bibnamefont {El-Ganainy}},
  \bibinfo {author} {\bibfnamefont {D.~N.}\ \bibnamefont {Christodoulides}}, \
  and\ \bibinfo {author} {\bibfnamefont {Z.~H.}\ \bibnamefont {Musslimani}},\
  }\href {\doibase 10.1103/PhysRevLett.100.103904} {\bibfield  {journal}
  {\bibinfo  {journal} {Phys. Rev. Lett.}\ }\textbf {\bibinfo {volume} {100}},\
  \bibinfo {pages} {103904} (\bibinfo {year} {2008})}\BibitemShut {NoStop}%
\bibitem [{\citenamefont {Graefe}\ \emph {et~al.}(2008)\citenamefont {Graefe},
  \citenamefont {Korsch},\ and\ \citenamefont {Niederle}}]{Graefe08}%
  \BibitemOpen
  \bibfield  {author} {\bibinfo {author} {\bibfnamefont {E.~M.}\ \bibnamefont
  {Graefe}}, \bibinfo {author} {\bibfnamefont {H.~J.}\ \bibnamefont {Korsch}},
  \ and\ \bibinfo {author} {\bibfnamefont {A.~E.}\ \bibnamefont {Niederle}},\
  }\href {\doibase 10.1103/PhysRevLett.101.150408} {\bibfield  {journal}
  {\bibinfo  {journal} {Phys. Rev. Lett.}\ }\textbf {\bibinfo {volume} {101}},\
  \bibinfo {pages} {150408} (\bibinfo {year} {2008})}\BibitemShut {NoStop}%
\bibitem [{\citenamefont {Musslimani}\ \emph {et~al.}(2008)\citenamefont
  {Musslimani}, \citenamefont {Makris}, \citenamefont {El-Ganainy},\ and\
  \citenamefont {Christodoulides}}]{Musslimani08}%
  \BibitemOpen
  \bibfield  {author} {\bibinfo {author} {\bibfnamefont {Z.~H.}\ \bibnamefont
  {Musslimani}}, \bibinfo {author} {\bibfnamefont {K.~G.}\ \bibnamefont
  {Makris}}, \bibinfo {author} {\bibfnamefont {R.}~\bibnamefont {El-Ganainy}},
  \ and\ \bibinfo {author} {\bibfnamefont {D.~N.}\ \bibnamefont
  {Christodoulides}},\ }\href {\doibase 10.1103/PhysRevLett.100.030402}
  {\bibfield  {journal} {\bibinfo  {journal} {Phys. Rev. Lett.}\ }\textbf
  {\bibinfo {volume} {100}},\ \bibinfo {pages} {030402} (\bibinfo {year}
  {2008})}\BibitemShut {NoStop}%
\bibitem [{\citenamefont {Longhi}(2009)}]{Longhi09}%
  \BibitemOpen
  \bibfield  {author} {\bibinfo {author} {\bibfnamefont {S.}~\bibnamefont
  {Longhi}},\ }\href {\doibase 10.1103/PhysRevLett.103.123601} {\bibfield
  {journal} {\bibinfo  {journal} {Phys. Rev. Lett.}\ }\textbf {\bibinfo
  {volume} {103}},\ \bibinfo {pages} {123601} (\bibinfo {year}
  {2009})}\BibitemShut {NoStop}%
\bibitem [{\citenamefont {Klaiman}\ \emph {et~al.}(2008)\citenamefont
  {Klaiman}, \citenamefont {G\"unther},\ and\ \citenamefont
  {Moiseyev}}]{Klaiman08}%
  \BibitemOpen
  \bibfield  {author} {\bibinfo {author} {\bibfnamefont {S.}~\bibnamefont
  {Klaiman}}, \bibinfo {author} {\bibfnamefont {U.}~\bibnamefont {G\"unther}},
  \ and\ \bibinfo {author} {\bibfnamefont {N.}~\bibnamefont {Moiseyev}},\
  }\href {\doibase 10.1103/PhysRevLett.101.080402} {\bibfield  {journal}
  {\bibinfo  {journal} {Phys. Rev. Lett.}\ }\textbf {\bibinfo {volume} {101}},\
  \bibinfo {pages} {080402} (\bibinfo {year} {2008})}\BibitemShut {NoStop}%
\bibitem [{\citenamefont {Lin}\ \emph {et~al.}(2011)\citenamefont {Lin},
  \citenamefont {Ramezani}, \citenamefont {Eichelkraut}, \citenamefont
  {Kottos}, \citenamefont {Cao},\ and\ \citenamefont
  {Christodoulides}}]{Lin11}%
  \BibitemOpen
  \bibfield  {author} {\bibinfo {author} {\bibfnamefont {Z.}~\bibnamefont
  {Lin}}, \bibinfo {author} {\bibfnamefont {H.}~\bibnamefont {Ramezani}},
  \bibinfo {author} {\bibfnamefont {T.}~\bibnamefont {Eichelkraut}}, \bibinfo
  {author} {\bibfnamefont {T.}~\bibnamefont {Kottos}}, \bibinfo {author}
  {\bibfnamefont {H.}~\bibnamefont {Cao}}, \ and\ \bibinfo {author}
  {\bibfnamefont {D.~N.}\ \bibnamefont {Christodoulides}},\ }\href {\doibase
  10.1103/PhysRevLett.106.213901} {\bibfield  {journal} {\bibinfo  {journal}
  {Phys. Rev. Lett.}\ }\textbf {\bibinfo {volume} {106}},\ \bibinfo {pages}
  {213901} (\bibinfo {year} {2011})}\BibitemShut {NoStop}%
\bibitem [{\citenamefont {Malzard}\ \emph {et~al.}(2015)\citenamefont
  {Malzard}, \citenamefont {Poli},\ and\ \citenamefont
  {Schomerus}}]{Malzard15}%
  \BibitemOpen
  \bibfield  {author} {\bibinfo {author} {\bibfnamefont {S.}~\bibnamefont
  {Malzard}}, \bibinfo {author} {\bibfnamefont {C.}~\bibnamefont {Poli}}, \
  and\ \bibinfo {author} {\bibfnamefont {H.}~\bibnamefont {Schomerus}},\ }\href
  {\doibase 10.1103/PhysRevLett.115.200402} {\bibfield  {journal} {\bibinfo
  {journal} {Phys. Rev. Lett.}\ }\textbf {\bibinfo {volume} {115}},\ \bibinfo
  {pages} {200402} (\bibinfo {year} {2015})}\BibitemShut {NoStop}%
\bibitem [{\citenamefont {Kawabata}\ \emph {et~al.}(2018)\citenamefont
  {Kawabata}, \citenamefont {Ashida}, \citenamefont {Katsura},\ and\
  \citenamefont {Ueda}}]{Kawabata18P}%
  \BibitemOpen
  \bibfield  {author} {\bibinfo {author} {\bibfnamefont {K.}~\bibnamefont
  {Kawabata}}, \bibinfo {author} {\bibfnamefont {Y.}~\bibnamefont {Ashida}},
  \bibinfo {author} {\bibfnamefont {H.}~\bibnamefont {Katsura}}, \ and\
  \bibinfo {author} {\bibfnamefont {M.}~\bibnamefont {Ueda}},\ }\href {\doibase
  10.1103/PhysRevB.98.085116} {\bibfield  {journal} {\bibinfo  {journal} {Phys.
  Rev. B}\ }\textbf {\bibinfo {volume} {98}},\ \bibinfo {pages} {085116}
  (\bibinfo {year} {2018})}\BibitemShut {NoStop}%
\bibitem [{\citenamefont {Jin}\ and\ \citenamefont {Song}(2013)}]{Jin13}%
  \BibitemOpen
  \bibfield  {author} {\bibinfo {author} {\bibfnamefont {L.}~\bibnamefont
  {Jin}}\ and\ \bibinfo {author} {\bibfnamefont {Z.}~\bibnamefont {Song}},\
  }\href {\doibase 10.1016/j.aop.2012.11.017} {\bibfield  {journal} {\bibinfo
  {journal} {Annals of Physics}\ }\textbf {\bibinfo {volume} {330}},\ \bibinfo
  {pages} {142} (\bibinfo {year} {2013})}\BibitemShut {NoStop}%
\bibitem [{\citenamefont {Tripathi}\ \emph {et~al.}(2016)\citenamefont
  {Tripathi}, \citenamefont {Galda}, \citenamefont {Barman},\ and\
  \citenamefont {Vinokur}}]{Tripathi16}%
  \BibitemOpen
  \bibfield  {author} {\bibinfo {author} {\bibfnamefont {V.}~\bibnamefont
  {Tripathi}}, \bibinfo {author} {\bibfnamefont {A.}~\bibnamefont {Galda}},
  \bibinfo {author} {\bibfnamefont {H.}~\bibnamefont {Barman}}, \ and\ \bibinfo
  {author} {\bibfnamefont {V.~M.}\ \bibnamefont {Vinokur}},\ }\href {\doibase
  10.1103/PhysRevB.94.041104} {\bibfield  {journal} {\bibinfo  {journal}
  {Physical Review B}\ }\textbf {\bibinfo {volume} {94}},\ \bibinfo {pages}
  {041104(R)} (\bibinfo {year} {2016})}\BibitemShut {NoStop}%
\bibitem [{\citenamefont {Ashida}\ \emph {et~al.}(2017)\citenamefont {Ashida},
  \citenamefont {Furukawa},\ and\ \citenamefont {Ueda}}]{Ashida17}%
  \BibitemOpen
  \bibfield  {author} {\bibinfo {author} {\bibfnamefont {Y.}~\bibnamefont
  {Ashida}}, \bibinfo {author} {\bibfnamefont {S.}~\bibnamefont {Furukawa}}, \
  and\ \bibinfo {author} {\bibfnamefont {M.}~\bibnamefont {Ueda}},\ }\href
  {\doibase 10.1038/ncomms15791} {\bibfield  {journal} {\bibinfo  {journal}
  {Nature Communications}\ }\textbf {\bibinfo {volume} {8}},\ \bibinfo {pages}
  {15791} (\bibinfo {year} {2017})}\BibitemShut {NoStop}%
\bibitem [{\citenamefont {Lourenco}\ \emph {et~al.}(2018)\citenamefont
  {Lourenco}, \citenamefont {Eneias},\ and\ \citenamefont
  {Pereira}}]{Lourenco18}%
  \BibitemOpen
  \bibfield  {author} {\bibinfo {author} {\bibfnamefont {J.~A.~S.}\
  \bibnamefont {Lourenco}}, \bibinfo {author} {\bibfnamefont {R.~L.}\
  \bibnamefont {Eneias}}, \ and\ \bibinfo {author} {\bibfnamefont {R.~G.}\
  \bibnamefont {Pereira}},\ }\href {\doibase 10.1103/PhysRevB.98.085126}
  {\bibfield  {journal} {\bibinfo  {journal} {Phys. Rev. B}\ }\textbf {\bibinfo
  {volume} {98}},\ \bibinfo {pages} {085126} (\bibinfo {year}
  {2018})}\BibitemShut {NoStop}%
\bibitem [{\citenamefont {Nakagawa}\ \emph {et~al.}(2018)\citenamefont
  {Nakagawa}, \citenamefont {Kawakami},\ and\ \citenamefont
  {Ueda}}]{Nakagawa18}%
  \BibitemOpen
  \bibfield  {author} {\bibinfo {author} {\bibfnamefont {M.}~\bibnamefont
  {Nakagawa}}, \bibinfo {author} {\bibfnamefont {N.}~\bibnamefont {Kawakami}},
  \ and\ \bibinfo {author} {\bibfnamefont {M.}~\bibnamefont {Ueda}},\ }\href
  {\doibase 10.1103/PhysRevLett.121.203001} {\bibfield  {journal} {\bibinfo
  {journal} {Phys. Rev. Lett.}\ }\textbf {\bibinfo {volume} {121}},\ \bibinfo
  {pages} {203001} (\bibinfo {year} {2018})}\BibitemShut {NoStop}%
\bibitem [{\citenamefont {Guo}\ \emph {et~al.}(2009)\citenamefont {Guo},
  \citenamefont {Salamo}, \citenamefont {Duchesne}, \citenamefont {Morandotti},
  \citenamefont {Volatier-Ravat}, \citenamefont {Aimez}, \citenamefont
  {Siviloglou},\ and\ \citenamefont {Christodoulides}}]{Guo09}%
  \BibitemOpen
  \bibfield  {author} {\bibinfo {author} {\bibfnamefont {A.}~\bibnamefont
  {Guo}}, \bibinfo {author} {\bibfnamefont {G.~J.}\ \bibnamefont {Salamo}},
  \bibinfo {author} {\bibfnamefont {D.}~\bibnamefont {Duchesne}}, \bibinfo
  {author} {\bibfnamefont {R.}~\bibnamefont {Morandotti}}, \bibinfo {author}
  {\bibfnamefont {M.}~\bibnamefont {Volatier-Ravat}}, \bibinfo {author}
  {\bibfnamefont {V.}~\bibnamefont {Aimez}}, \bibinfo {author} {\bibfnamefont
  {G.~A.}\ \bibnamefont {Siviloglou}}, \ and\ \bibinfo {author} {\bibfnamefont
  {D.~N.}\ \bibnamefont {Christodoulides}},\ }\href {\doibase
  10.1103/PhysRevLett.103.093902} {\bibfield  {journal} {\bibinfo  {journal}
  {Phys. Rev. Lett.}\ }\textbf {\bibinfo {volume} {103}},\ \bibinfo {pages}
  {093902} (\bibinfo {year} {2009})}\BibitemShut {NoStop}%
\bibitem [{\citenamefont {R{\"u}ter}\ \emph {et~al.}(2010)\citenamefont
  {R{\"u}ter}, \citenamefont {Makris}, \citenamefont {El-Ganainy},
  \citenamefont {Christodoulides}, \citenamefont {Segev},\ and\ \citenamefont
  {Kip}}]{Ruter10}%
  \BibitemOpen
  \bibfield  {author} {\bibinfo {author} {\bibfnamefont {C.~E.}\ \bibnamefont
  {R{\"u}ter}}, \bibinfo {author} {\bibfnamefont {K.~G.}\ \bibnamefont
  {Makris}}, \bibinfo {author} {\bibfnamefont {R.}~\bibnamefont {El-Ganainy}},
  \bibinfo {author} {\bibfnamefont {D.~N.}\ \bibnamefont {Christodoulides}},
  \bibinfo {author} {\bibfnamefont {M.}~\bibnamefont {Segev}}, \ and\ \bibinfo
  {author} {\bibfnamefont {D.}~\bibnamefont {Kip}},\ }\href {\doibase
  10.1038/nphys1515} {\bibfield  {journal} {\bibinfo  {journal} {Nature
  Physics}\ }\textbf {\bibinfo {volume} {6}},\ \bibinfo {pages} {192} (\bibinfo
  {year} {2010})}\BibitemShut {NoStop}%
\bibitem [{\citenamefont {Regensburger}\ \emph {et~al.}(2012)\citenamefont
  {Regensburger}, \citenamefont {Bersch}, \citenamefont {Miri}, \citenamefont
  {Onishchukov}, \citenamefont {Christodoulides},\ and\ \citenamefont
  {Peschel}}]{Regensburger12}%
  \BibitemOpen
  \bibfield  {author} {\bibinfo {author} {\bibfnamefont {A.}~\bibnamefont
  {Regensburger}}, \bibinfo {author} {\bibfnamefont {C.}~\bibnamefont
  {Bersch}}, \bibinfo {author} {\bibfnamefont {M.-A.}\ \bibnamefont {Miri}},
  \bibinfo {author} {\bibfnamefont {G.}~\bibnamefont {Onishchukov}}, \bibinfo
  {author} {\bibfnamefont {D.~N.}\ \bibnamefont {Christodoulides}}, \ and\
  \bibinfo {author} {\bibfnamefont {U.}~\bibnamefont {Peschel}},\ }\href
  {\doibase 10.1038/nature11298} {\bibfield  {journal} {\bibinfo  {journal}
  {Nature}\ }\textbf {\bibinfo {volume} {488}},\ \bibinfo {pages} {167}
  (\bibinfo {year} {2012})}\BibitemShut {NoStop}%
\bibitem [{\citenamefont {Peng}\ \emph {et~al.}(2014)\citenamefont {Peng},
  \citenamefont {{\"O}zdemir}, \citenamefont {Lei}, \citenamefont {Monifi},
  \citenamefont {Gianfreda}, \citenamefont {Long}, \citenamefont {Fan},
  \citenamefont {Nori}, \citenamefont {Bender},\ and\ \citenamefont
  {Yang}}]{Peng14}%
  \BibitemOpen
  \bibfield  {author} {\bibinfo {author} {\bibfnamefont {B.}~\bibnamefont
  {Peng}}, \bibinfo {author} {\bibfnamefont {{\c{S}}.~K.}\ \bibnamefont
  {{\"O}zdemir}}, \bibinfo {author} {\bibfnamefont {F.}~\bibnamefont {Lei}},
  \bibinfo {author} {\bibfnamefont {F.}~\bibnamefont {Monifi}}, \bibinfo
  {author} {\bibfnamefont {M.}~\bibnamefont {Gianfreda}}, \bibinfo {author}
  {\bibfnamefont {G.~L.}\ \bibnamefont {Long}}, \bibinfo {author}
  {\bibfnamefont {S.}~\bibnamefont {Fan}}, \bibinfo {author} {\bibfnamefont
  {F.}~\bibnamefont {Nori}}, \bibinfo {author} {\bibfnamefont {C.~M.}\
  \bibnamefont {Bender}}, \ and\ \bibinfo {author} {\bibfnamefont
  {L.}~\bibnamefont {Yang}},\ }\href {\doibase 10.1038/nphys2927} {\bibfield
  {journal} {\bibinfo  {journal} {Nature Physics}\ }\textbf {\bibinfo {volume}
  {10}},\ \bibinfo {pages} {394} (\bibinfo {year} {2014})}\BibitemShut
  {NoStop}%
\bibitem [{\citenamefont {Feng}\ \emph {et~al.}(2014)\citenamefont {Feng},
  \citenamefont {Wong}, \citenamefont {Ma}, \citenamefont {Wang},\ and\
  \citenamefont {Zhang}}]{Feng14}%
  \BibitemOpen
  \bibfield  {author} {\bibinfo {author} {\bibfnamefont {L.}~\bibnamefont
  {Feng}}, \bibinfo {author} {\bibfnamefont {Z.~J.}\ \bibnamefont {Wong}},
  \bibinfo {author} {\bibfnamefont {R.-M.}\ \bibnamefont {Ma}}, \bibinfo
  {author} {\bibfnamefont {Y.}~\bibnamefont {Wang}}, \ and\ \bibinfo {author}
  {\bibfnamefont {X.}~\bibnamefont {Zhang}},\ }\href {\doibase
  10.1126/science.1258479} {\bibfield  {journal} {\bibinfo  {journal}
  {Science}\ }\textbf {\bibinfo {volume} {346}},\ \bibinfo {pages} {972}
  (\bibinfo {year} {2014})}\BibitemShut {NoStop}%
\bibitem [{\citenamefont {Hodaei}\ \emph {et~al.}(2014)\citenamefont {Hodaei},
  \citenamefont {Miri}, \citenamefont {Heinrich}, \citenamefont
  {Christodoulides},\ and\ \citenamefont {Khajavikhan}}]{Hodaei14}%
  \BibitemOpen
  \bibfield  {author} {\bibinfo {author} {\bibfnamefont {H.}~\bibnamefont
  {Hodaei}}, \bibinfo {author} {\bibfnamefont {M.-A.}\ \bibnamefont {Miri}},
  \bibinfo {author} {\bibfnamefont {M.}~\bibnamefont {Heinrich}}, \bibinfo
  {author} {\bibfnamefont {D.~N.}\ \bibnamefont {Christodoulides}}, \ and\
  \bibinfo {author} {\bibfnamefont {M.}~\bibnamefont {Khajavikhan}},\ }\href
  {\doibase 10.1126/science.1258480} {\bibfield  {journal} {\bibinfo  {journal}
  {Science}\ }\textbf {\bibinfo {volume} {346}},\ \bibinfo {pages} {975}
  (\bibinfo {year} {2014})}\BibitemShut {NoStop}%
\bibitem [{\citenamefont {Zhen}\ \emph {et~al.}(2015)\citenamefont {Zhen},
  \citenamefont {Hsu}, \citenamefont {Igarashi}, \citenamefont {Lu},
  \citenamefont {Kaminer}, \citenamefont {Pick}, \citenamefont {Chua},
  \citenamefont {Joannopoulos},\ and\ \citenamefont
  {Solja{\v{c}}i{\'c}}}]{Zhen15}%
  \BibitemOpen
  \bibfield  {author} {\bibinfo {author} {\bibfnamefont {B.}~\bibnamefont
  {Zhen}}, \bibinfo {author} {\bibfnamefont {C.~W.}\ \bibnamefont {Hsu}},
  \bibinfo {author} {\bibfnamefont {Y.}~\bibnamefont {Igarashi}}, \bibinfo
  {author} {\bibfnamefont {L.}~\bibnamefont {Lu}}, \bibinfo {author}
  {\bibfnamefont {I.}~\bibnamefont {Kaminer}}, \bibinfo {author} {\bibfnamefont
  {A.}~\bibnamefont {Pick}}, \bibinfo {author} {\bibfnamefont {S.-L.}\
  \bibnamefont {Chua}}, \bibinfo {author} {\bibfnamefont {J.~D.}\ \bibnamefont
  {Joannopoulos}}, \ and\ \bibinfo {author} {\bibfnamefont {M.}~\bibnamefont
  {Solja{\v{c}}i{\'c}}},\ }\href {\doibase 10.1038/nature14889} {\bibfield
  {journal} {\bibinfo  {journal} {Nature}\ }\textbf {\bibinfo {volume} {525}},\
  \bibinfo {pages} {354} (\bibinfo {year} {2015})}\BibitemShut {NoStop}%
\bibitem [{\citenamefont {Zeuner}\ \emph {et~al.}(2015)\citenamefont {Zeuner},
  \citenamefont {Rechtsman}, \citenamefont {Plotnik}, \citenamefont {Lumer},
  \citenamefont {Nolte}, \citenamefont {Rudner}, \citenamefont {Segev},\ and\
  \citenamefont {Szameit}}]{Zeuner15}%
  \BibitemOpen
  \bibfield  {author} {\bibinfo {author} {\bibfnamefont {J.~M.}\ \bibnamefont
  {Zeuner}}, \bibinfo {author} {\bibfnamefont {M.~C.}\ \bibnamefont
  {Rechtsman}}, \bibinfo {author} {\bibfnamefont {Y.}~\bibnamefont {Plotnik}},
  \bibinfo {author} {\bibfnamefont {Y.}~\bibnamefont {Lumer}}, \bibinfo
  {author} {\bibfnamefont {S.}~\bibnamefont {Nolte}}, \bibinfo {author}
  {\bibfnamefont {M.~S.}\ \bibnamefont {Rudner}}, \bibinfo {author}
  {\bibfnamefont {M.}~\bibnamefont {Segev}}, \ and\ \bibinfo {author}
  {\bibfnamefont {A.}~\bibnamefont {Szameit}},\ }\href {\doibase
  10.1103/PhysRevLett.115.040402} {\bibfield  {journal} {\bibinfo  {journal}
  {Phys. Rev. Lett.}\ }\textbf {\bibinfo {volume} {115}},\ \bibinfo {pages}
  {040402} (\bibinfo {year} {2015})}\BibitemShut {NoStop}%
\bibitem [{\citenamefont {Poli}\ \emph {et~al.}(2015)\citenamefont {Poli},
  \citenamefont {Bellec}, \citenamefont {Kuhl}, \citenamefont {Mortessagne},\
  and\ \citenamefont {Schomerus}}]{Poli15}%
  \BibitemOpen
  \bibfield  {author} {\bibinfo {author} {\bibfnamefont {C.}~\bibnamefont
  {Poli}}, \bibinfo {author} {\bibfnamefont {M.}~\bibnamefont {Bellec}},
  \bibinfo {author} {\bibfnamefont {U.}~\bibnamefont {Kuhl}}, \bibinfo {author}
  {\bibfnamefont {F.}~\bibnamefont {Mortessagne}}, \ and\ \bibinfo {author}
  {\bibfnamefont {H.}~\bibnamefont {Schomerus}},\ }\href {\doibase
  10.1038/ncomms7710} {\bibfield  {journal} {\bibinfo  {journal} {Nature
  Communications}\ }\textbf {\bibinfo {volume} {6}},\ \bibinfo {pages} {6710}
  (\bibinfo {year} {2015})}\BibitemShut {NoStop}%
\bibitem [{\citenamefont {Li}\ \emph {et~al.}(2019)\citenamefont {Li},
  \citenamefont {Harter}, \citenamefont {Liu}, \citenamefont {de~Melo},
  \citenamefont {Joglekar},\ and\ \citenamefont {Luo}}]{Li19}%
  \BibitemOpen
  \bibfield  {author} {\bibinfo {author} {\bibfnamefont {J.}~\bibnamefont
  {Li}}, \bibinfo {author} {\bibfnamefont {A.~K.}\ \bibnamefont {Harter}},
  \bibinfo {author} {\bibfnamefont {J.}~\bibnamefont {Liu}}, \bibinfo {author}
  {\bibfnamefont {L.}~\bibnamefont {de~Melo}}, \bibinfo {author} {\bibfnamefont
  {Y.~N.}\ \bibnamefont {Joglekar}}, \ and\ \bibinfo {author} {\bibfnamefont
  {L.}~\bibnamefont {Luo}},\ }\href
  {https://www.nature.com/articles/s41467-019-08596-1} {\bibfield  {journal}
  {\bibinfo  {journal} {Nature communications}\ }\textbf {\bibinfo {volume}
  {10}},\ \bibinfo {pages} {855} (\bibinfo {year} {2019})}\BibitemShut
  {NoStop}%
\bibitem [{\citenamefont {Doppler}\ \emph {et~al.}(2016)\citenamefont
  {Doppler}, \citenamefont {Mailybaev}, \citenamefont {B{\"o}hm}, \citenamefont
  {Kuhl}, \citenamefont {Girschik}, \citenamefont {Libisch}, \citenamefont
  {Milburn}, \citenamefont {Rabl}, \citenamefont {Moiseyev},\ and\
  \citenamefont {Rotter}}]{Doppler16}%
  \BibitemOpen
  \bibfield  {author} {\bibinfo {author} {\bibfnamefont {J.}~\bibnamefont
  {Doppler}}, \bibinfo {author} {\bibfnamefont {A.~A.}\ \bibnamefont
  {Mailybaev}}, \bibinfo {author} {\bibfnamefont {J.}~\bibnamefont {B{\"o}hm}},
  \bibinfo {author} {\bibfnamefont {U.}~\bibnamefont {Kuhl}}, \bibinfo {author}
  {\bibfnamefont {A.}~\bibnamefont {Girschik}}, \bibinfo {author}
  {\bibfnamefont {F.}~\bibnamefont {Libisch}}, \bibinfo {author} {\bibfnamefont
  {T.~J.}\ \bibnamefont {Milburn}}, \bibinfo {author} {\bibfnamefont
  {P.}~\bibnamefont {Rabl}}, \bibinfo {author} {\bibfnamefont {N.}~\bibnamefont
  {Moiseyev}}, \ and\ \bibinfo {author} {\bibfnamefont {S.}~\bibnamefont
  {Rotter}},\ }\href {https://www.nature.com/articles/nature18605} {\bibfield
  {journal} {\bibinfo  {journal} {Nature}\ }\textbf {\bibinfo {volume} {537}},\
  \bibinfo {pages} {76} (\bibinfo {year} {2016})}\BibitemShut {NoStop}%
\bibitem [{\citenamefont {Weimann}\ \emph {et~al.}(2017)\citenamefont
  {Weimann}, \citenamefont {Kremer}, \citenamefont {Plotnik}, \citenamefont
  {Lumer}, \citenamefont {Nolte}, \citenamefont {Makris}, \citenamefont
  {Segev}, \citenamefont {Rechtsman},\ and\ \citenamefont
  {Szameit}}]{Weimann17}%
  \BibitemOpen
  \bibfield  {author} {\bibinfo {author} {\bibfnamefont {S.}~\bibnamefont
  {Weimann}}, \bibinfo {author} {\bibfnamefont {M.}~\bibnamefont {Kremer}},
  \bibinfo {author} {\bibfnamefont {Y.}~\bibnamefont {Plotnik}}, \bibinfo
  {author} {\bibfnamefont {Y.}~\bibnamefont {Lumer}}, \bibinfo {author}
  {\bibfnamefont {S.}~\bibnamefont {Nolte}}, \bibinfo {author} {\bibfnamefont
  {K.}~\bibnamefont {Makris}}, \bibinfo {author} {\bibfnamefont
  {M.}~\bibnamefont {Segev}}, \bibinfo {author} {\bibfnamefont
  {M.}~\bibnamefont {Rechtsman}}, \ and\ \bibinfo {author} {\bibfnamefont
  {A.}~\bibnamefont {Szameit}},\ }\href {\doibase 10.1038/nmat4811} {\bibfield
  {journal} {\bibinfo  {journal} {Nature Materials}\ }\textbf {\bibinfo
  {volume} {16}},\ \bibinfo {pages} {433} (\bibinfo {year} {2017})}\BibitemShut
  {NoStop}%
\bibitem [{\citenamefont {Hodaei}\ \emph {et~al.}(2017)\citenamefont {Hodaei},
  \citenamefont {Hassan}, \citenamefont {Wittek}, \citenamefont
  {Garcia-Gracia}, \citenamefont {El-Ganainy}, \citenamefont
  {Christodoulides},\ and\ \citenamefont {Khajavikhan}}]{Hodaei17}%
  \BibitemOpen
  \bibfield  {author} {\bibinfo {author} {\bibfnamefont {H.}~\bibnamefont
  {Hodaei}}, \bibinfo {author} {\bibfnamefont {A.~U.}\ \bibnamefont {Hassan}},
  \bibinfo {author} {\bibfnamefont {S.}~\bibnamefont {Wittek}}, \bibinfo
  {author} {\bibfnamefont {H.}~\bibnamefont {Garcia-Gracia}}, \bibinfo {author}
  {\bibfnamefont {R.}~\bibnamefont {El-Ganainy}}, \bibinfo {author}
  {\bibfnamefont {D.~N.}\ \bibnamefont {Christodoulides}}, \ and\ \bibinfo
  {author} {\bibfnamefont {M.}~\bibnamefont {Khajavikhan}},\ }\href {\doibase
  10.1038/nature23280} {\bibfield  {journal} {\bibinfo  {journal} {Nature}\
  }\textbf {\bibinfo {volume} {548}},\ \bibinfo {pages} {187} (\bibinfo {year}
  {2017})}\BibitemShut {NoStop}%
\bibitem [{\citenamefont {Xiao}\ \emph {et~al.}(2017)\citenamefont {Xiao},
  \citenamefont {Zhan}, \citenamefont {Bian}, \citenamefont {Wang},
  \citenamefont {Zhang}, \citenamefont {Wang}, \citenamefont {Li},
  \citenamefont {Mochizuki}, \citenamefont {Kim}, \citenamefont {Kawakami},
  \citenamefont {Yi}, \citenamefont {Obuse}, \citenamefont {Sanders},\ and\
  \citenamefont {Xue}}]{Xiao17}%
  \BibitemOpen
  \bibfield  {author} {\bibinfo {author} {\bibfnamefont {L.}~\bibnamefont
  {Xiao}}, \bibinfo {author} {\bibfnamefont {X.}~\bibnamefont {Zhan}}, \bibinfo
  {author} {\bibfnamefont {Z.}~\bibnamefont {Bian}}, \bibinfo {author}
  {\bibfnamefont {K.}~\bibnamefont {Wang}}, \bibinfo {author} {\bibfnamefont
  {X.}~\bibnamefont {Zhang}}, \bibinfo {author} {\bibfnamefont
  {X.}~\bibnamefont {Wang}}, \bibinfo {author} {\bibfnamefont {J.}~\bibnamefont
  {Li}}, \bibinfo {author} {\bibfnamefont {K.}~\bibnamefont {Mochizuki}},
  \bibinfo {author} {\bibfnamefont {D.}~\bibnamefont {Kim}}, \bibinfo {author}
  {\bibfnamefont {N.}~\bibnamefont {Kawakami}}, \bibinfo {author}
  {\bibfnamefont {W.}~\bibnamefont {Yi}}, \bibinfo {author} {\bibfnamefont
  {H.}~\bibnamefont {Obuse}}, \bibinfo {author} {\bibfnamefont {B.~C.}\
  \bibnamefont {Sanders}}, \ and\ \bibinfo {author} {\bibfnamefont
  {P.}~\bibnamefont {Xue}},\ }\href {\doibase 10.1038/nphys4204} {\bibfield
  {journal} {\bibinfo  {journal} {Nature Physics}\ }\textbf {\bibinfo {volume}
  {13}},\ \bibinfo {pages} {1117} (\bibinfo {year} {2017})}\BibitemShut
  {NoStop}%
\bibitem [{\citenamefont {Chen}\ \emph {et~al.}(2017)\citenamefont {Chen},
  \citenamefont {{\"O}zdemir}, \citenamefont {Zhao}, \citenamefont {Wiersig},\
  and\ \citenamefont {Yang}}]{Chen17}%
  \BibitemOpen
  \bibfield  {author} {\bibinfo {author} {\bibfnamefont {W.}~\bibnamefont
  {Chen}}, \bibinfo {author} {\bibfnamefont {{\c{S}}.~K.}\ \bibnamefont
  {{\"O}zdemir}}, \bibinfo {author} {\bibfnamefont {G.}~\bibnamefont {Zhao}},
  \bibinfo {author} {\bibfnamefont {J.}~\bibnamefont {Wiersig}}, \ and\
  \bibinfo {author} {\bibfnamefont {L.}~\bibnamefont {Yang}},\ }\href {\doibase
  10.1038/nature23281} {\bibfield  {journal} {\bibinfo  {journal} {Nature}\
  }\textbf {\bibinfo {volume} {548}},\ \bibinfo {pages} {192} (\bibinfo {year}
  {2017})}\BibitemShut {NoStop}%
\bibitem [{\citenamefont {Zhou}\ \emph {et~al.}(2018)\citenamefont {Zhou},
  \citenamefont {Peng}, \citenamefont {Yoon}, \citenamefont {Hsu},
  \citenamefont {Nelson}, \citenamefont {Fu}, \citenamefont {Joannopoulos},
  \citenamefont {Solja{\v{c}}i{\'c}},\ and\ \citenamefont {Zhen}}]{Zhou18}%
  \BibitemOpen
  \bibfield  {author} {\bibinfo {author} {\bibfnamefont {H.}~\bibnamefont
  {Zhou}}, \bibinfo {author} {\bibfnamefont {C.}~\bibnamefont {Peng}}, \bibinfo
  {author} {\bibfnamefont {Y.}~\bibnamefont {Yoon}}, \bibinfo {author}
  {\bibfnamefont {C.~W.}\ \bibnamefont {Hsu}}, \bibinfo {author} {\bibfnamefont
  {K.~A.}\ \bibnamefont {Nelson}}, \bibinfo {author} {\bibfnamefont
  {L.}~\bibnamefont {Fu}}, \bibinfo {author} {\bibfnamefont {J.~D.}\
  \bibnamefont {Joannopoulos}}, \bibinfo {author} {\bibfnamefont
  {M.}~\bibnamefont {Solja{\v{c}}i{\'c}}}, \ and\ \bibinfo {author}
  {\bibfnamefont {B.}~\bibnamefont {Zhen}},\ }\href {\doibase
  10.1126/science.aap9859} {\bibfield  {journal} {\bibinfo  {journal}
  {Science}\ }\textbf {\bibinfo {volume} {359}},\ \bibinfo {pages} {1009}
  (\bibinfo {year} {2018})}\BibitemShut {NoStop}%
\bibitem [{\citenamefont {Bandres}\ \emph {et~al.}(2018)\citenamefont
  {Bandres}, \citenamefont {Wittek}, \citenamefont {Harari}, \citenamefont
  {Parto}, \citenamefont {Ren}, \citenamefont {Segev}, \citenamefont
  {Christodoulides},\ and\ \citenamefont {Khajavikhan}}]{Bandres18}%
  \BibitemOpen
  \bibfield  {author} {\bibinfo {author} {\bibfnamefont {M.~A.}\ \bibnamefont
  {Bandres}}, \bibinfo {author} {\bibfnamefont {S.}~\bibnamefont {Wittek}},
  \bibinfo {author} {\bibfnamefont {G.}~\bibnamefont {Harari}}, \bibinfo
  {author} {\bibfnamefont {M.}~\bibnamefont {Parto}}, \bibinfo {author}
  {\bibfnamefont {J.}~\bibnamefont {Ren}}, \bibinfo {author} {\bibfnamefont
  {M.}~\bibnamefont {Segev}}, \bibinfo {author} {\bibfnamefont {D.~N.}\
  \bibnamefont {Christodoulides}}, \ and\ \bibinfo {author} {\bibfnamefont
  {M.}~\bibnamefont {Khajavikhan}},\ }\href {\doibase 10.1126/science.aar4005}
  {\bibfield  {journal} {\bibinfo  {journal} {Science}\ }\textbf {\bibinfo
  {volume} {359}},\ \bibinfo {pages} {eaar4005} (\bibinfo {year}
  {2018})}\BibitemShut {NoStop}%
\bibitem [{\citenamefont {Konotop}\ \emph {et~al.}(2016)\citenamefont
  {Konotop}, \citenamefont {Yang},\ and\ \citenamefont {Zezyulin}}]{Konotop16}%
  \BibitemOpen
  \bibfield  {author} {\bibinfo {author} {\bibfnamefont {V.~V.}\ \bibnamefont
  {Konotop}}, \bibinfo {author} {\bibfnamefont {J.}~\bibnamefont {Yang}}, \
  and\ \bibinfo {author} {\bibfnamefont {D.~A.}\ \bibnamefont {Zezyulin}},\
  }\href {\doibase 10.1103/RevModPhys.88.035002} {\bibfield  {journal}
  {\bibinfo  {journal} {Rev. Mod. Phys.}\ }\textbf {\bibinfo {volume} {88}},\
  \bibinfo {pages} {035002} (\bibinfo {year} {2016})}\BibitemShut {NoStop}%
\bibitem [{\citenamefont {El-Ganainy}\ \emph {et~al.}(2018)\citenamefont
  {El-Ganainy}, \citenamefont {Makris}, \citenamefont {Khajavikhan},
  \citenamefont {Musslimani}, \citenamefont {Rotter},\ and\ \citenamefont
  {Christodoulides}}]{ElGanainy18}%
  \BibitemOpen
  \bibfield  {author} {\bibinfo {author} {\bibfnamefont {R.}~\bibnamefont
  {El-Ganainy}}, \bibinfo {author} {\bibfnamefont {K.~G.}\ \bibnamefont
  {Makris}}, \bibinfo {author} {\bibfnamefont {M.}~\bibnamefont {Khajavikhan}},
  \bibinfo {author} {\bibfnamefont {Z.~H.}\ \bibnamefont {Musslimani}},
  \bibinfo {author} {\bibfnamefont {S.}~\bibnamefont {Rotter}}, \ and\ \bibinfo
  {author} {\bibfnamefont {D.~N.}\ \bibnamefont {Christodoulides}},\ }\href
  {\doibase 10.1038/nphys4323} {\bibfield  {journal} {\bibinfo  {journal}
  {Nature Physics}\ }\textbf {\bibinfo {volume} {14}},\ \bibinfo {pages} {11}
  (\bibinfo {year} {2018})}\BibitemShut {NoStop}%
\bibitem [{\citenamefont {Hatano}\ and\ \citenamefont
  {Nelson}(1996)}]{Hatano96}%
  \BibitemOpen
  \bibfield  {author} {\bibinfo {author} {\bibfnamefont {N.}~\bibnamefont
  {Hatano}}\ and\ \bibinfo {author} {\bibfnamefont {D.~R.}\ \bibnamefont
  {Nelson}},\ }\href {\doibase 10.1103/PhysRevLett.77.570} {\bibfield
  {journal} {\bibinfo  {journal} {Phys. Rev. Lett.}\ }\textbf {\bibinfo
  {volume} {77}},\ \bibinfo {pages} {570} (\bibinfo {year} {1996})}\BibitemShut
  {NoStop}%
\bibitem [{\citenamefont {Hatano}\ and\ \citenamefont
  {Nelson}(1997)}]{Hatano97}%
  \BibitemOpen
  \bibfield  {author} {\bibinfo {author} {\bibfnamefont {N.}~\bibnamefont
  {Hatano}}\ and\ \bibinfo {author} {\bibfnamefont {D.~R.}\ \bibnamefont
  {Nelson}},\ }\href {\doibase 10.1103/PhysRevB.56.8651} {\bibfield  {journal}
  {\bibinfo  {journal} {Phys. Rev. B}\ }\textbf {\bibinfo {volume} {56}},\
  \bibinfo {pages} {8651} (\bibinfo {year} {1997})}\BibitemShut {NoStop}%
\bibitem [{\citenamefont {Hatano}\ and\ \citenamefont
  {Nelson}(1998)}]{Hatano98}%
  \BibitemOpen
  \bibfield  {author} {\bibinfo {author} {\bibfnamefont {N.}~\bibnamefont
  {Hatano}}\ and\ \bibinfo {author} {\bibfnamefont {D.~R.}\ \bibnamefont
  {Nelson}},\ }\href {\doibase 10.1103/PhysRevB.58.8384} {\bibfield  {journal}
  {\bibinfo  {journal} {Phys. Rev. B}\ }\textbf {\bibinfo {volume} {58}},\
  \bibinfo {pages} {8384} (\bibinfo {year} {1998})}\BibitemShut {NoStop}%
\bibitem [{\citenamefont {Nelson}\ and\ \citenamefont
  {Shnerb}(1998)}]{Nelson98}%
  \BibitemOpen
  \bibfield  {author} {\bibinfo {author} {\bibfnamefont {D.~R.}\ \bibnamefont
  {Nelson}}\ and\ \bibinfo {author} {\bibfnamefont {N.~M.}\ \bibnamefont
  {Shnerb}},\ }\href {\doibase 10.1103/PhysRevE.58.1383} {\bibfield  {journal}
  {\bibinfo  {journal} {Phys. Rev. E}\ }\textbf {\bibinfo {volume} {58}},\
  \bibinfo {pages} {1383} (\bibinfo {year} {1998})}\BibitemShut {NoStop}%
\bibitem [{\citenamefont {Efetov}(1997)}]{Efetov97}%
  \BibitemOpen
  \bibfield  {author} {\bibinfo {author} {\bibfnamefont {K.~B.}\ \bibnamefont
  {Efetov}},\ }\href {\doibase 10.1103/PhysRevLett.79.491} {\bibfield
  {journal} {\bibinfo  {journal} {Phys. Rev. Lett.}\ }\textbf {\bibinfo
  {volume} {79}},\ \bibinfo {pages} {491} (\bibinfo {year} {1997})}\BibitemShut
  {NoStop}%
\bibitem [{\citenamefont {Feinberg}\ and\ \citenamefont
  {Zee}(1997)}]{Feinberg97}%
  \BibitemOpen
  \bibfield  {author} {\bibinfo {author} {\bibfnamefont {J.}~\bibnamefont
  {Feinberg}}\ and\ \bibinfo {author} {\bibfnamefont {A.}~\bibnamefont {Zee}},\
  }\href {\doibase 10.1016/S0550-3213(97)00419-7} {\bibfield  {journal}
  {\bibinfo  {journal} {Nuclear Physics B}\ }\textbf {\bibinfo {volume}
  {501}},\ \bibinfo {pages} {643} (\bibinfo {year} {1997})}\BibitemShut
  {NoStop}%
\bibitem [{\citenamefont {Goldsheid}\ and\ \citenamefont
  {Khoruzhenko}(1998)}]{Goldsheid98}%
  \BibitemOpen
  \bibfield  {author} {\bibinfo {author} {\bibfnamefont {I.~Y.}\ \bibnamefont
  {Goldsheid}}\ and\ \bibinfo {author} {\bibfnamefont {B.~A.}\ \bibnamefont
  {Khoruzhenko}},\ }\href {\doibase 10.1103/PhysRevLett.80.2897} {\bibfield
  {journal} {\bibinfo  {journal} {Phys. Rev. Lett.}\ }\textbf {\bibinfo
  {volume} {80}},\ \bibinfo {pages} {2897} (\bibinfo {year}
  {1998})}\BibitemShut {NoStop}%
\bibitem [{\citenamefont {Markum}\ \emph {et~al.}(1999)\citenamefont {Markum},
  \citenamefont {Pullirsch},\ and\ \citenamefont {Wettig}}]{Markum99}%
  \BibitemOpen
  \bibfield  {author} {\bibinfo {author} {\bibfnamefont {H.}~\bibnamefont
  {Markum}}, \bibinfo {author} {\bibfnamefont {R.}~\bibnamefont {Pullirsch}}, \
  and\ \bibinfo {author} {\bibfnamefont {T.}~\bibnamefont {Wettig}},\ }\href
  {\doibase 10.1103/PhysRevLett.83.484} {\bibfield  {journal} {\bibinfo
  {journal} {Phys. Rev. Lett.}\ }\textbf {\bibinfo {volume} {83}},\ \bibinfo
  {pages} {484} (\bibinfo {year} {1999})}\BibitemShut {NoStop}%
\bibitem [{\citenamefont {Feinberg}\ and\ \citenamefont
  {Zee}(1999)}]{Feinberg99}%
  \BibitemOpen
  \bibfield  {author} {\bibinfo {author} {\bibfnamefont {J.}~\bibnamefont
  {Feinberg}}\ and\ \bibinfo {author} {\bibfnamefont {A.}~\bibnamefont {Zee}},\
  }\href {\doibase 10.1103/PhysRevE.59.6433} {\bibfield  {journal} {\bibinfo
  {journal} {Phys. Rev. E}\ }\textbf {\bibinfo {volume} {59}},\ \bibinfo
  {pages} {6433} (\bibinfo {year} {1999})}\BibitemShut {NoStop}%
\bibitem [{\citenamefont {Longhi}\ \emph
  {et~al.}(2015{\natexlab{a}})\citenamefont {Longhi}, \citenamefont {Gatti},\
  and\ \citenamefont {Della~Valle}}]{Longhi15}%
  \BibitemOpen
  \bibfield  {author} {\bibinfo {author} {\bibfnamefont {S.}~\bibnamefont
  {Longhi}}, \bibinfo {author} {\bibfnamefont {D.}~\bibnamefont {Gatti}}, \
  and\ \bibinfo {author} {\bibfnamefont {G.}~\bibnamefont {Della~Valle}},\
  }\href {\doibase 10.1103/PhysRevB.92.094204} {\bibfield  {journal} {\bibinfo
  {journal} {Phys. Rev. B}\ }\textbf {\bibinfo {volume} {92}},\ \bibinfo
  {pages} {094204} (\bibinfo {year} {2015}{\natexlab{a}})}\BibitemShut
  {NoStop}%
\bibitem [{\citenamefont {Longhi}\ \emph
  {et~al.}(2015{\natexlab{b}})\citenamefont {Longhi}, \citenamefont {Gatti},\
  and\ \citenamefont {Della~Valle}}]{Longhi15R}%
  \BibitemOpen
  \bibfield  {author} {\bibinfo {author} {\bibfnamefont {S.}~\bibnamefont
  {Longhi}}, \bibinfo {author} {\bibfnamefont {D.}~\bibnamefont {Gatti}}, \
  and\ \bibinfo {author} {\bibfnamefont {G.}~\bibnamefont {Della~Valle}},\
  }\href {\doibase 10.1038/srep13376} {\bibfield  {journal} {\bibinfo
  {journal} {Scientific Reports}\ }\textbf {\bibinfo {volume} {5}},\ \bibinfo
  {pages} {13376} (\bibinfo {year} {2015}{\natexlab{b}})}\BibitemShut {NoStop}%
\bibitem [{\citenamefont {Kim}\ and\ \citenamefont {Nelson}(2001)}]{Kim01}%
  \BibitemOpen
  \bibfield  {author} {\bibinfo {author} {\bibfnamefont {K.}~\bibnamefont
  {Kim}}\ and\ \bibinfo {author} {\bibfnamefont {D.~R.}\ \bibnamefont
  {Nelson}},\ }\href {\doibase 10.1103/PhysRevB.64.054508} {\bibfield
  {journal} {\bibinfo  {journal} {Phys. Rev. B}\ }\textbf {\bibinfo {volume}
  {64}},\ \bibinfo {pages} {054508} (\bibinfo {year} {2001})}\BibitemShut
  {NoStop}%
\bibitem [{\citenamefont {Basko}\ \emph {et~al.}(2006)\citenamefont {Basko},
  \citenamefont {Aleiner},\ and\ \citenamefont {Altshuler}}]{Basko06}%
  \BibitemOpen
  \bibfield  {author} {\bibinfo {author} {\bibfnamefont {D.}~\bibnamefont
  {Basko}}, \bibinfo {author} {\bibfnamefont {I.}~\bibnamefont {Aleiner}}, \
  and\ \bibinfo {author} {\bibfnamefont {B.}~\bibnamefont {Altshuler}},\ }\href
  {\doibase 10.1016/j.aop.2005.11.014} {\bibfield  {journal} {\bibinfo
  {journal} {Annals of physics}\ }\textbf {\bibinfo {volume} {321}},\ \bibinfo
  {pages} {1126} (\bibinfo {year} {2006})}\BibitemShut {NoStop}%
\bibitem [{\citenamefont {\ifmmode \check{Z}\else
  \v{Z}\fi{}nidari\ifmmode~\check{c}\else \v{c}\fi{}}\ \emph
  {et~al.}(2008)\citenamefont {\ifmmode \check{Z}\else
  \v{Z}\fi{}nidari\ifmmode~\check{c}\else \v{c}\fi{}}, \citenamefont {Prosen},\
  and\ \citenamefont {Prelov\ifmmode~\check{s}\else
  \v{s}\fi{}ek}}]{Zunidaric08}%
  \BibitemOpen
  \bibfield  {author} {\bibinfo {author} {\bibfnamefont {M.}~\bibnamefont
  {\ifmmode \check{Z}\else \v{Z}\fi{}nidari\ifmmode~\check{c}\else
  \v{c}\fi{}}}, \bibinfo {author} {\bibfnamefont {T.}~\bibnamefont {Prosen}}, \
  and\ \bibinfo {author} {\bibfnamefont {P.}~\bibnamefont
  {Prelov\ifmmode~\check{s}\else \v{s}\fi{}ek}},\ }\href {\doibase
  10.1103/PhysRevB.77.064426} {\bibfield  {journal} {\bibinfo  {journal} {Phys.
  Rev. B}\ }\textbf {\bibinfo {volume} {77}},\ \bibinfo {pages} {064426}
  (\bibinfo {year} {2008})}\BibitemShut {NoStop}%
\bibitem [{\citenamefont {Pal}\ and\ \citenamefont {Huse}(2010)}]{Pal10}%
  \BibitemOpen
  \bibfield  {author} {\bibinfo {author} {\bibfnamefont {A.}~\bibnamefont
  {Pal}}\ and\ \bibinfo {author} {\bibfnamefont {D.~A.}\ \bibnamefont {Huse}},\
  }\href {\doibase 10.1103/PhysRevB.82.174411} {\bibfield  {journal} {\bibinfo
  {journal} {Phys. Rev. B}\ }\textbf {\bibinfo {volume} {82}},\ \bibinfo
  {pages} {174411} (\bibinfo {year} {2010})}\BibitemShut {NoStop}%
\bibitem [{\citenamefont {Gogolin}\ \emph {et~al.}(2011)\citenamefont
  {Gogolin}, \citenamefont {M\"uller},\ and\ \citenamefont
  {Eisert}}]{Gogolin11}%
  \BibitemOpen
  \bibfield  {author} {\bibinfo {author} {\bibfnamefont {C.}~\bibnamefont
  {Gogolin}}, \bibinfo {author} {\bibfnamefont {M.~P.}\ \bibnamefont
  {M\"uller}}, \ and\ \bibinfo {author} {\bibfnamefont {J.}~\bibnamefont
  {Eisert}},\ }\href {\doibase 10.1103/PhysRevLett.106.040401} {\bibfield
  {journal} {\bibinfo  {journal} {Phys. Rev. Lett.}\ }\textbf {\bibinfo
  {volume} {106}},\ \bibinfo {pages} {040401} (\bibinfo {year}
  {2011})}\BibitemShut {NoStop}%
\bibitem [{\citenamefont {Bardarson}\ \emph {et~al.}(2012)\citenamefont
  {Bardarson}, \citenamefont {Pollmann},\ and\ \citenamefont
  {Moore}}]{Bardarson12}%
  \BibitemOpen
  \bibfield  {author} {\bibinfo {author} {\bibfnamefont {J.~H.}\ \bibnamefont
  {Bardarson}}, \bibinfo {author} {\bibfnamefont {F.}~\bibnamefont {Pollmann}},
  \ and\ \bibinfo {author} {\bibfnamefont {J.~E.}\ \bibnamefont {Moore}},\
  }\href {\doibase 10.1103/PhysRevLett.109.017202} {\bibfield  {journal}
  {\bibinfo  {journal} {Phys. Rev. Lett.}\ }\textbf {\bibinfo {volume} {109}},\
  \bibinfo {pages} {017202} (\bibinfo {year} {2012})}\BibitemShut {NoStop}%
\bibitem [{\citenamefont {Iyer}\ \emph {et~al.}(2013)\citenamefont {Iyer},
  \citenamefont {Oganesyan}, \citenamefont {Refael},\ and\ \citenamefont
  {Huse}}]{Iyer13}%
  \BibitemOpen
  \bibfield  {author} {\bibinfo {author} {\bibfnamefont {S.}~\bibnamefont
  {Iyer}}, \bibinfo {author} {\bibfnamefont {V.}~\bibnamefont {Oganesyan}},
  \bibinfo {author} {\bibfnamefont {G.}~\bibnamefont {Refael}}, \ and\ \bibinfo
  {author} {\bibfnamefont {D.~A.}\ \bibnamefont {Huse}},\ }\href {\doibase
  10.1103/PhysRevB.87.134202} {\bibfield  {journal} {\bibinfo  {journal} {Phys.
  Rev. B}\ }\textbf {\bibinfo {volume} {87}},\ \bibinfo {pages} {134202}
  (\bibinfo {year} {2013})}\BibitemShut {NoStop}%
\bibitem [{\citenamefont {Huse}\ \emph {et~al.}(2013)\citenamefont {Huse},
  \citenamefont {Nandkishore}, \citenamefont {Oganesyan}, \citenamefont {Pal},\
  and\ \citenamefont {Sondhi}}]{Huse13}%
  \BibitemOpen
  \bibfield  {author} {\bibinfo {author} {\bibfnamefont {D.~A.}\ \bibnamefont
  {Huse}}, \bibinfo {author} {\bibfnamefont {R.}~\bibnamefont {Nandkishore}},
  \bibinfo {author} {\bibfnamefont {V.}~\bibnamefont {Oganesyan}}, \bibinfo
  {author} {\bibfnamefont {A.}~\bibnamefont {Pal}}, \ and\ \bibinfo {author}
  {\bibfnamefont {S.~L.}\ \bibnamefont {Sondhi}},\ }\href {\doibase
  10.1103/PhysRevB.88.014206} {\bibfield  {journal} {\bibinfo  {journal} {Phys.
  Rev. B}\ }\textbf {\bibinfo {volume} {88}},\ \bibinfo {pages} {014206}
  (\bibinfo {year} {2013})}\BibitemShut {NoStop}%
\bibitem [{\citenamefont {Serbyn}\ \emph {et~al.}(2013)\citenamefont {Serbyn},
  \citenamefont {Papi\ifmmode~\acute{c}\else \'{c}\fi{}},\ and\ \citenamefont
  {Abanin}}]{Serbyn13}%
  \BibitemOpen
  \bibfield  {author} {\bibinfo {author} {\bibfnamefont {M.}~\bibnamefont
  {Serbyn}}, \bibinfo {author} {\bibfnamefont {Z.}~\bibnamefont
  {Papi\ifmmode~\acute{c}\else \'{c}\fi{}}}, \ and\ \bibinfo {author}
  {\bibfnamefont {D.~A.}\ \bibnamefont {Abanin}},\ }\href {\doibase
  10.1103/PhysRevLett.111.127201} {\bibfield  {journal} {\bibinfo  {journal}
  {Phys. Rev. Lett.}\ }\textbf {\bibinfo {volume} {111}},\ \bibinfo {pages}
  {127201} (\bibinfo {year} {2013})}\BibitemShut {NoStop}%
\bibitem [{\citenamefont {Huse}\ \emph {et~al.}(2014)\citenamefont {Huse},
  \citenamefont {Nandkishore},\ and\ \citenamefont {Oganesyan}}]{Huse14}%
  \BibitemOpen
  \bibfield  {author} {\bibinfo {author} {\bibfnamefont {D.~A.}\ \bibnamefont
  {Huse}}, \bibinfo {author} {\bibfnamefont {R.}~\bibnamefont {Nandkishore}}, \
  and\ \bibinfo {author} {\bibfnamefont {V.}~\bibnamefont {Oganesyan}},\ }\href
  {\doibase 10.1103/PhysRevB.90.174202} {\bibfield  {journal} {\bibinfo
  {journal} {Phys. Rev. B}\ }\textbf {\bibinfo {volume} {90}},\ \bibinfo
  {pages} {174202} (\bibinfo {year} {2014})}\BibitemShut {NoStop}%
\bibitem [{\citenamefont {Nandkishore}\ and\ \citenamefont
  {Huse}(2015)}]{Nandkishore14}%
  \BibitemOpen
  \bibfield  {author} {\bibinfo {author} {\bibfnamefont {R.}~\bibnamefont
  {Nandkishore}}\ and\ \bibinfo {author} {\bibfnamefont {D.~A.}\ \bibnamefont
  {Huse}},\ }\href {\doibase 10.1146/annurev-conmatphys-031214-014726}
  {\bibfield  {journal} {\bibinfo  {journal} {Annual Review of Condensed Matter
  Physics}\ }\textbf {\bibinfo {volume} {6}},\ \bibinfo {pages} {15} (\bibinfo
  {year} {2015})}\BibitemShut {NoStop}%
\bibitem [{\citenamefont {Bera}\ \emph {et~al.}(2015)\citenamefont {Bera},
  \citenamefont {Schomerus}, \citenamefont {Heidrich-Meisner},\ and\
  \citenamefont {Bardarson}}]{Bera15}%
  \BibitemOpen
  \bibfield  {author} {\bibinfo {author} {\bibfnamefont {S.}~\bibnamefont
  {Bera}}, \bibinfo {author} {\bibfnamefont {H.}~\bibnamefont {Schomerus}},
  \bibinfo {author} {\bibfnamefont {F.}~\bibnamefont {Heidrich-Meisner}}, \
  and\ \bibinfo {author} {\bibfnamefont {J.~H.}\ \bibnamefont {Bardarson}},\
  }\href {\doibase 10.1103/PhysRevLett.115.046603} {\bibfield  {journal}
  {\bibinfo  {journal} {Phys. Rev. Lett.}\ }\textbf {\bibinfo {volume} {115}},\
  \bibinfo {pages} {046603} (\bibinfo {year} {2015})}\BibitemShut {NoStop}%
\bibitem [{\citenamefont {Ponte}\ \emph {et~al.}(2015)\citenamefont {Ponte},
  \citenamefont {Papi\ifmmode~\acute{c}\else \'{c}\fi{}}, \citenamefont
  {Huveneers},\ and\ \citenamefont {Abanin}}]{Ponte15}%
  \BibitemOpen
  \bibfield  {author} {\bibinfo {author} {\bibfnamefont {P.}~\bibnamefont
  {Ponte}}, \bibinfo {author} {\bibfnamefont {Z.}~\bibnamefont
  {Papi\ifmmode~\acute{c}\else \'{c}\fi{}}}, \bibinfo {author} {\bibfnamefont
  {F.}~\bibnamefont {Huveneers}}, \ and\ \bibinfo {author} {\bibfnamefont
  {D.~A.}\ \bibnamefont {Abanin}},\ }\href {\doibase
  10.1103/PhysRevLett.114.140401} {\bibfield  {journal} {\bibinfo  {journal}
  {Phys. Rev. Lett.}\ }\textbf {\bibinfo {volume} {114}},\ \bibinfo {pages}
  {140401} (\bibinfo {year} {2015})}\BibitemShut {NoStop}%
\bibitem [{\citenamefont {Schreiber}\ \emph {et~al.}(2015)\citenamefont
  {Schreiber}, \citenamefont {Hodgman}, \citenamefont {Bordia}, \citenamefont
  {L\"{u}schen}, \citenamefont {Fischer}, \citenamefont {Vosk}, \citenamefont
  {Altman}, \citenamefont {Schneider},\ and\ \citenamefont
  {Bloch}}]{Schreiber15}%
  \BibitemOpen
  \bibfield  {author} {\bibinfo {author} {\bibfnamefont {M.}~\bibnamefont
  {Schreiber}}, \bibinfo {author} {\bibfnamefont {S.~S.}\ \bibnamefont
  {Hodgman}}, \bibinfo {author} {\bibfnamefont {P.}~\bibnamefont {Bordia}},
  \bibinfo {author} {\bibfnamefont {H.~P.}\ \bibnamefont {L\"{u}schen}},
  \bibinfo {author} {\bibfnamefont {M.~H.}\ \bibnamefont {Fischer}}, \bibinfo
  {author} {\bibfnamefont {R.}~\bibnamefont {Vosk}}, \bibinfo {author}
  {\bibfnamefont {E.}~\bibnamefont {Altman}}, \bibinfo {author} {\bibfnamefont
  {U.}~\bibnamefont {Schneider}}, \ and\ \bibinfo {author} {\bibfnamefont
  {I.}~\bibnamefont {Bloch}},\ }\href
  {http://www.sciencemag.org/content/349/6250/842.abstract} {\bibfield
  {journal} {\bibinfo  {journal} {Science}\ }\textbf {\bibinfo {volume}
  {349}},\ \bibinfo {pages} {842} (\bibinfo {year} {2015})}\BibitemShut
  {NoStop}%
\bibitem [{\citenamefont {Luitz}\ \emph {et~al.}(2015)\citenamefont {Luitz},
  \citenamefont {Laflorencie},\ and\ \citenamefont {Alet}}]{Luitz15}%
  \BibitemOpen
  \bibfield  {author} {\bibinfo {author} {\bibfnamefont {D.~J.}\ \bibnamefont
  {Luitz}}, \bibinfo {author} {\bibfnamefont {N.}~\bibnamefont {Laflorencie}},
  \ and\ \bibinfo {author} {\bibfnamefont {F.}~\bibnamefont {Alet}},\ }\href
  {\doibase 10.1103/PhysRevB.91.081103} {\bibfield  {journal} {\bibinfo
  {journal} {Physical Review B}\ }\textbf {\bibinfo {volume} {91}},\ \bibinfo
  {pages} {081103(R)} (\bibinfo {year} {2015})}\BibitemShut {NoStop}%
\bibitem [{\citenamefont {Smith}\ \emph {et~al.}(2016)\citenamefont {Smith},
  \citenamefont {Lee}, \citenamefont {Richerme}, \citenamefont {Neyenhuis},
  \citenamefont {Hess}, \citenamefont {Hauke}, \citenamefont {Heyl},
  \citenamefont {Huse},\ and\ \citenamefont {Monroe}}]{Smith16}%
  \BibitemOpen
  \bibfield  {author} {\bibinfo {author} {\bibfnamefont {J.}~\bibnamefont
  {Smith}}, \bibinfo {author} {\bibfnamefont {A.}~\bibnamefont {Lee}}, \bibinfo
  {author} {\bibfnamefont {P.}~\bibnamefont {Richerme}}, \bibinfo {author}
  {\bibfnamefont {B.}~\bibnamefont {Neyenhuis}}, \bibinfo {author}
  {\bibfnamefont {P.~W.}\ \bibnamefont {Hess}}, \bibinfo {author}
  {\bibfnamefont {P.}~\bibnamefont {Hauke}}, \bibinfo {author} {\bibfnamefont
  {M.}~\bibnamefont {Heyl}}, \bibinfo {author} {\bibfnamefont {D.~A.}\
  \bibnamefont {Huse}}, \ and\ \bibinfo {author} {\bibfnamefont
  {C.}~\bibnamefont {Monroe}},\ }\href {\doibase 10.1038/nphys3783} {\bibfield
  {journal} {\bibinfo  {journal} {Nature Physics}\ }\textbf {\bibinfo {volume}
  {12}},\ \bibinfo {pages} {907} (\bibinfo {year} {2016})}\BibitemShut
  {NoStop}%
\bibitem [{\citenamefont {Choi}\ \emph {et~al.}(2016)\citenamefont {Choi},
  \citenamefont {Hild}, \citenamefont {Zeiher}, \citenamefont {Schau{\ss}},
  \citenamefont {Rubio-Abadal}, \citenamefont {Yefsah}, \citenamefont
  {Khemani}, \citenamefont {Huse}, \citenamefont {Bloch},\ and\ \citenamefont
  {Gross}}]{Choi16}%
  \BibitemOpen
  \bibfield  {author} {\bibinfo {author} {\bibfnamefont {J.-y.}\ \bibnamefont
  {Choi}}, \bibinfo {author} {\bibfnamefont {S.}~\bibnamefont {Hild}}, \bibinfo
  {author} {\bibfnamefont {J.}~\bibnamefont {Zeiher}}, \bibinfo {author}
  {\bibfnamefont {P.}~\bibnamefont {Schau{\ss}}}, \bibinfo {author}
  {\bibfnamefont {A.}~\bibnamefont {Rubio-Abadal}}, \bibinfo {author}
  {\bibfnamefont {T.}~\bibnamefont {Yefsah}}, \bibinfo {author} {\bibfnamefont
  {V.}~\bibnamefont {Khemani}}, \bibinfo {author} {\bibfnamefont {D.~A.}\
  \bibnamefont {Huse}}, \bibinfo {author} {\bibfnamefont {I.}~\bibnamefont
  {Bloch}}, \ and\ \bibinfo {author} {\bibfnamefont {C.}~\bibnamefont
  {Gross}},\ }\href {\doibase 10.1126/science.aaf8834} {\bibfield  {journal}
  {\bibinfo  {journal} {Science}\ }\textbf {\bibinfo {volume} {352}},\ \bibinfo
  {pages} {1547} (\bibinfo {year} {2016})}\BibitemShut {NoStop}%
\bibitem [{\citenamefont {Imbrie}(2016{\natexlab{a}})}]{Imbrie16D}%
  \BibitemOpen
  \bibfield  {author} {\bibinfo {author} {\bibfnamefont {J.~Z.}\ \bibnamefont
  {Imbrie}},\ }\href {\doibase 10.1103/PhysRevLett.117.027201} {\bibfield
  {journal} {\bibinfo  {journal} {Phys. Rev. Lett.}\ }\textbf {\bibinfo
  {volume} {117}},\ \bibinfo {pages} {027201} (\bibinfo {year}
  {2016}{\natexlab{a}})}\BibitemShut {NoStop}%
\bibitem [{\citenamefont {Imbrie}(2016{\natexlab{b}})}]{Imbrie16O}%
  \BibitemOpen
  \bibfield  {author} {\bibinfo {author} {\bibfnamefont {J.~Z.}\ \bibnamefont
  {Imbrie}},\ }\href {\doibase 10.1007/s10955-016-1508-x} {\bibfield  {journal}
  {\bibinfo  {journal} {Journal of Statistical Physics}\ }\textbf {\bibinfo
  {volume} {163}},\ \bibinfo {pages} {998} (\bibinfo {year}
  {2016}{\natexlab{b}})}\BibitemShut {NoStop}%
\bibitem [{\citenamefont {Daley}(2014)}]{Daley14}%
  \BibitemOpen
  \bibfield  {author} {\bibinfo {author} {\bibfnamefont {A.~J.}\ \bibnamefont
  {Daley}},\ }\href
  {https://www.tandfonline.com/doi/abs/10.1080/00018732.2014.933502?casa_token=mPTgI8wWblgAAAAA:V5hDaaHmZD49Uqxn5FhqxjzIQgYb3f8p-QxK1p8Yyc_I_9pFaDRj4Yarb9hlfBgbfCoiPLuKuhGa}
  {\bibfield  {journal} {\bibinfo  {journal} {Advances in Physics}\ }\textbf
  {\bibinfo {volume} {63}},\ \bibinfo {pages} {77} (\bibinfo {year}
  {2014})}\BibitemShut {NoStop}%
\bibitem [{\citenamefont {Fischer}\ \emph {et~al.}(2016)\citenamefont
  {Fischer}, \citenamefont {Maksymenko},\ and\ \citenamefont
  {Altman}}]{Fischer16}%
  \BibitemOpen
  \bibfield  {author} {\bibinfo {author} {\bibfnamefont {M.~H.}\ \bibnamefont
  {Fischer}}, \bibinfo {author} {\bibfnamefont {M.}~\bibnamefont {Maksymenko}},
  \ and\ \bibinfo {author} {\bibfnamefont {E.}~\bibnamefont {Altman}},\ }\href
  {\doibase 10.1103/PhysRevLett.116.160401} {\bibfield  {journal} {\bibinfo
  {journal} {Phys. Rev. Lett.}\ }\textbf {\bibinfo {volume} {116}},\ \bibinfo
  {pages} {160401} (\bibinfo {year} {2016})}\BibitemShut {NoStop}%
\bibitem [{\citenamefont {Levi}\ \emph {et~al.}(2016)\citenamefont {Levi},
  \citenamefont {Heyl}, \citenamefont {Lesanovsky},\ and\ \citenamefont
  {Garrahan}}]{Levy16}%
  \BibitemOpen
  \bibfield  {author} {\bibinfo {author} {\bibfnamefont {E.}~\bibnamefont
  {Levi}}, \bibinfo {author} {\bibfnamefont {M.}~\bibnamefont {Heyl}}, \bibinfo
  {author} {\bibfnamefont {I.}~\bibnamefont {Lesanovsky}}, \ and\ \bibinfo
  {author} {\bibfnamefont {J.~P.}\ \bibnamefont {Garrahan}},\ }\href {\doibase
  10.1103/PhysRevLett.116.237203} {\bibfield  {journal} {\bibinfo  {journal}
  {Phys. Rev. Lett.}\ }\textbf {\bibinfo {volume} {116}},\ \bibinfo {pages}
  {237203} (\bibinfo {year} {2016})}\BibitemShut {NoStop}%
\bibitem [{\citenamefont {Medvedyeva}\ \emph
  {et~al.}(2016{\natexlab{a}})\citenamefont {Medvedyeva}, \citenamefont
  {Prosen},\ and\ \citenamefont {Znidaric}}]{Medvedyeva16I}%
  \BibitemOpen
  \bibfield  {author} {\bibinfo {author} {\bibfnamefont {M.~V.}\ \bibnamefont
  {Medvedyeva}}, \bibinfo {author} {\bibfnamefont {T.}~\bibnamefont {Prosen}},
  \ and\ \bibinfo {author} {\bibfnamefont {M.}~\bibnamefont {Znidaric}},\
  }\href {\doibase 10.1103/PhysRevB.93.094205} {\bibfield  {journal} {\bibinfo
  {journal} {Phys. Rev. B}\ }\textbf {\bibinfo {volume} {93}},\ \bibinfo
  {pages} {094205} (\bibinfo {year} {2016}{\natexlab{a}})}\BibitemShut
  {NoStop}%
\bibitem [{\citenamefont {L\"uschen}\ \emph {et~al.}(2017)\citenamefont
  {L\"uschen}, \citenamefont {Bordia}, \citenamefont {Hodgman}, \citenamefont
  {Schreiber}, \citenamefont {Sarkar}, \citenamefont {Daley}, \citenamefont
  {Fischer}, \citenamefont {Altman}, \citenamefont {Bloch},\ and\ \citenamefont
  {Schneider}}]{Luschen16}%
  \BibitemOpen
  \bibfield  {author} {\bibinfo {author} {\bibfnamefont {H.~P.}\ \bibnamefont
  {L\"uschen}}, \bibinfo {author} {\bibfnamefont {P.}~\bibnamefont {Bordia}},
  \bibinfo {author} {\bibfnamefont {S.~S.}\ \bibnamefont {Hodgman}}, \bibinfo
  {author} {\bibfnamefont {M.}~\bibnamefont {Schreiber}}, \bibinfo {author}
  {\bibfnamefont {S.}~\bibnamefont {Sarkar}}, \bibinfo {author} {\bibfnamefont
  {A.~J.}\ \bibnamefont {Daley}}, \bibinfo {author} {\bibfnamefont {M.~H.}\
  \bibnamefont {Fischer}}, \bibinfo {author} {\bibfnamefont {E.}~\bibnamefont
  {Altman}}, \bibinfo {author} {\bibfnamefont {I.}~\bibnamefont {Bloch}}, \
  and\ \bibinfo {author} {\bibfnamefont {U.}~\bibnamefont {Schneider}},\ }\href
  {\doibase 10.1103/PhysRevX.7.011034} {\bibfield  {journal} {\bibinfo
  {journal} {Phys. Rev. X}\ }\textbf {\bibinfo {volume} {7}},\ \bibinfo {pages}
  {011034} (\bibinfo {year} {2017})}\BibitemShut {NoStop}%
\bibitem [{\citenamefont {van Nieuwenburg}\ \emph {et~al.}(2017)\citenamefont
  {van Nieuwenburg}, \citenamefont {Malo}, \citenamefont {Daley},\ and\
  \citenamefont {Fischer}}]{Nieuwenburg17}%
  \BibitemOpen
  \bibfield  {author} {\bibinfo {author} {\bibfnamefont {E.~P.}\ \bibnamefont
  {van Nieuwenburg}}, \bibinfo {author} {\bibfnamefont {J.~Y.}\ \bibnamefont
  {Malo}}, \bibinfo {author} {\bibfnamefont {A.~J.}\ \bibnamefont {Daley}}, \
  and\ \bibinfo {author} {\bibfnamefont {M.~H.}\ \bibnamefont {Fischer}},\
  }\href {https://iopscience.iop.org/article/10.1088/2058-9565/aa9a02/meta}
  {\bibfield  {journal} {\bibinfo  {journal} {Quantum Science and Technology}\
  }\textbf {\bibinfo {volume} {3}},\ \bibinfo {pages} {01LT02} (\bibinfo {year}
  {2017})}\BibitemShut {NoStop}%
\bibitem [{\citenamefont {Brody}(2013)}]{Brody13}%
  \BibitemOpen
  \bibfield  {author} {\bibinfo {author} {\bibfnamefont {D.~C.}\ \bibnamefont
  {Brody}},\ }\href {\doibase 10.1088/1751-8113/47/3/035305} {\bibfield
  {journal} {\bibinfo  {journal} {Journal of Physics A: Mathematical and
  Theoretical}\ }\textbf {\bibinfo {volume} {47}},\ \bibinfo {pages} {035305}
  (\bibinfo {year} {2013})}\BibitemShut {NoStop}%
\bibitem [{TRS()}]{TRS-NHMBL}%
  \BibitemOpen
  \href@noop {} {}\bibinfo {note} {{In the presence of time-reversal symmetry
  $\hat{T}$ ($[\hat{H}_\mr{0},\hat{T}]=[\hat{V}_\mr{NH},\hat{T}]=0$), we can
  choose eigenstates with real eigenenergies such that
  $\hat{T}\ket{\mc{E}_a^{R/L}}=\ket{\mc{E}_a^{R/L}}$. Then, we can show that
  $\braket{\mc{E}_a^L|\hat{V}_\mr{NH}|\mc{E}_a^R}={(\ket{\mc{E}_a^L},\hat{V}_\mr{NH}\ket{\mc{E}_a^R})}={(\hat{T}\ket{\mc{E}_a^L},\hat{T}\hat{V}_\mr{NH}\ket{\mc{E}_a^R})^*}={(\ket{\mc{E}_a^L},\hat{V}_\mr{NH}\ket{\mc{E}_a^R})^*}=\braket{\mc{E}_a^L|\hat{V}_\mr{NH}|\mc{E}_a^R}^*$.}}\BibitemShut
  {Stop}%
\bibitem [{\citenamefont {Serbyn}\ \emph {et~al.}(2015)\citenamefont {Serbyn},
  \citenamefont {Papi\ifmmode~\acute{c}\else \'{c}\fi{}},\ and\ \citenamefont
  {Abanin}}]{Serbyn15}%
  \BibitemOpen
  \bibfield  {author} {\bibinfo {author} {\bibfnamefont {M.}~\bibnamefont
  {Serbyn}}, \bibinfo {author} {\bibfnamefont {Z.}~\bibnamefont
  {Papi\ifmmode~\acute{c}\else \'{c}\fi{}}}, \ and\ \bibinfo {author}
  {\bibfnamefont {D.~A.}\ \bibnamefont {Abanin}},\ }\href {\doibase
  10.1103/PhysRevX.5.041047} {\bibfield  {journal} {\bibinfo  {journal} {Phys.
  Rev. X}\ }\textbf {\bibinfo {volume} {5}},\ \bibinfo {pages} {041047}
  (\bibinfo {year} {2015})}\BibitemShut {NoStop}%
\bibitem [{exc()}]{exception-NHMBL}%
  \BibitemOpen
  \href@noop {} {}\bibinfo {note} {{An exceptional case is realized by
  tailoring the Hamiltonian such that
  $\hat{H}=\hat{\mc{V}}^{-1}\hat{H}_\mr{H}\hat{\mc{V}}$, where $\hat{H}_\mr{H}$
  is Hermitian and $\hat{\mc{V}}$ is a nonunitary invertible operator, since
  $\hat{H}$ and $\hat{H}_\mr{H}$ have the same eigenspectrum~\cite{footNHMBL1}.
  On the other hand, our discussion applies to general non-Hermitian
  Hamiltonians without such a specific structure. This is consistent with
  previous numerical simulations for the Lindblad
  operators~\cite{Prosen12,Medvedyeva16,Gong18D,Gong18T,Caspel18}, which can be
  mapped to non-Hermitian matrices.}}\BibitemShut {Stop}%
\bibitem [{\citenamefont {Gong}\ \emph
  {et~al.}(2018{\natexlab{a}})\citenamefont {Gong}, \citenamefont {Ashida},
  \citenamefont {Kawabata}, \citenamefont {Takasan}, \citenamefont
  {Higashikawa},\ and\ \citenamefont {Ueda}}]{Gong18T}%
  \BibitemOpen
  \bibfield  {author} {\bibinfo {author} {\bibfnamefont {Z.}~\bibnamefont
  {Gong}}, \bibinfo {author} {\bibfnamefont {Y.}~\bibnamefont {Ashida}},
  \bibinfo {author} {\bibfnamefont {K.}~\bibnamefont {Kawabata}}, \bibinfo
  {author} {\bibfnamefont {K.}~\bibnamefont {Takasan}}, \bibinfo {author}
  {\bibfnamefont {S.}~\bibnamefont {Higashikawa}}, \ and\ \bibinfo {author}
  {\bibfnamefont {M.}~\bibnamefont {Ueda}},\ }\href {\doibase
  10.1103/PhysRevX.8.031079} {\bibfield  {journal} {\bibinfo  {journal} {Phys.
  Rev. X}\ }\textbf {\bibinfo {volume} {8}},\ \bibinfo {pages} {031079}
  (\bibinfo {year} {2018}{\natexlab{a}})}\BibitemShut {NoStop}%
\bibitem [{foo()}]{footNHMBL1}%
  \BibitemOpen
  \href@noop {} {}\bibinfo {note} {See Supplemental Material for the results of
  the other parameters, quantities and models, the detail of time evolutions,
  and the discussions on the perturbation theory and the similarity
  transformation.}\BibitemShut {Stop}%
\bibitem [{ave()}]{average-NHMBL}%
  \BibitemOpen
  \href@noop {} {}\bibinfo {note} {{Data show ensemble-averaged values over
  $N_{s}$ samples with different disorder realizations, where $N_{s} = 10000$
  for $L = 6, 8, 10,12$, $N_{s} = 1000$ for $L = 14$, and $N_{s} = 100$ for $L
  = 16$.}}\BibitemShut {Stop}%
\bibitem [{den()}]{density-NHMBL}%
  \BibitemOpen
  \href@noop {} {}\bibinfo {note} {{While the density of states depends on
  energy, we here consider the average over eigenenergies for the entire energy
  range for simplicity. Thus, for sufficiently large systems, we essentially
  consider the energy range that corresponds to the infinite temperature, where
  the density of states is maximal.}}\BibitemShut {Stop}%
\bibitem [{the()}]{therm-NHMBL}%
  \BibitemOpen
  \href@noop {} {}\bibinfo {note} {{ There are no well-defined definitions for
  ergodicity and thermalization in non-Hermitian systems, since stationary
  states are not thermal. On the other hand, while the initial information of
  the state is lost for the complex-eigenergy phase, the initial information is
  retained in the real-eigenenergy phase. The latter case is reminiscent of the
  non-ergodic behavior in the Hermitian MBL phase. As another example, the
  entanglement dynamics suggests that the long-time behavior of entanglement is
  larger for the delocalized phase than the MBL phase, just as thermalization
  of Hermitian systems. This can be related to the entanglement entropy of
  eigenstates, which we find exhibits the volume/area law for the
  delocalized/MBL phase even for non-Hermitian systems}}\BibitemShut {NoStop}%
\bibitem [{\citenamefont {Haake}(2010)}]{Haake}%
  \BibitemOpen
  \bibfield  {author} {\bibinfo {author} {\bibfnamefont {F.}~\bibnamefont
  {Haake}},\ }\href {https://www.springer.com/us/book/9783642054273} {\emph
  {\bibinfo {title} {Quantum signatures of chaos}}},\ Vol.~\bibinfo {volume}
  {54}\ (\bibinfo  {publisher} {Springer Science \& Business Media},\ \bibinfo
  {year} {2010})\BibitemShut {NoStop}%
\bibitem [{rar()}]{rare-NHMBL}%
  \BibitemOpen
  \href@noop {} {}\bibinfo {note} {{We omit the results of real eigenenergies
  for weak disorder and those of complex eigenenergies for strong disorder,
  since their fractions are small}}\BibitemShut {NoStop}%
\bibitem [{\citenamefont {Schomerus}(2017)}]{Schomerus17}%
  \BibitemOpen
  \bibfield  {author} {\bibinfo {author} {\bibfnamefont {H.}~\bibnamefont
  {Schomerus}},\ }\href
  {https://global.oup.com/academic/product/stochastic-processes-and-random-matrices-9780198797319?cc=jp&lang=en&#}
  {\bibfield  {journal} {\bibinfo  {journal} {Stochastic Processes and Random
  Matrices: Lecture Notes of the Les Houches Summer School: Volume 104, July
  2015}\ }\textbf {\bibinfo {volume} {104}},\ \bibinfo {pages} {409} (\bibinfo
  {year} {2017})}\BibitemShut {NoStop}%
\bibitem [{\citenamefont {Mehta}(2004)}]{Mehta}%
  \BibitemOpen
  \bibfield  {author} {\bibinfo {author} {\bibfnamefont {M.~L.}\ \bibnamefont
  {Mehta}},\ }\href
  {https://www.elsevier.com/books/random-matrices/lal-mehta/978-0-12-088409-4}
  {\emph {\bibinfo {title} {Random matrices}}},\ Vol.\ \bibinfo {volume} {142}\
  (\bibinfo  {publisher} {Academic press},\ \bibinfo {year} {2004})\BibitemShut
  {NoStop}%
\bibitem [{\citenamefont {Bohigas}\ \emph {et~al.}(1984)\citenamefont
  {Bohigas}, \citenamefont {Giannoni},\ and\ \citenamefont
  {Schmit}}]{Bohigas84}%
  \BibitemOpen
  \bibfield  {author} {\bibinfo {author} {\bibfnamefont {O.}~\bibnamefont
  {Bohigas}}, \bibinfo {author} {\bibfnamefont {M.~J.}\ \bibnamefont
  {Giannoni}}, \ and\ \bibinfo {author} {\bibfnamefont {C.}~\bibnamefont
  {Schmit}},\ }\href {\doibase 10.1103/PhysRevLett.52.1} {\bibfield  {journal}
  {\bibinfo  {journal} {Phys. Rev. Lett.}\ }\textbf {\bibinfo {volume} {52}},\
  \bibinfo {pages} {1} (\bibinfo {year} {1984})}\BibitemShut {NoStop}%
\bibitem [{lef()}]{left-NHMBL}%
  \BibitemOpen
  \href@noop {} {}\bibinfo {note} {{Similar results can be obtained for left
  eigenstates $\ket{E_\alpha^L}$.}}\BibitemShut {Stop}%
\bibitem [{ent()}]{entanglement-NHMBL}%
  \BibitemOpen
  \href@noop {} {}\bibinfo {note} {{Entanglement entropy is unambiguously
  calculated even for non-Hermitian systems as long as we use a normalized pure
  state.}}\BibitemShut {Stop}%
\bibitem [{mid()}]{middle-NHMBL}%
  \BibitemOpen
  \href@noop {} {}\bibinfo {note} {{This result is obtained for eigenstates
  whose eigenenergies can be either real or complex. We also find that near a
  critical point eigenstates with complex eigenenergies tend to have larger
  entropy than those with real eigenenergies.}}\BibitemShut {Stop}%
\bibitem [{\citenamefont {Bauer}\ and\ \citenamefont {Nayak}(2013)}]{Bauer13}%
  \BibitemOpen
  \bibfield  {author} {\bibinfo {author} {\bibfnamefont {B.}~\bibnamefont
  {Bauer}}\ and\ \bibinfo {author} {\bibfnamefont {C.}~\bibnamefont {Nayak}},\
  }\href {\doibase 10.1088/1742-5468/2013/09/P09005} {\bibfield  {journal}
  {\bibinfo  {journal} {Journal of Statistical Mechanics: Theory and
  Experiment}\ }\textbf {\bibinfo {volume} {2013}},\ \bibinfo {pages} {P09005}
  (\bibinfo {year} {2013})}\BibitemShut {NoStop}%
\bibitem [{\citenamefont {Srednicki}(1999)}]{Srednicki99}%
  \BibitemOpen
  \bibfield  {author} {\bibinfo {author} {\bibfnamefont {M.}~\bibnamefont
  {Srednicki}},\ }\href {\doibase 10.1088/0305-4470/32/7/007} {\bibfield
  {journal} {\bibinfo  {journal} {Journal of Physics A: Mathematical and
  General}\ }\textbf {\bibinfo {volume} {32}},\ \bibinfo {pages} {1163}
  (\bibinfo {year} {1999})}\BibitemShut {NoStop}%
\bibitem [{\citenamefont {Khatami}\ \emph {et~al.}(2013)\citenamefont
  {Khatami}, \citenamefont {Pupillo}, \citenamefont {Srednicki},\ and\
  \citenamefont {Rigol}}]{Khatami13}%
  \BibitemOpen
  \bibfield  {author} {\bibinfo {author} {\bibfnamefont {E.}~\bibnamefont
  {Khatami}}, \bibinfo {author} {\bibfnamefont {G.}~\bibnamefont {Pupillo}},
  \bibinfo {author} {\bibfnamefont {M.}~\bibnamefont {Srednicki}}, \ and\
  \bibinfo {author} {\bibfnamefont {M.}~\bibnamefont {Rigol}},\ }\href
  {\doibase 10.1103/PhysRevLett.111.050403} {\bibfield  {journal} {\bibinfo
  {journal} {Phys. Rev. Lett.}\ }\textbf {\bibinfo {volume} {111}},\ \bibinfo
  {pages} {050403} (\bibinfo {year} {2013})}\BibitemShut {NoStop}%
\bibitem [{\citenamefont {Beugeling}\ \emph {et~al.}(2015)\citenamefont
  {Beugeling}, \citenamefont {Moessner},\ and\ \citenamefont
  {Haque}}]{Beugeling15}%
  \BibitemOpen
  \bibfield  {author} {\bibinfo {author} {\bibfnamefont {W.}~\bibnamefont
  {Beugeling}}, \bibinfo {author} {\bibfnamefont {R.}~\bibnamefont {Moessner}},
  \ and\ \bibinfo {author} {\bibfnamefont {M.}~\bibnamefont {Haque}},\ }\href
  {\doibase 10.1103/PhysRevE.91.012144} {\bibfield  {journal} {\bibinfo
  {journal} {Phys. Rev. E}\ }\textbf {\bibinfo {volume} {91}},\ \bibinfo
  {pages} {012144} (\bibinfo {year} {2015})}\BibitemShut {NoStop}%
\bibitem [{\citenamefont {Mondaini}\ and\ \citenamefont
  {Rigol}(2017)}]{Mondaini17}%
  \BibitemOpen
  \bibfield  {author} {\bibinfo {author} {\bibfnamefont {R.}~\bibnamefont
  {Mondaini}}\ and\ \bibinfo {author} {\bibfnamefont {M.}~\bibnamefont
  {Rigol}},\ }\href {\doibase 10.1103/PhysRevE.96.012157} {\bibfield  {journal}
  {\bibinfo  {journal} {Phys. Rev. E}\ }\textbf {\bibinfo {volume} {96}},\
  \bibinfo {pages} {012157} (\bibinfo {year} {2017})}\BibitemShut {NoStop}%
\bibitem [{\citenamefont {Hamazaki}\ and\ \citenamefont
  {Ueda}(2018)}]{Hamazaki18A}%
  \BibitemOpen
  \bibfield  {author} {\bibinfo {author} {\bibfnamefont {R.}~\bibnamefont
  {Hamazaki}}\ and\ \bibinfo {author} {\bibfnamefont {M.}~\bibnamefont
  {Ueda}},\ }\href {\doibase 10.1103/PhysRevLett.120.080603} {\bibfield
  {journal} {\bibinfo  {journal} {Phys. Rev. Lett.}\ }\textbf {\bibinfo
  {volume} {120}},\ \bibinfo {pages} {080603} (\bibinfo {year}
  {2018})}\BibitemShut {NoStop}%
\bibitem [{\citenamefont {Hamazaki}\ \emph {et~al.}(2019)\citenamefont
  {Hamazaki}, \citenamefont {Kawabata}, \citenamefont {Kura},\ and\
  \citenamefont {Ueda}}]{hamazaki19T}%
  \BibitemOpen
  \bibfield  {author} {\bibinfo {author} {\bibfnamefont {R.}~\bibnamefont
  {Hamazaki}}, \bibinfo {author} {\bibfnamefont {K.}~\bibnamefont {Kawabata}},
  \bibinfo {author} {\bibfnamefont {N.}~\bibnamefont {Kura}}, \ and\ \bibinfo
  {author} {\bibfnamefont {M.}~\bibnamefont {Ueda}},\ }\href
  {https://arxiv.org/abs/1904.13082} {\bibfield  {journal} {\bibinfo  {journal}
  {arXiv preprint arXiv:1904.13082}\ } (\bibinfo {year} {2019})}\BibitemShut
  {NoStop}%
\bibitem [{\citenamefont {Lee}\ and\ \citenamefont {Chan}(2014)}]{Lee14}%
  \BibitemOpen
  \bibfield  {author} {\bibinfo {author} {\bibfnamefont {T.~E.}\ \bibnamefont
  {Lee}}\ and\ \bibinfo {author} {\bibfnamefont {C.-K.}\ \bibnamefont {Chan}},\
  }\href {\doibase 10.1103/PhysRevX.4.041001} {\bibfield  {journal} {\bibinfo
  {journal} {Phys. Rev. X}\ }\textbf {\bibinfo {volume} {4}},\ \bibinfo {pages}
  {041001} (\bibinfo {year} {2014})}\BibitemShut {NoStop}%
\bibitem [{\citenamefont {Lapp}\ \emph {et~al.}(2019)\citenamefont {Lapp},
  \citenamefont {Ang'ong'a}, \citenamefont {An},\ and\ \citenamefont
  {Gadway}}]{Lapp19}%
  \BibitemOpen
  \bibfield  {author} {\bibinfo {author} {\bibfnamefont {S.}~\bibnamefont
  {Lapp}}, \bibinfo {author} {\bibfnamefont {J.}~\bibnamefont {Ang'ong'a}},
  \bibinfo {author} {\bibfnamefont {F.~A.}\ \bibnamefont {An}}, \ and\ \bibinfo
  {author} {\bibfnamefont {B.}~\bibnamefont {Gadway}},\ }\href {\doibase
  10.1088/1367-2630/ab1147} {\bibfield  {journal} {\bibinfo  {journal} {New
  Journal of Physics}\ }\textbf {\bibinfo {volume} {21}},\ \bibinfo {pages}
  {045006} (\bibinfo {year} {2019})}\BibitemShut {NoStop}%
\bibitem [{\citenamefont {Grobe}\ \emph {et~al.}(1988)\citenamefont {Grobe},
  \citenamefont {Haake},\ and\ \citenamefont {Sommers}}]{Grobe88}%
  \BibitemOpen
  \bibfield  {author} {\bibinfo {author} {\bibfnamefont {R.}~\bibnamefont
  {Grobe}}, \bibinfo {author} {\bibfnamefont {F.}~\bibnamefont {Haake}}, \ and\
  \bibinfo {author} {\bibfnamefont {H.-J.}\ \bibnamefont {Sommers}},\ }\href
  {\doibase 10.1103/PhysRevLett.61.1899} {\bibfield  {journal} {\bibinfo
  {journal} {Phys. Rev. Lett.}\ }\textbf {\bibinfo {volume} {61}},\ \bibinfo
  {pages} {1899} (\bibinfo {year} {1988})}\BibitemShut {NoStop}%
\bibitem [{\citenamefont {D'Alessio}\ \emph {et~al.}(2016)\citenamefont
  {D'Alessio}, \citenamefont {Kafri}, \citenamefont {Polkovnikov},\ and\
  \citenamefont {Rigol}}]{DAlessio16}%
  \BibitemOpen
  \bibfield  {author} {\bibinfo {author} {\bibfnamefont {L.}~\bibnamefont
  {D'Alessio}}, \bibinfo {author} {\bibfnamefont {Y.}~\bibnamefont {Kafri}},
  \bibinfo {author} {\bibfnamefont {A.}~\bibnamefont {Polkovnikov}}, \ and\
  \bibinfo {author} {\bibfnamefont {M.}~\bibnamefont {Rigol}},\ }\href
  {https://www.tandfonline.com/doi/abs/10.1080/00018732.2016.1198134}
  {\bibfield  {journal} {\bibinfo  {journal} {Advances in Physics}\ }\textbf
  {\bibinfo {volume} {65}},\ \bibinfo {pages} {239} (\bibinfo {year}
  {2016})}\BibitemShut {NoStop}%
\bibitem [{\citenamefont {Prosen}(2012)}]{Prosen12}%
  \BibitemOpen
  \bibfield  {author} {\bibinfo {author} {\bibfnamefont {T.}~\bibnamefont
  {Prosen}},\ }\href {\doibase 10.1103/PhysRevLett.109.090404} {\bibfield
  {journal} {\bibinfo  {journal} {Phys. Rev. Lett.}\ }\textbf {\bibinfo
  {volume} {109}},\ \bibinfo {pages} {090404} (\bibinfo {year}
  {2012})}\BibitemShut {NoStop}%
\bibitem [{\citenamefont {Medvedyeva}\ \emph
  {et~al.}(2016{\natexlab{b}})\citenamefont {Medvedyeva}, \citenamefont
  {Essler},\ and\ \citenamefont {Prosen}}]{Medvedyeva16}%
  \BibitemOpen
  \bibfield  {author} {\bibinfo {author} {\bibfnamefont {M.~V.}\ \bibnamefont
  {Medvedyeva}}, \bibinfo {author} {\bibfnamefont {F.~H.~L.}\ \bibnamefont
  {Essler}}, \ and\ \bibinfo {author} {\bibfnamefont {T.}~\bibnamefont
  {Prosen}},\ }\href {\doibase 10.1103/PhysRevLett.117.137202} {\bibfield
  {journal} {\bibinfo  {journal} {Phys. Rev. Lett.}\ }\textbf {\bibinfo
  {volume} {117}},\ \bibinfo {pages} {137202} (\bibinfo {year}
  {2016}{\natexlab{b}})}\BibitemShut {NoStop}%
\bibitem [{\citenamefont {Gong}\ \emph
  {et~al.}(2018{\natexlab{b}})\citenamefont {Gong}, \citenamefont {Hamazaki},\
  and\ \citenamefont {Ueda}}]{Gong18D}%
  \BibitemOpen
  \bibfield  {author} {\bibinfo {author} {\bibfnamefont {Z.}~\bibnamefont
  {Gong}}, \bibinfo {author} {\bibfnamefont {R.}~\bibnamefont {Hamazaki}}, \
  and\ \bibinfo {author} {\bibfnamefont {M.}~\bibnamefont {Ueda}},\ }\href
  {\doibase 10.1103/PhysRevLett.120.040404} {\bibfield  {journal} {\bibinfo
  {journal} {Phys. Rev. Lett.}\ }\textbf {\bibinfo {volume} {120}},\ \bibinfo
  {pages} {040404} (\bibinfo {year} {2018}{\natexlab{b}})}\BibitemShut
  {NoStop}%
\bibitem [{\citenamefont {van Caspel}\ and\ \citenamefont
  {Gritsev}(2018)}]{Caspel18}%
  \BibitemOpen
  \bibfield  {author} {\bibinfo {author} {\bibfnamefont {M.}~\bibnamefont {van
  Caspel}}\ and\ \bibinfo {author} {\bibfnamefont {V.}~\bibnamefont
  {Gritsev}},\ }\href {\doibase 10.1103/PhysRevA.97.052106} {\bibfield
  {journal} {\bibinfo  {journal} {Phys. Rev. A}\ }\textbf {\bibinfo {volume}
  {97}},\ \bibinfo {pages} {052106} (\bibinfo {year} {2018})}\BibitemShut
  {NoStop}%
\end{thebibliography}%


\begin{thebibliography}{25}%
\makeatletter
\providecommand \@ifxundefined [1]{%
 \@ifx{#1\undefined}
}%
\providecommand \@ifnum [1]{%
 \ifnum #1\expandafter \@firstoftwo
 \else \expandafter \@secondoftwo
 \fi
}%
\providecommand \@ifx [1]{%
 \ifx #1\expandafter \@firstoftwo
 \else \expandafter \@secondoftwo
 \fi
}%
\providecommand \natexlab [1]{#1}%
\providecommand \enquote  [1]{``#1''}%
\providecommand \bibnamefont  [1]{#1}%
\providecommand \bibfnamefont [1]{#1}%
\providecommand \citenamefont [1]{#1}%
\providecommand \href@noop [0]{\@secondoftwo}%
\providecommand \href [0]{\begingroup \@sanitize@url \@href}%
\providecommand \@href[1]{\@@startlink{#1}\@@href}%
\providecommand \@@href[1]{\endgroup#1\@@endlink}%
\providecommand \@sanitize@url [0]{\catcode `\\12\catcode `\$12\catcode
  `\&12\catcode `\#12\catcode `\^12\catcode `\_12\catcode `\%12\relax}%
\providecommand \@@startlink[1]{}%
\providecommand \@@endlink[0]{}%
\providecommand \url  [0]{\begingroup\@sanitize@url \@url }%
\providecommand \@url [1]{\endgroup\@href {#1}{\urlprefix }}%
\providecommand \urlprefix  [0]{URL }%
\providecommand \Eprint [0]{\href }%
\providecommand \doibase [0]{http://dx.doi.org/}%
\providecommand \selectlanguage [0]{\@gobble}%
\providecommand \bibinfo  [0]{\@secondoftwo}%
\providecommand \bibfield  [0]{\@secondoftwo}%
\providecommand \translation [1]{[#1]}%
\providecommand \BibitemOpen [0]{}%
\providecommand \bibitemStop [0]{}%
\providecommand \bibitemNoStop [0]{.\EOS\space}%
\providecommand \EOS [0]{\spacefactor3000\relax}%
\providecommand \BibitemShut  [1]{\csname bibitem#1\endcsname}%
\let\auto@bib@innerbib\@empty
\bibitem [{\citenamefont {Haake}(2010)}]{Haake}%
  \BibitemOpen
  \bibfield  {author} {\bibinfo {author} {\bibfnamefont {F.}~\bibnamefont
  {Haake}},\ }\href {https://www.springer.com/us/book/9783642054273} {\emph
  {\bibinfo {title} {Quantum signatures of chaos}}},\ Vol.~\bibinfo {volume}
  {54}\ (\bibinfo  {publisher} {Springer Science \& Business Media},\ \bibinfo
  {year} {2010})\BibitemShut {NoStop}%
\bibitem [{\citenamefont {Kim}\ and\ \citenamefont {Nelson}(2001)}]{Kim01}%
  \BibitemOpen
  \bibfield  {author} {\bibinfo {author} {\bibfnamefont {K.}~\bibnamefont
  {Kim}}\ and\ \bibinfo {author} {\bibfnamefont {D.~R.}\ \bibnamefont
  {Nelson}},\ }\href {\doibase 10.1103/PhysRevB.64.054508} {\bibfield
  {journal} {\bibinfo  {journal} {Phys. Rev. B}\ }\textbf {\bibinfo {volume}
  {64}},\ \bibinfo {pages} {054508} (\bibinfo {year} {2001})}\BibitemShut
  {NoStop}%
\bibitem [{\citenamefont {Kj\"all}\ \emph {et~al.}(2014)\citenamefont
  {Kj\"all}, \citenamefont {Bardarson},\ and\ \citenamefont
  {Pollmann}}]{Kjall14}%
  \BibitemOpen
  \bibfield  {author} {\bibinfo {author} {\bibfnamefont {J.~A.}\ \bibnamefont
  {Kj\"all}}, \bibinfo {author} {\bibfnamefont {J.~H.}\ \bibnamefont
  {Bardarson}}, \ and\ \bibinfo {author} {\bibfnamefont {F.}~\bibnamefont
  {Pollmann}},\ }\href {\doibase 10.1103/PhysRevLett.113.107204} {\bibfield
  {journal} {\bibinfo  {journal} {Phys. Rev. Lett.}\ }\textbf {\bibinfo
  {volume} {113}},\ \bibinfo {pages} {107204} (\bibinfo {year}
  {2014})}\BibitemShut {NoStop}%
\bibitem [{\citenamefont {Daley}(2014)}]{Daley14}%
  \BibitemOpen
  \bibfield  {author} {\bibinfo {author} {\bibfnamefont {A.~J.}\ \bibnamefont
  {Daley}},\ }\href
  {https://www.tandfonline.com/doi/abs/10.1080/00018732.2014.933502?casa_token=mPTgI8wWblgAAAAA:V5hDaaHmZD49Uqxn5FhqxjzIQgYb3f8p-QxK1p8Yyc_I_9pFaDRj4Yarb9hlfBgbfCoiPLuKuhGa}
  {\bibfield  {journal} {\bibinfo  {journal} {Advances in Physics}\ }\textbf
  {\bibinfo {volume} {63}},\ \bibinfo {pages} {77} (\bibinfo {year}
  {2014})}\BibitemShut {NoStop}%
\bibitem [{\citenamefont {\ifmmode \check{Z}\else
  \v{Z}\fi{}nidari\ifmmode~\check{c}\else \v{c}\fi{}}\ \emph
  {et~al.}(2008)\citenamefont {\ifmmode \check{Z}\else
  \v{Z}\fi{}nidari\ifmmode~\check{c}\else \v{c}\fi{}}, \citenamefont {Prosen},\
  and\ \citenamefont {Prelov\ifmmode~\check{s}\else
  \v{s}\fi{}ek}}]{Zunidaric08}%
  \BibitemOpen
  \bibfield  {author} {\bibinfo {author} {\bibfnamefont {M.}~\bibnamefont
  {\ifmmode \check{Z}\else \v{Z}\fi{}nidari\ifmmode~\check{c}\else
  \v{c}\fi{}}}, \bibinfo {author} {\bibfnamefont {T.}~\bibnamefont {Prosen}}, \
  and\ \bibinfo {author} {\bibfnamefont {P.}~\bibnamefont
  {Prelov\ifmmode~\check{s}\else \v{s}\fi{}ek}},\ }\href {\doibase
  10.1103/PhysRevB.77.064426} {\bibfield  {journal} {\bibinfo  {journal} {Phys.
  Rev. B}\ }\textbf {\bibinfo {volume} {77}},\ \bibinfo {pages} {064426}
  (\bibinfo {year} {2008})}\BibitemShut {NoStop}%
\bibitem [{\citenamefont {Bardarson}\ \emph {et~al.}(2012)\citenamefont
  {Bardarson}, \citenamefont {Pollmann},\ and\ \citenamefont
  {Moore}}]{Bardarson12}%
  \BibitemOpen
  \bibfield  {author} {\bibinfo {author} {\bibfnamefont {J.~H.}\ \bibnamefont
  {Bardarson}}, \bibinfo {author} {\bibfnamefont {F.}~\bibnamefont {Pollmann}},
  \ and\ \bibinfo {author} {\bibfnamefont {J.~E.}\ \bibnamefont {Moore}},\
  }\href {\doibase 10.1103/PhysRevLett.109.017202} {\bibfield  {journal}
  {\bibinfo  {journal} {Phys. Rev. Lett.}\ }\textbf {\bibinfo {volume} {109}},\
  \bibinfo {pages} {017202} (\bibinfo {year} {2012})}\BibitemShut {NoStop}%
\bibitem [{\citenamefont {Serbyn}\ \emph {et~al.}(2013)\citenamefont {Serbyn},
  \citenamefont {Papi\ifmmode~\acute{c}\else \'{c}\fi{}},\ and\ \citenamefont
  {Abanin}}]{Serbyn13}%
  \BibitemOpen
  \bibfield  {author} {\bibinfo {author} {\bibfnamefont {M.}~\bibnamefont
  {Serbyn}}, \bibinfo {author} {\bibfnamefont {Z.}~\bibnamefont
  {Papi\ifmmode~\acute{c}\else \'{c}\fi{}}}, \ and\ \bibinfo {author}
  {\bibfnamefont {D.~A.}\ \bibnamefont {Abanin}},\ }\href {\doibase
  10.1103/PhysRevLett.111.127201} {\bibfield  {journal} {\bibinfo  {journal}
  {Phys. Rev. Lett.}\ }\textbf {\bibinfo {volume} {111}},\ \bibinfo {pages}
  {127201} (\bibinfo {year} {2013})}\BibitemShut {NoStop}%
\bibitem [{\citenamefont {Huse}\ \emph {et~al.}(2014)\citenamefont {Huse},
  \citenamefont {Nandkishore},\ and\ \citenamefont {Oganesyan}}]{Huse14}%
  \BibitemOpen
  \bibfield  {author} {\bibinfo {author} {\bibfnamefont {D.~A.}\ \bibnamefont
  {Huse}}, \bibinfo {author} {\bibfnamefont {R.}~\bibnamefont {Nandkishore}}, \
  and\ \bibinfo {author} {\bibfnamefont {V.}~\bibnamefont {Oganesyan}},\ }\href
  {\doibase 10.1103/PhysRevB.90.174202} {\bibfield  {journal} {\bibinfo
  {journal} {Phys. Rev. B}\ }\textbf {\bibinfo {volume} {90}},\ \bibinfo
  {pages} {174202} (\bibinfo {year} {2014})}\BibitemShut {NoStop}%
\bibitem [{\citenamefont {Serbyn}\ \emph {et~al.}(2015)\citenamefont {Serbyn},
  \citenamefont {Papi\ifmmode~\acute{c}\else \'{c}\fi{}},\ and\ \citenamefont
  {Abanin}}]{Serbyn15}%
  \BibitemOpen
  \bibfield  {author} {\bibinfo {author} {\bibfnamefont {M.}~\bibnamefont
  {Serbyn}}, \bibinfo {author} {\bibfnamefont {Z.}~\bibnamefont
  {Papi\ifmmode~\acute{c}\else \'{c}\fi{}}}, \ and\ \bibinfo {author}
  {\bibfnamefont {D.~A.}\ \bibnamefont {Abanin}},\ }\href {\doibase
  10.1103/PhysRevX.5.041047} {\bibfield  {journal} {\bibinfo  {journal} {Phys.
  Rev. X}\ }\textbf {\bibinfo {volume} {5}},\ \bibinfo {pages} {041047}
  (\bibinfo {year} {2015})}\BibitemShut {NoStop}%
\bibitem [{\citenamefont {Srednicki}(1999)}]{Srednicki99}%
  \BibitemOpen
  \bibfield  {author} {\bibinfo {author} {\bibfnamefont {M.}~\bibnamefont
  {Srednicki}},\ }\href {\doibase 10.1088/0305-4470/32/7/007} {\bibfield
  {journal} {\bibinfo  {journal} {Journal of Physics A: Mathematical and
  General}\ }\textbf {\bibinfo {volume} {32}},\ \bibinfo {pages} {1163}
  (\bibinfo {year} {1999})}\BibitemShut {NoStop}%
\bibitem [{\citenamefont {Khatami}\ \emph {et~al.}(2013)\citenamefont
  {Khatami}, \citenamefont {Pupillo}, \citenamefont {Srednicki},\ and\
  \citenamefont {Rigol}}]{Khatami13}%
  \BibitemOpen
  \bibfield  {author} {\bibinfo {author} {\bibfnamefont {E.}~\bibnamefont
  {Khatami}}, \bibinfo {author} {\bibfnamefont {G.}~\bibnamefont {Pupillo}},
  \bibinfo {author} {\bibfnamefont {M.}~\bibnamefont {Srednicki}}, \ and\
  \bibinfo {author} {\bibfnamefont {M.}~\bibnamefont {Rigol}},\ }\href
  {\doibase 10.1103/PhysRevLett.111.050403} {\bibfield  {journal} {\bibinfo
  {journal} {Phys. Rev. Lett.}\ }\textbf {\bibinfo {volume} {111}},\ \bibinfo
  {pages} {050403} (\bibinfo {year} {2013})}\BibitemShut {NoStop}%
\bibitem [{\citenamefont {Beugeling}\ \emph {et~al.}(2015)\citenamefont
  {Beugeling}, \citenamefont {Moessner},\ and\ \citenamefont
  {Haque}}]{Beugeling15}%
  \BibitemOpen
  \bibfield  {author} {\bibinfo {author} {\bibfnamefont {W.}~\bibnamefont
  {Beugeling}}, \bibinfo {author} {\bibfnamefont {R.}~\bibnamefont {Moessner}},
  \ and\ \bibinfo {author} {\bibfnamefont {M.}~\bibnamefont {Haque}},\ }\href
  {\doibase 10.1103/PhysRevE.91.012144} {\bibfield  {journal} {\bibinfo
  {journal} {Phys. Rev. E}\ }\textbf {\bibinfo {volume} {91}},\ \bibinfo
  {pages} {012144} (\bibinfo {year} {2015})}\BibitemShut {NoStop}%
\bibitem [{\citenamefont {Mondaini}\ and\ \citenamefont
  {Rigol}(2017)}]{Mondaini17}%
  \BibitemOpen
  \bibfield  {author} {\bibinfo {author} {\bibfnamefont {R.}~\bibnamefont
  {Mondaini}}\ and\ \bibinfo {author} {\bibfnamefont {M.}~\bibnamefont
  {Rigol}},\ }\href {\doibase 10.1103/PhysRevE.96.012157} {\bibfield  {journal}
  {\bibinfo  {journal} {Phys. Rev. E}\ }\textbf {\bibinfo {volume} {96}},\
  \bibinfo {pages} {012157} (\bibinfo {year} {2017})}\BibitemShut {NoStop}%
\bibitem [{\citenamefont {Hamazaki}\ and\ \citenamefont
  {Ueda}(2018)}]{Hamazaki18A}%
  \BibitemOpen
  \bibfield  {author} {\bibinfo {author} {\bibfnamefont {R.}~\bibnamefont
  {Hamazaki}}\ and\ \bibinfo {author} {\bibfnamefont {M.}~\bibnamefont
  {Ueda}},\ }\href {\doibase 10.1103/PhysRevLett.120.080603} {\bibfield
  {journal} {\bibinfo  {journal} {Phys. Rev. Lett.}\ }\textbf {\bibinfo
  {volume} {120}},\ \bibinfo {pages} {080603} (\bibinfo {year}
  {2018})}\BibitemShut {NoStop}%
\bibitem [{\citenamefont {Imbrie}(2016{\natexlab{a}})}]{Imbrie16D}%
  \BibitemOpen
  \bibfield  {author} {\bibinfo {author} {\bibfnamefont {J.~Z.}\ \bibnamefont
  {Imbrie}},\ }\href {\doibase 10.1103/PhysRevLett.117.027201} {\bibfield
  {journal} {\bibinfo  {journal} {Phys. Rev. Lett.}\ }\textbf {\bibinfo
  {volume} {117}},\ \bibinfo {pages} {027201} (\bibinfo {year}
  {2016}{\natexlab{a}})}\BibitemShut {NoStop}%
\bibitem [{\citenamefont {Imbrie}(2016{\natexlab{b}})}]{Imbrie16O}%
  \BibitemOpen
  \bibfield  {author} {\bibinfo {author} {\bibfnamefont {J.~Z.}\ \bibnamefont
  {Imbrie}},\ }\href {\doibase 10.1007/s10955-016-1508-x} {\bibfield  {journal}
  {\bibinfo  {journal} {Journal of Statistical Physics}\ }\textbf {\bibinfo
  {volume} {163}},\ \bibinfo {pages} {998} (\bibinfo {year}
  {2016}{\natexlab{b}})}\BibitemShut {NoStop}%
\bibitem [{\citenamefont {Vosk}\ \emph {et~al.}(2015)\citenamefont {Vosk},
  \citenamefont {Huse},\ and\ \citenamefont {Altman}}]{Vosk15}%
  \BibitemOpen
  \bibfield  {author} {\bibinfo {author} {\bibfnamefont {R.}~\bibnamefont
  {Vosk}}, \bibinfo {author} {\bibfnamefont {D.~A.}\ \bibnamefont {Huse}}, \
  and\ \bibinfo {author} {\bibfnamefont {E.}~\bibnamefont {Altman}},\ }\href
  {\doibase 10.1103/PhysRevX.5.031032} {\bibfield  {journal} {\bibinfo
  {journal} {Phys. Rev. X}\ }\textbf {\bibinfo {volume} {5}},\ \bibinfo {pages}
  {031032} (\bibinfo {year} {2015})}\BibitemShut {NoStop}%
\bibitem [{\citenamefont {Potter}\ \emph {et~al.}(2015)\citenamefont {Potter},
  \citenamefont {Vasseur},\ and\ \citenamefont {Parameswaran}}]{Potter15}%
  \BibitemOpen
  \bibfield  {author} {\bibinfo {author} {\bibfnamefont {A.~C.}\ \bibnamefont
  {Potter}}, \bibinfo {author} {\bibfnamefont {R.}~\bibnamefont {Vasseur}}, \
  and\ \bibinfo {author} {\bibfnamefont {S.~A.}\ \bibnamefont {Parameswaran}},\
  }\href {\doibase 10.1103/PhysRevX.5.031033} {\bibfield  {journal} {\bibinfo
  {journal} {Phys. Rev. X}\ }\textbf {\bibinfo {volume} {5}},\ \bibinfo {pages}
  {031033} (\bibinfo {year} {2015})}\BibitemShut {NoStop}%
\bibitem [{\citenamefont {Hatano}\ and\ \citenamefont
  {Nelson}(1996)}]{Hatano96}%
  \BibitemOpen
  \bibfield  {author} {\bibinfo {author} {\bibfnamefont {N.}~\bibnamefont
  {Hatano}}\ and\ \bibinfo {author} {\bibfnamefont {D.~R.}\ \bibnamefont
  {Nelson}},\ }\href {\doibase 10.1103/PhysRevLett.77.570} {\bibfield
  {journal} {\bibinfo  {journal} {Phys. Rev. Lett.}\ }\textbf {\bibinfo
  {volume} {77}},\ \bibinfo {pages} {570} (\bibinfo {year} {1996})}\BibitemShut
  {NoStop}%
\bibitem [{\citenamefont {Hatano}\ and\ \citenamefont
  {Nelson}(1997)}]{Hatano97}%
  \BibitemOpen
  \bibfield  {author} {\bibinfo {author} {\bibfnamefont {N.}~\bibnamefont
  {Hatano}}\ and\ \bibinfo {author} {\bibfnamefont {D.~R.}\ \bibnamefont
  {Nelson}},\ }\href {\doibase 10.1103/PhysRevB.56.8651} {\bibfield  {journal}
  {\bibinfo  {journal} {Phys. Rev. B}\ }\textbf {\bibinfo {volume} {56}},\
  \bibinfo {pages} {8651} (\bibinfo {year} {1997})}\BibitemShut {NoStop}%
\bibitem [{\citenamefont {Hatano}\ and\ \citenamefont
  {Nelson}(1998)}]{Hatano98}%
  \BibitemOpen
  \bibfield  {author} {\bibinfo {author} {\bibfnamefont {N.}~\bibnamefont
  {Hatano}}\ and\ \bibinfo {author} {\bibfnamefont {D.~R.}\ \bibnamefont
  {Nelson}},\ }\href {\doibase 10.1103/PhysRevB.58.8384} {\bibfield  {journal}
  {\bibinfo  {journal} {Phys. Rev. B}\ }\textbf {\bibinfo {volume} {58}},\
  \bibinfo {pages} {8384} (\bibinfo {year} {1998})}\BibitemShut {NoStop}%
\bibitem [{\citenamefont {Longhi}\ \emph
  {et~al.}(2015{\natexlab{a}})\citenamefont {Longhi}, \citenamefont {Gatti},\
  and\ \citenamefont {Della~Valle}}]{Longhi15}%
  \BibitemOpen
  \bibfield  {author} {\bibinfo {author} {\bibfnamefont {S.}~\bibnamefont
  {Longhi}}, \bibinfo {author} {\bibfnamefont {D.}~\bibnamefont {Gatti}}, \
  and\ \bibinfo {author} {\bibfnamefont {G.}~\bibnamefont {Della~Valle}},\
  }\href {\doibase 10.1103/PhysRevB.92.094204} {\bibfield  {journal} {\bibinfo
  {journal} {Phys. Rev. B}\ }\textbf {\bibinfo {volume} {92}},\ \bibinfo
  {pages} {094204} (\bibinfo {year} {2015}{\natexlab{a}})}\BibitemShut
  {NoStop}%
\bibitem [{\citenamefont {Longhi}\ \emph
  {et~al.}(2015{\natexlab{b}})\citenamefont {Longhi}, \citenamefont {Gatti},\
  and\ \citenamefont {Della~Valle}}]{Longhi15R}%
  \BibitemOpen
  \bibfield  {author} {\bibinfo {author} {\bibfnamefont {S.}~\bibnamefont
  {Longhi}}, \bibinfo {author} {\bibfnamefont {D.}~\bibnamefont {Gatti}}, \
  and\ \bibinfo {author} {\bibfnamefont {G.}~\bibnamefont {Della~Valle}},\
  }\href {\doibase 10.1038/srep13376} {\bibfield  {journal} {\bibinfo
  {journal} {Scientific Reports}\ }\textbf {\bibinfo {volume} {5}},\ \bibinfo
  {pages} {13376} (\bibinfo {year} {2015}{\natexlab{b}})}\BibitemShut {NoStop}%
\bibitem [{Note1()}]{Note1}%
  \BibitemOpen
  \bibinfo {note} {Some work~\cite {Potter15,Khemani17} proposed that the
  resonant regions spread over the whole sites, even when they are rare. We
  expect that our discussion is qualitatively valid in that situation as
  well.}\BibitemShut {Stop}%
\bibitem [{\citenamefont {Khemani}\ \emph {et~al.}(2017)\citenamefont
  {Khemani}, \citenamefont {Lim}, \citenamefont {Sheng},\ and\ \citenamefont
  {Huse}}]{Khemani17}%
  \BibitemOpen
  \bibfield  {author} {\bibinfo {author} {\bibfnamefont {V.}~\bibnamefont
  {Khemani}}, \bibinfo {author} {\bibfnamefont {S.~P.}\ \bibnamefont {Lim}},
  \bibinfo {author} {\bibfnamefont {D.~N.}\ \bibnamefont {Sheng}}, \ and\
  \bibinfo {author} {\bibfnamefont {D.~A.}\ \bibnamefont {Huse}},\ }\href
  {\doibase 10.1103/PhysRevX.7.021013} {\bibfield  {journal} {\bibinfo
  {journal} {Phys. Rev. X}\ }\textbf {\bibinfo {volume} {7}},\ \bibinfo {pages}
  {021013} (\bibinfo {year} {2017})}\BibitemShut {NoStop}%
\end{thebibliography}%

\end{document}


\title{
Supplemental Material for ``Non-Hermitian Many-Body Localization"}
\date{\today}
\author{Ryusuke Hamazaki}
\affiliation{
Department of Physics, University of Tokyo, 7-3-1 Hongo, Bunkyo-ku, Tokyo 113-0033, Japan
}
\author{Kohei Kawabata}
\affiliation{
Department of Physics, University of Tokyo, 7-3-1 Hongo, Bunkyo-ku, Tokyo 113-0033, Japan
}
\author{Masahito Ueda}
\affiliation{
Department of Physics, University of Tokyo, 7-3-1 Hongo, Bunkyo-ku, Tokyo 113-0033, Japan
}
\affiliation{
RIKEN Center for Emergent Matter Science (CEMS), Wako 351-0198, Japan
}

\maketitle
\section{Other parameters and models}
We here discuss real-complex transitions and non-Hermitian MBL for parameters and models different from those in the main text.

\subsection{Real-complex transition and non-Hermitian MBL for strong non-Hermiticity}

\begin{figure}
\begin{center}
\includegraphics[width=11.5cm]{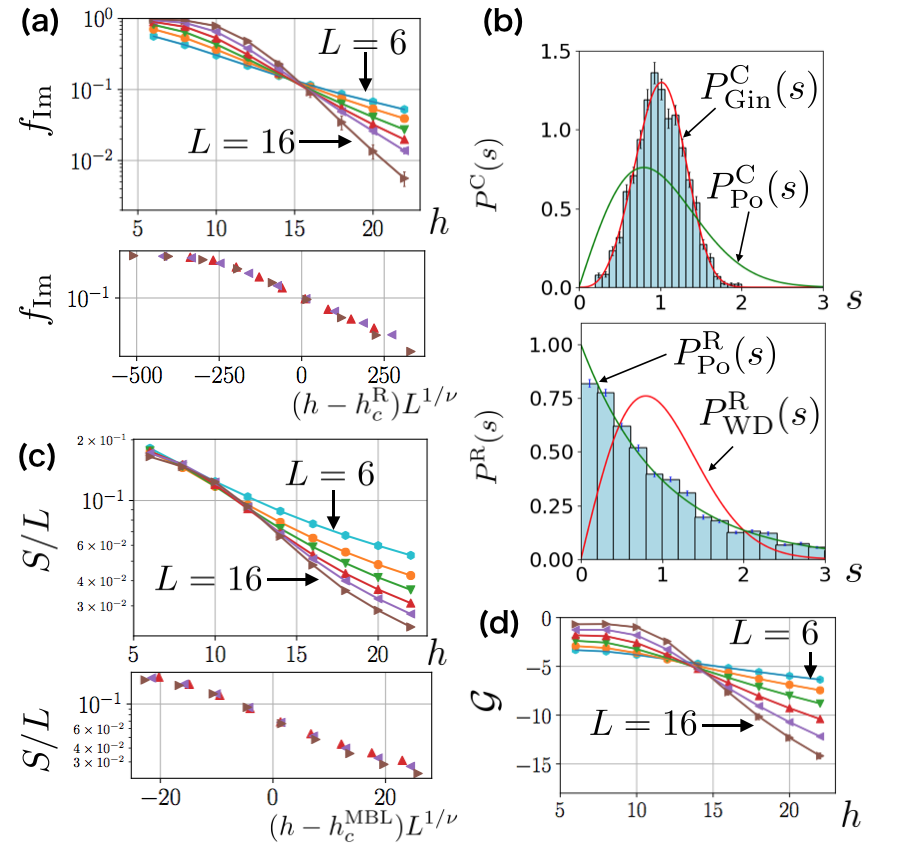}
\caption{
Real-complex transition and non-Hermitian MBL indicators in the asymmetric hopping model with $g=1$.
(a) Dependence of $f_\mr{Im}$ on $h$ for different $L$, where we find the transition point $h_c^\mr{R} \simeq 16$.
We also show the critical scaling collapse, where we choose $h_c^\mr{R}=15.7$ and $\nu=0.7$.
(b) Nearest-level-spacing distribution of (unfolded) eigenenergies on the complex plane for $h=8$ and that on the real axis for $h=20$ (single disorder realization, $L=16$).
For $h=8$, the distribution is a Ginibre distribution $P_\mr{Gin}^\mr{C}(s)$.
For $h=20$, the level-spacing distribution becomes a Poisson distribution on the real axis $P_\mr{Po}^\mr{R}(s)=e^{-s}$.
Statistics are taken for eigenstates in the middle of the spectrum (we take eigenstates whose energies lie within $\pm 10\% $ from the middle of the  real and imaginary parts of the eigenspectrum).
(c) Half-chain entanglement entropy $S/L$  in the middle of the spectrum (we take eigenstates whose energies lie within $\pm 2\% $ from the middle of the  real part of the  eigenspectrum).
For weak $h$, $S/L$ is almost constant, and
for stronger $h$, $S/L$ decreases with increasing $L$.
We also show the critical scaling collapse, where $h_c^\mr{MBL}=13.5$ and $\nu=2.5$.
(d) Stability $\mc{G}$ of eigenstates with real eigenenergies for varying disorder strength $h$ against the perturbation $\hat{V}_{\rm NH} =\hat{b}_i^\dag\hat{b}_{i+1}$ (see the main text for the definition).
With increasing $h$, $\mc{G}$ changes from $\sim \alpha L$ to $\sim -\beta L$ ($\alpha,\beta>0$), where the transition occurs at $h_c^\mr{MBL}\simeq 14$.
For (a), (c), and (d), we use $N_{s} = 10000$ for $L = 6, 8, 10, 12$, $N_{s} = 1000$ for $L = 14$, and $N_{s} = 100$ for $L = 16$.
}
\label{appfig4}
\end{center}
\end{figure}

The real-complex transition occurs also for strong non-Hermiticity, e.g., $g=1$, in the model in Eq.~(1) in the main text.
Figure \ref{appfig4}(a) shows
the $h$-dependence of $f_\mr{Im}$ for different values of $L$ with $g=1$.
As the system size increases, $f_\mr{Im}$ grows for $h\lesssim h_c^\mr{R}\simeq 16$ and decreases for $h\gtrsim h_c^\mr{R}$.
We also find the critical scaling collapse.
Note that this transition point is far from that of the Hermitian case ($g=0$) and depends on all of non-Hermiticity $g$, interaction $U$, and disorder strength $h$.

As in the case of $g=0.1$ discussed in the main text, the non-Hermitian MBL transition also takes place for $g=1$.
Figure \ref{appfig4}(b) shows the nearest-level-spacing distribution of (unfolded) eigenenergies~\cite{Haake} on the complex plane for sufficiently small $h$ ($h=8$) and that of eigenenergies on the real axis for large $h$ ($h=20$).
For weak disorder, we find that the distribution is a Ginibre distribution $P^\mr{C}_\mr{Gin}(s)$ (see the main text for the definition).
On the other hand, for large $h$, the level-spacing distribution becomes a Poisson distribution on the real axis $P_\mr{Po}^\mr{R}(s)=e^{-s}$.
This is similar to the case for $g=0.1$.

We next discuss the half-chain entanglement entropy $S$ for the right eigenstates (see the main text for the definition).
In Fig.~\ref{appfig4}(c), we show the $L$-dependence of the ratio $S/L$ averaged over the eigenstates around the middle of the real part of the spectrum for different $h$.
We find that the entanglement entropy decreases as we enter the localized regime, and that delocalized and MBL phases can be distinguished for $g=1$ as well.
We also find the critical scaling collapse.

In Fig.~\ref{appfig4}(d), we show the $h$-dependence of $\mc{G}$ for the perturbation $\hat{V}_\mr{NH}=\hat{b}_i^\dag\hat{b}_{i+1}$ (see the main text for the definition).
We see $\mc{G} \sim \alpha L\:(\alpha>0)$ in the delocalized phase and $\mc{G}\sim -\beta L\:(\beta>0)$  in the localized phase.
We can identify the non-Hermitian MBL transition point to be the one at which $\mc{G}$ is independent of $L$, which is $h_c^\mr{MBL}\simeq 14$, and close to $h_c^\mr{R}$ for $g=1$, where the small deviation is presumably due to a finite-size effect.

\subsection{Real-complex transition with varying non-Hermiticity $g$}
\begin{figure}
\begin{center}
\includegraphics[width=\linewidth]{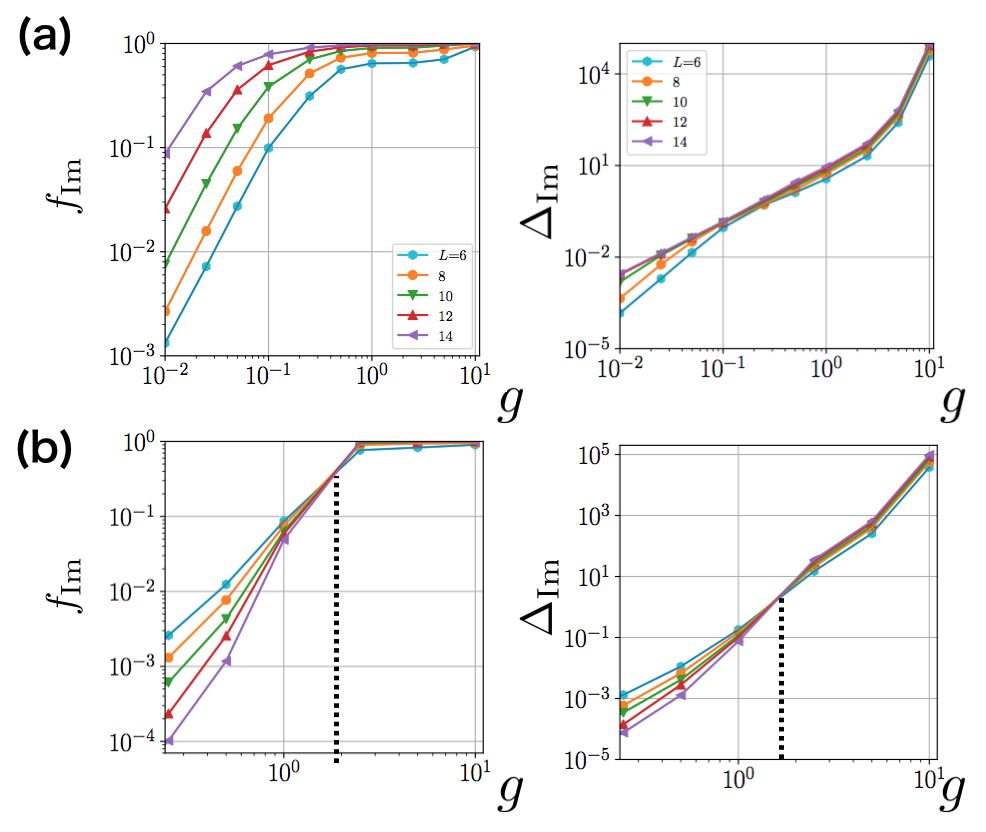}
\caption{
Dependences  of $f_{\rm Im}$ and $\Delta_{\rm Im}$ on the non-Hermiticity $g$ in the model described by Eq.~(1) in the main text.
(a)
With weak disorder $h=2<h_{0c}^\mr{MBL}$,
both $f_\mr{Im}$ and $\Delta_\mr{Im}$ increase for all $g$ with increasing the system size.
(b)
With strong disorder $h=18>h_{0c}^\mr{MBL}$,
both $f_\mr{Im}$ and $\Delta_\mr{Im}$ decrease for $g \lesssim 2$ and increase for $g\gtrsim 2$ with increasing the system size.
}
\label{appfig3}
\end{center}
\end{figure}

Here by fixing $h$ and varying $g$ for the model in Eq.~(1) in the main text, we investigate the transition point $g_c^\mr{R}$ of the real-complex transition.
We first show the $g$-dependences of $f_\mr{Im}$ and $\Delta_\mr{Im}$ for $h=2<h_{0c}^\mr{MBL}$ in Fig.~\ref{appfig3}(a).
Here $\Delta_{\rm Im}$ is the maximum value of imaginary parts of eigenenergies defined by Eq.~\eqref{eqmax} (see Appendix~\ref{maxv} for details).
Both $f_\mr{Im}$ and $\Delta_\mr{Im}$ increase for all $g$ with increasing the  system size.
The figure also implies that this behavior holds for infinitesimal $g$, which means  $g_c^\mr{R}=0$ for $h=2$.

Next, we show the $g$-dependences of $f_\mr{Im}$ and $\Delta_\mr{Im}$ for $h=18>h_{0c}^\mr{MBL}$ in Fig.~\ref{appfig3}(b).
Both $f_\mr{Im}$ and $\Delta_\mr{Im}$ decrease for $g \lesssim 2$ and increase for $g\gtrsim 2$ with increasing the size of the system.
This means that the real-complex transition occurs at $g_c^\mr{R}\simeq 2$ for $h=18$ in the thermodynamic limit.
In general, we have $g_c^\mr{R}>0$ only for $h>h_{0c}^\mr{MBL}$ in this model.

\subsection{Quarter-filling case}
\begin{figure}
\begin{center}
\includegraphics[width=\linewidth]{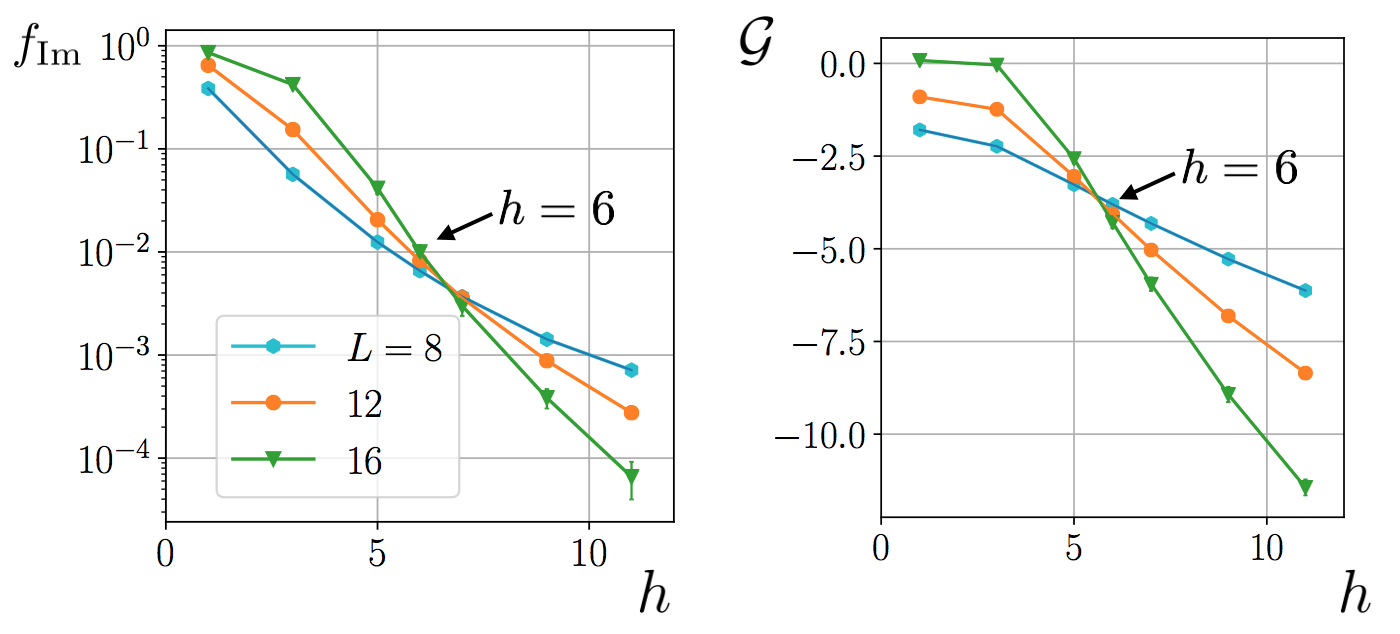}
\caption{
Dependences of $f_{\rm Im}$ and $\mc{G}$ on  $h$ in the model described by Eq.~(1) in the main text for the quarter-filling case ($L=4M$).
Both the real-complex phase transition and the MBL transition occur also for this filling. The two transition points are close to each other (around $h\simeq 6$), although they are slightly different in this system size.
We use $N_{s} = 10000$ for $L = 8, 12$, and $N_{s} = 100$ for $L = 16$.
}
\label{appfigquar}
\end{center}
\end{figure}

We here demonstrate that the model in Eq.~(1) in the main text for quarter-filling ($L=4M$) also exhibits the real-complex and MBL transitions.
Figure~\ref{appfigquar} shows that both real-complex phase transition (about $f_\mr{Im}$) and MBL transition (about $\mc{G}$) do occur for this case. 
The two transition points are close (around $h\simeq 6$), where a slight deviation is presumably due to a finite-size effect.

\subsection{Real-complex transition for the non-Hermitian Bose-Hubbard model}
While we investigate the model with hard-core bosons in the main text, we here show that the real-complex phase transition of many-body eigenenergies occurs in the  disordered Bose-Hubbard model~\cite{Kim01} with asymmetric hopping.
The Hamiltonian can be written as
\aln{\label{model}
\hat{H}_\mr{BH}=-J\sum_{i=1}^L(e^{-g}\ha_{i+1}^\dag\ha_i+e^{g}\ha_{i}^\dag\ha_{i+1})+\frac{U}{2}\sum_{i=1}^L\hat{n}'_i(\hat{n}'_{i}-1)+\sum_{i=1}^Lh_i\hat{n}'_i,
}
where $\hat{n}'_i =\hat{a}_i^\dag\hat{a}_i$ is the particle-number operator at site $i$ with $\hat{a}_i$ being the annihilation operator of a  boson at site $i$.
Here the prime ($'$) is used to distinguish $\hat{n}'_i$ from the particle-number operator $\hat{n}_i$ of hard-core bosons in the main text.
We again consider $J=1$, $U=2$, and the half-filling case with the periodic boundary condition.

\begin{figure}
\begin{center}
\includegraphics[width=\linewidth]{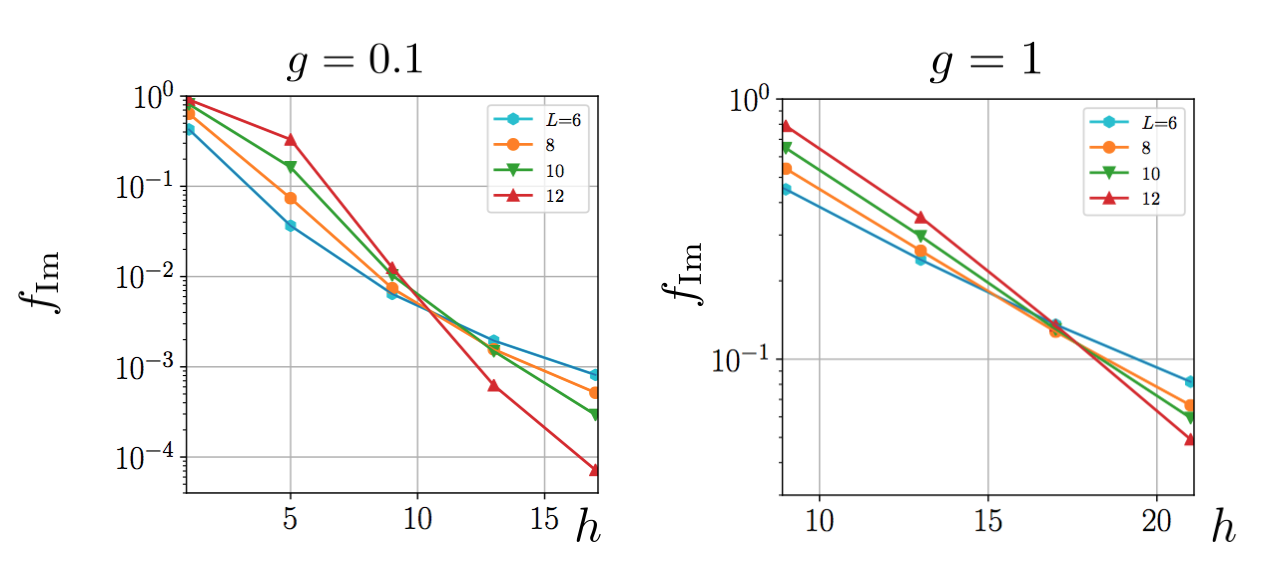}
\caption{
Real-complex transition of the non-Hermitian Bose-Hubbard model described by Eq.~(S-\ref{model}) as a function of the disorder strength $h$ for $g=0.1$ (left) and 1 (right).
As the system size $L$ increases, the fraction $f_\mr{Im}$ of complex eigenenergies increases for $h\lesssim h_c^\mr{R}$ and decreases for $h\gtrsim h_c^\mr{R}$, where the critical disorder strength is $h_c^\mr{R}\simeq 10$ for $g=0.1$ and $h_c^\mr{R}\simeq 17$ for $g=1$.
Data show the  average values of $f_\mathrm{Im}$ over $N_{s}$ samples, where $N_{s} = 10000$ for $L = 6, 8$, $N_{s} = 1000$ for $L = 10$, and $N_{s} = 100$ for $L=12$.
}
\label{appfig1}
\end{center}
\end{figure}

Figure \ref{appfig1} shows the fraction $f_\mr{Im}$ of complex eigenenergies as a function of the disorder strength $h$.
The real-complex transition occurs at a critical disorder strength $h_c^\mr{R}$, as in the case of hard-core bosons discussed in the main text.
Note that the numerically achievable size of the system is smaller for the Bose-Hubbard model than the case of hard-core bosons.

\section{Other measures to determine the transitions}
Here we discuss other measures to determine the transitions.

\subsection{Maximum values of imaginary parts}\label{maxv}
While we discuss the fraction $f_{\rm Im}$ of complex eigenenergies in the main text, we can also consider the maximum value $\Delta_{\rm Im}$ of imaginary parts of eigenenergies as yet another indicator of the real-complex transition, which is defined as
\aln{\label{eqmax}
\Delta_\mr{Im}=\av{\max_\alpha|\mr{Im}[E_\alpha]|},
}
where $E_\alpha$ is an eigenenergy of $\hat{H}$.
Note that $\max_\alpha|\mr{Im}[E_\alpha]|$ governs the dynamical instability of non-Hermitian systems:
since the imaginary part of eigenenergies describes the rate of amplification or attenuation of that mode, the system is shown to be stable for $t\lesssim [\max_\alpha|\mr{Im}[E_\alpha]|]^{-1}$ (see also Fig. 2(c) in the main text and Appendix III in this Supplemental Material).
Note that this real-complex transition point can, in general, be different from the transition point of $f_\mr{Im}$ and that of the non-Hermitian many-body localization (MBL).

\begin{figure}
\begin{center}
\includegraphics[width=\linewidth]{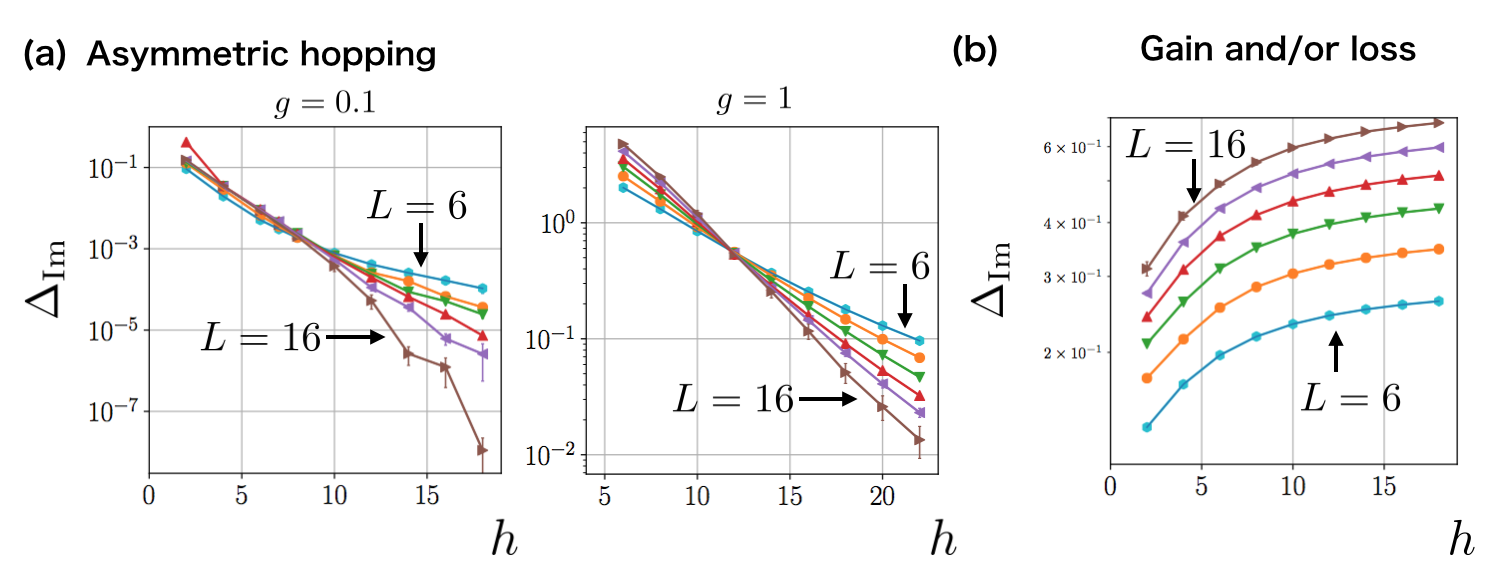}
\caption{
Maximum values of imaginary parts of eigenenergies for
(a) the model with asymmetric hopping ($g=0.1$ and 1) described by Eq.~(1) in the main text and (b) the model with gain and loss ($\gamma=0.1$) described by Eq.~(2) in the main text.
In (a), $\Delta_\mr{Im}$ slowly increases and saturates with increasing $L$ for $h\lesssim 8 \:(g=0.1)$ or $h\lesssim  12\: (g=1)$, and rapidly decreases for $h\gtrsim 8 \:(g=0.1)$ or $h\gtrsim 12\: (g=1)$ with increasing $L$.
In (b), $\Delta_\mr{Im}$ increases for any $h$ with increasing the system size, which means the absence of the real-complex transition.
We use $N_{s} = 10000$ for $L = 6, 8, 10, 12$, $N_{s} = 1000$ for $L = 14$, and $N_{s} = 100$ for $L=16$ for both (a) and (b).
}
\label{appfig2}
\end{center}
\end{figure}

In Fig.~\ref{appfig2}, we show the $h$-dependence of $\Delta_\mr{Im}$ for different values of $L$ with asymmetric hopping (see Eq.~(1) in the main text) or with gain and loss (see Eq.~(2) in the main text).
For the former case, $\Delta_\mr{Im}$ slowly increases and saturates with increasing $L$ for $h\lesssim 8 \:(g=0.1)$ or $h\lesssim  12\: (g=1)$, whereas it rapidly decreases for $h\gtrsim 8\: (g=0.1)$ or $h\gtrsim 12\: (g=1)$.
This indicates the presence of a real-complex transition also for the measure of $\Delta_\mr{Im}$.
On the other hand, for the latter case, $\Delta_\mr{Im}$ monotonically decreases with increasing $L$ for all $h$, which indicates the absence of the real-complex transition.
Note that the transition point of $\Delta_\mr{Im}$ for the asymmetric-hopping model is different from $h_c^\mr{R}$ and $h_c^\mr{MBL}$ for this finite system size, especially for $g=1$.
We leave the precise evaluations of the transition points for larger system sizes for a future work.

\subsection{Entanglement entropy and stability indicator as functions of $L$}
\begin{figure}
\begin{center}
\includegraphics[width=\linewidth]{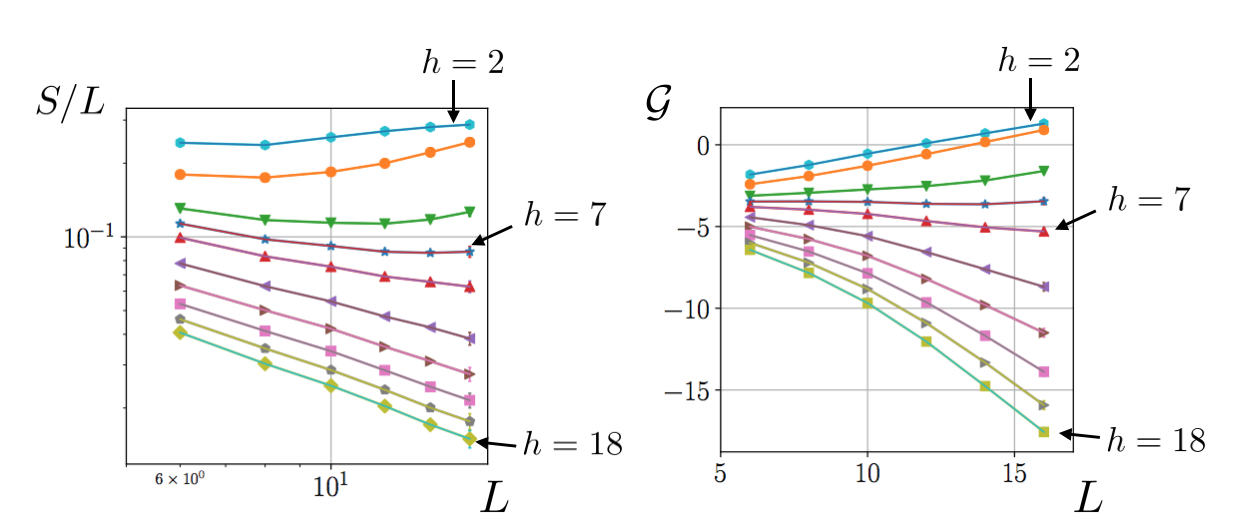}
\caption{
Half-chain entanglement entropy $S/L$ and stability indicator $\mc{G}$ as functions of $L$ for different values of $h$.
The same data in Figs.~3(c) and (d) in the main text are used.
We can identify the MBL transition point ($h_c^\mr{MBL}\simeq 7$) for both quantities.
}
\label{appfigzu}
\end{center}
\end{figure}

In the main text, we have demonstrated the results of the entanglement entropy $S/L$ and the stability indicator $\mc{G}$ as functions of $h$ for different $L$.
In Fig.~\ref{appfigzu}, we show $S/L$ and $\mc{G}$ as functions of $L$ for different $h$.
We can read out the MBL transition point ($h_c^\mr{MBL}\simeq 7$) by the change from the volume law ($S/L\sim \mr{const}$) to the area law ($S/L\propto 1/L$) for the entanglement entropy and from $\mc{G}\sim \alpha L $ to $\sim -\beta L$ ($\alpha,\beta>0$) for the stability indicator.

\subsection{Standard deviation of entanglement entropy}
\begin{figure}
\begin{center}
\includegraphics[width=\linewidth]{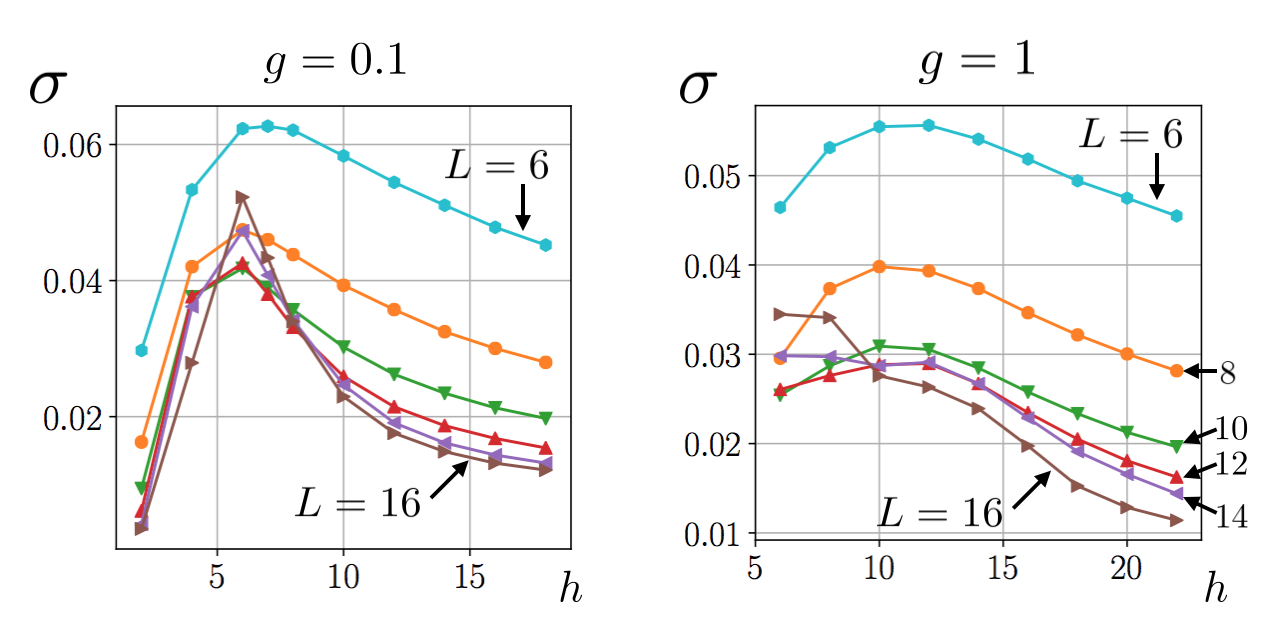}
\caption{
Standard deviation $\sigma$ of the half-chain entanglement entropy per system size over disorder realizations as a function of $h$ for different values of $L$ ($=6,8,10,12,14,16$).
While  we find a peak at a disorder strength $h\simeq 6$ for $g=0.1$, which is not far from the MBL transition point, the peak vanishes for a stronger non-Hermiticity $g=1$.
We use $N_{s} = 10000$ for $L = 6, 8$, $N_{s} = 1000$ for $L = 10$, and $N_{s} = 100$ for $L=12$.
}
\label{appfigvar}
\end{center}
\end{figure}

We here show the standard deviation $\sigma$ of the entanglement entropy per system size $S/L$  over disorder realizations, which is known to have a peak at the critical point for the Hermitian case $g=0$~\cite{Kjall14}. 
As shown in Fig.~\ref{appfigvar}, for weak non-Hermiticity ($g=0.1$), we find a peak at a disorder strength that is not far from the MBL transition point. 
On the other hand, for the stronger non-Hermiticity ($g=1$), we find that no peak arises, in contrast with the Hermitian case. 
The investigation of the origin of this unusual behavior of the standard deviation for the large non-Hermiticity is beyond the scope of the present work, but it implies that the standard deviation of the entanglement entropy cannot be used as a probe of the MBL transition especially for large non-Hermiticity.

\section{Dynamics in the non-Hermitian system with asymmetric hopping}
We address the nonequilibrium dynamics of the non-Hermitian model described in Eq.~(1) in the main text.
We assume that the dynamics of the non-Hermitian Hamiltonian is described by
\aln{
\ket{\psi(t)}=\frac{e^{-i\hat{H}t}\ket{\psi_0}}{||e^{-i\hat{H}t}\ket{\psi_0}||},
}
which corresponds to the dynamics of quantum trajectories with the no-jump condition for continuously measured quantum systems~\cite{Daley14}.
Here the initial state $\ket{\psi_0}$ is taken as a charge-density-wave state
\aln{
\ket{\psi_0}=\ket{1010\cdots10}.
}

\begin{figure}
\begin{center}
\includegraphics[width=8cm]{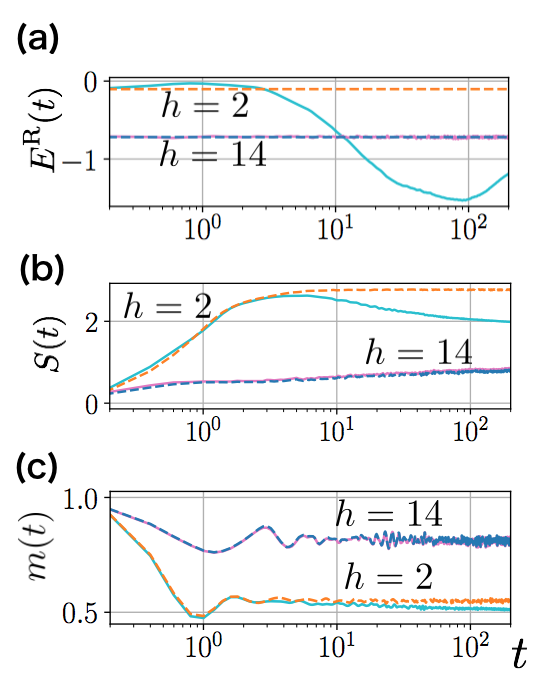}
\caption{
(a) Time evolution of the real part of the energy $E^\mr{R}(t)$ in Eq.~(S-\ref{renergy}) for different values of $g$.
For the Hermitian case ($g=0$, dashed lines), $E^\mr{R}(t)$ is constant due to the energy conservation.
For weak disorder with $h=2$ and $g=0.1$ (solid lines), $E^\mr{R}(t)$ changes significantly in time.
This means that the system is unstable in the delocalized phase.
For strong disorder with $h=14$ and $g=0.1$, the energy is kept constant except for  negligibly small oscillations.
This means that the system is stable in the localized phase.
(b) Time evolution of the half-chain entanglement entropy $S(t)$ defined in Eq.~(S-\ref{halfe}).
In the delocalized regime with weak disorder $h=2$, $S(t)$ exhibits similar growth in both the Hermitian ($g=0$) and non-Hermitian ($g=0.1$) models for $t\lesssim 4$.
On the other hand, in the longer times ($t\gtrsim 10$), while $S(t)$ saturates in the Hermitian case, it gradually decreases in the non-Hermitian case.
In the localized regime with strong disorder $h=14$, $S(t)$ in non-Hermitian model exhibits a logarithmic growth similar to the Hermitian model in a long-time regime.
This indicates that the non-Hermitian MBL phase is dynamically stable despite its non-Hermiticity, in contrast to the delocalized phase.
(c) Time evolution of the local particle density $m(t)$ in Eq.~(S-\ref{lpd}).
In the delocalized regime with weak disorder $h=2$, $m(t)$ saturates for $g=0$ and keeps decaying for $g=0.1$ in the long-time regime.
In the localized regime with strong disorder $h=14$, $m(t)$ behaves similarly for $g=0$ and $g=0.1$ with a smaller decay.
All the data show the averages over $N_s=100$ samples for the system with  $L=12$.
}
\label{appfig5}
\end{center}
\end{figure}

Figure~\ref{appfig5}(a) shows the real part of the energy discussed in Fig. 2 in the main text:
\aln{\label{renergy}
E^\mr{R}(t)=\mr{Re}[\av{\braket{\psi(t)|\hat{H}|\psi(t)}}],
}
where $\hat{H}$ is the non-Hermitian Hamiltonian in Eq.~(1) in the main text and $\av{\cdots}$ denotes the disorder average.
We first note that $E^\mr{R}(t)$ is constant due to the energy conservation for the Hermitian case ($g=0$).
On the other hand, $E^\mr{R}(t)$ is no longer constant for weak disorder with $h=2$ in the presence of the non-Hermitian perturbation, however small it may be ($g=0.1$).
This means that the system is unstable in the delocalized phase against energy absorption and emission, which is due to the delocalized eigenstates with nonzero imaginary parts.
However, for strong disorder with $h=14$, the energy is kept constant except for  negligibly small oscillations.
This means that the system is stable in the localized phase due to the reality of  almost all eigenenergies.

Next, Fig.~\ref{appfig5}(b) (also shown in Fig. 2 in the main text) shows the time evolution of the half-chain entanglement entropy
\aln{\label{halfe}
S(t):=\av{\Tr_{L/2}[\ket{\psi(t)}\bra{\psi(t)}]}.
}
In the delocalized regime with weak disorder $h=2$, $S(t)$ grows similarly for  both the Hermitian ($g=0$) and non-Hermitian ($g=0.1$) models for $t\lesssim 4$.
On the other hand, in the longer time ($t\gtrsim 10$), while $S(t)$ saturates in the Hermitian case, it gradually decreases in the non-Hermitian case.
The decay again shows that this non-Hermitian model is unstable due to the nonzero imaginary parts of eigenenergies.
In the localized regime with strong disorder $h=14$, $S(t)$  exhibits a logarithmic growth similar to the Hermitian model~\cite{Zunidaric08,Bardarson12} for a long time.
This again indicates that the non-Hermitian MBL phase is dynamically stable despite its non-Hermiticity, in contrast to the delocalized phase.

Finally, we show the dynamics of the local particle density in Fig.~\ref{appfig5}(c):
\aln{\label{lpd}
m(t)=\av{\braket{\psi(t)|\hat{n}_1|\psi(t)}}.
}
In the delocalized regime with weak disorder $h=2$, $m(t)$ saturates for $g=0$ and  decays for a long time for $g=0.1$.
On the other hand, in the localized regime with strong disorder $h=14$, $m(t)$ behaves similarly for $g=0$ and $g=0.1$.
Note that the stationary value of $m(t)$ depends sensitively on the initial states in this case due to the presence of local conserved quantities~\cite{Serbyn13,Huse14}.

\section{Perturbation theory for non-Hermitian Hamiltonians}
Here we justify the measure of eigenstate stability introduced in the main text, i.e.,
\aln{\label{sihyounh}
\mc{G}=\av{\ln\frac{|\braket{\mc{E}^L_{a+1}|\hat{V}_\mr{NH}|\mc{E}^R_{a}}|}{|\mc{E}'_{a+1}-\mc{E}'_a|}},
}
where $\ket{\mc{E}_a^R}\:(\ket{\mc{E}_a^L})$ is a right (left) eigenstate of $\hat{H}_0$ and $\mc{E}'_a=\mc{E}_a+{\braket{\mc{E}_a^L|\hat{V}_\mr{NH}|\mc{E}_a^R}}$ is a shifted eigenenergy of $\hat{H}_0$.
This measure  is justified from the perturbation theory as follows.
We consider the original Hamiltonian as
\aln{
\hat{H}=\hat{H}_0+\hat{V}_\mr{NH}={\sum_a(\mc{E}_a+\Delta E_a^{(0)})\ket{\mc{E}_a^R}\bra{\mc{E}_a^L} }+{\hat{V}_\mr{NH}-\sum_a\Delta E_a^{(0)}\ket{\mc{E}_a^R}\bra{\mc{E}_a^L}},
}
where $\Delta E_a^{(0)}={\braket{\mc{E}_a^L|\hat{V}_\mr{NH}|\mc{E}_a^R}}.$
We treat the second and third terms as a perturbation.
Then, the first-order perturbation about the eigenstate leads to
\aln{
\ket{\mc{E}_{a,\mr{perturbed}}^R}=\ket{\mc{E}_{a}^R}+\sum_{b(\neq a)}\frac{\braket{\mc{E}^L_{b}|\hat{V}_\mr{NH}|\mc{E}^R_{a}}}{\mc{E}'_{b}-\mc{E}'_a}\ket{\mc{E}_{b}^R}.
}
The absolute value of the coefficient in the second term should be sufficiently small for the perturbation series to converge.
We take $b=a+1$ to discuss the localization transition~\cite{Serbyn15} and define it as a measure of the stability of eigenstates as in the main text.

Note that $\mc{G}$ is a generalization of the stability measure for the Hermitian setting introduced in Ref.~\cite{Serbyn15}.
They consider the stability of eigenstates of a Hermitian $\hat{H}_0$ against a Hermitian perturbation $\hat{V}$ with its measure
$
\mc{G}_\mr{H}=\av{\ln\frac{|\braket{\mc{E}_{a+1}|\hat{V}|\mc{E}_{a}}|}{|\mc{E}'_{a+1}-\mc{E}'_a|}}.
$
Here $\mc{E}'_a=\mc{E}_a+{\braket{\mc{E}_a|\hat{V}|\mc{E}_a}}$ and the labels of eigenstates (eigenenergies) $a$'s are taken such that $\mc{E}'_1\leq\mc{E}'_2\leq\cdots$.
When $\hat{H}_\mr{0}$ belongs to the delocalized phase, the Srednicki ansatz on the eigenstate thermalization hypothesis (which states that $|\braket{\mc{E}_{a+1}|\hat{V}|\mc{E}_{a}}|\sim e^{-sL/2}$, where $s$ is the entropy density)~\cite{Srednicki99,Khatami13,Beugeling15,Mondaini17,Hamazaki18A} leads to $\mc{G}\sim \frac{s}{2}L$ since $|\mc{E}'_{a+1}-\mc{E}'_a|\simeq e^{-sL}$.
This means that two adjacent eigenstates are mixed by the perturbation.
On the other hand, when $\hat{H}_\mr{0}$ belongs to the MBL phase,
$\ket{\mc{E}_a}$ and $\ket{\mc{E}_{a+1}}$ are characterized by quasi-local conserved quantities~\cite{Serbyn13,Huse14,Imbrie16D,Imbrie16O} and local perturbations cannot mix these two states, resulting in $\mc{G}\sim -\kappa L\: (\kappa>0)$~\cite{Serbyn15}.
As we have seen in the main text, similar behavior can be seen for the stability measure of the non-Hermitian setting $\mc{G}$, which is relevant for the real-complex transition.

\section{Similarity transformation of the non-Hermitian many-body Hamiltonian with asymmetric hopping}
The absence of coalescence between adjacent eigenstates due to the non-Hermitian MBL does not necessarily lead to complete suppression of complex eigenenergies.
In fact, due to statistical fluctuations of disorder, some (but rare) spatial regions are susceptible (resonant) to local perturbations~\cite{Vosk15,Potter15,Imbrie16D,Imbrie16O}, leading to a mixing of non-adjacent eigenstates and the formation of complex-conjugate pairs.
In such resonant regions, non-Hermitian effects should be treated non-perturbatively.

On the other hand, for our model in Eq.~(1) in the main text, we can perform a similarity transformation~\cite{Hatano96,Hatano97,Hatano98,Longhi15,Longhi15R} such that the non-Hermitian perturbation only acts on nonresonant regions.
Through investigation of this transformed Hamiltonian, the mixing of eigenstates is found to be suppressed even for non-adjacent eigenstates and hence the entirely real spectrum emerges.
Below we explain these points in detail.

For simplicity, we focus on the Hamiltonian $\hat{H}$ in Eq.~(1) in the main text for  sufficiently small $g$.
We consider $\hat{H}=\hat{H}_0+\hat{V}_\mr{NH}$, where
\aln{
\hat{H}_0=\sum_{i=1}^L\lrl{-J(\hb_{i+1}^\dag\hb_i+\mr{h.c.})+U\hat{n}_i\hat{n}_{i+1}+h_i\hat{n}_i}
}
is Hermitian and
\aln{
\hat{V}_\mr{NH}=-J\sum_{i=1}^L\lrl{(e^{-g}-1)\hb_{i+1}^\dag\hb_i+(e^{g}-1)\hb_{i}^\dag\hb_{i+1}}=\sum_{i=1}^L\hat{v}_{i,i+1}
}
is non-Hermitian for $g\neq0$.
In the localized phase of $\hat{H}_0$ with large $h$, the local tunneling amplitude $|\braket{\mc{E}_b|\hat{v}_{i,i+1}|\mc{E}_a}|$ by the perturbation $\hat{v}_{i,i+1}$ is smaller than $|{\mc{E}_a}-{\mc{E}_b}|$ for most $i$'s, where we assume that the two eigenstates $\ket{\mc{E}_a}$ and $\ket{\mc{E}_b}$ are approximately product states and connected by the hopping terms $\hb_{i+1}^\dag\hb_i$ ($\hb_{i}^\dag\hb_{i+1}$).
This is because moving particles in the localized regions is energetically costly (i.e., $|{\mc{E}_a}-{\mc{E}_b}|\sim |h_{i+1}-h_i|$ is sufficiently large).
Thus, the effect of $\hat{v}_{i,i+1}$ can be treated perturbatively.
On the other hand, for some (but rare) $i$'s, localization becomes very weak due to the statistical fluctuation of $h_i$ and particles are relatively mobile, leading to
$|\braket{\mc{E}_b|\hat{v}_{i,i+1}|\mc{E}_a}|> |{\mc{E}_a}-{\mc{E}_b}|$.
Thus, for the original model, we cannot control the perturbation due to such resonant $i$'s.

Fortunately,
we can show that the Hamiltonian in Eq.~(1) in the main text can be transformed into a matrix whose non-Hermitian perturbation only acts on the non-resonant regions.
To see this, we consider $\hat{\mc{V}}_i=e^{g\theta_i\hat{n}_i}=1+(e^{g\theta_i}-1)\hat{n}_i$ (similar transformations called the imaginary gauge transformation are used in Refs.~\cite{Hatano96,Hatano97,Hatano98}).
Then
\aln{
\hat{\mc{V}}_i\hat{b}_i\hat{\mc{V}}_i^{-1}&=\hat{b}_i[1+(e^{-g\theta_i}-1)\hat{n}_i]\nonumber\\
&=\hat{b}_i+(e^{-g\theta_i}-1)(1-2\hat{n}_i)\hat{b}_i\nonumber\\
&=e^{-g\theta_i}\hat{b}_i,
}
where we have used $\{\hat{b}_i,\hat{b}_i^\dag\}=1$ and $\hat{b}_i^2=0$.
Similarly,
\aln{
\hat{\mc{V}}_i\hat{b}_i^\dag\hat{\mc{V}}_i^{-1}&=[1+(e^{g\theta_i}-1)\hat{n}_i]\hat{b}_i^\dag\nonumber\\
&=\hat{b}_i^\dag+(e^{g\theta_i}-1)\hat{b}_i^\dag(1-2\hat{n}_i)\nonumber\\
&=e^{g\theta_i}\hat{b}_i^\dag.
}
Thus we obtain
\aln{
\hat{\mc{V}}_{i+1}\hat{\mc{V}}_i\hat{b}_{i+1}^\dag\hb_i\hat{\mc{V}}_i^{-1}\hat{\mc{V}}_{i+1}^{-1}&=e^{g(\theta_{i+1}-\theta_i)}\hat{b}_{i+1}^\dag\hb_i,\\
\hat{\mc{V}}_{i+1}\hat{\mc{V}}_i\hat{b}_{i}^\dag\hb_{i+1}\hat{\mc{V}}_i^{-1}\hat{\mc{V}}_{i+1}^{-1}&=e^{-g(\theta_{i+1}-\theta_i)}\hat{b}_{i}^\dag\hb_{i+1},\\
\hat{\mc{V}}_i\hat{n}_i\hat{\mc{V}}_i^{-1}&=\hat{n}_{i},\\
\hat{\mc{V}}_{i+1}\hat{\mc{V}}_i\hat{n}_{i+1}\hat{n}_i\hat{\mc{V}}_i^{-1}\hat{\mc{V}}_{i+1}^{-1}&=\hat{n}_{i+1}\hat{n}_i.
}

Now, we consider a similarity transformation $\hat{H}'=\hat{\mc{V}}\hat{H}\hat{\mc{V}}^{-1}$, where $\hat{\mc{V}}=\bigotimes_{i=1}^L\hat{\mc{V}}_i$.
Note that $\hat{H}'$ and $\hat{H}$ have the same eigenenergies.
In fact, the eigenstate $\ket{E_\alpha^R}$ of $\hat{H}$ with the eigenenergy $E_\alpha$ is obtained from $\ket{E_\alpha^R}'$ of $\hat{H}'$ with the same eigenenergies  as $\ket{E_\alpha^R}=\hat{\mc{V}}^{-1}\ket{E_\alpha^R}'$,
since,
\aln{
\hat{H}\ket{E_\alpha^R}=\hat{H}\hat{\mc{V}}^{-1}\ket{E_\alpha^R}'=\hat{\mc{V}}^{-1}\hat{H}'\ket{E_\alpha^R}'=E_\alpha\hat{\mc{V}}^{-1}\ket{E_\alpha^R}'=E_\alpha\ket{E_\alpha^R}.
}
Then, we can investigate the energy spectrum of $\hat{H}'$ instead of $\hat{H}$.

By choosing $\theta_i$'s appropriately, we can obtain $\hat{H}'$ for which the non-Hermitian perturbations only act on non-resonant regions.
To see this, we consider a simplified situation, where sites from 1 to $x$ may have resonant regions and other sites are non-resonant.
We can assume that $x$ is much smaller than $L$ because resonant regions are rare in the localized phase~\footnote{Some work~\cite{Potter15,Khemani17} proposed that the resonant regions spread over the whole sites, even when they are rare. We expect that our discussion is qualitatively valid in that situation as well.}.
If we choose
\aln{
\theta_i&=i\:\:\:(1\leq i\leq x+1),\nonumber\\
\theta_i&=-i+2x+2\:\:\:(x+2\leq i\leq 2x+2),\nonumber\\
\theta_i&=0\:\:\:(2x+3\leq i\leq L),
}
then
\aln{
\hat{H}'=&
\hat{\mc{V}}\hat{H}\hat{\mc{V}}^{-1}\nonumber\\
=&-J\sum_{i=1}^{L}(e^{-gz_i}\hb_{i+1}^\dag\hb_i+e^{gz_i}\hb_{i}^\dag\hb_{i+1})
+\sum_{i=1}^LU\hat{n}_i\hat{n}_{i+1}+\sum_{i=1}^Lh_i\hat{n}_i\nonumber\\
=&\hat{H}_0+\hat{V}_\mr{NH}',
}
where
\aln{
z_i&=0\:\:\:(1\leq i\leq x,i=L),\nonumber\\
z_i&=2\:\:\:(x+2\leq i\leq 2x+1),\nonumber\\
z_i&=1\:\:\:(2x+2\leq i\leq L-1).
}
Since $\hat{V}_\mr{NH}'$ acts only on the non-resonant regions, it does not mix the eigenstates, leading to further suppression of complex eigenenergies that are, in general, non-adjacent.

\bibliography{../../../../refer_them}